\documentclass[longauth]{aa}
\usepackage{euclid}
\usepackage{graphicx}
\usepackage{txfonts}
\usepackage{natbib}
\usepackage{hyperref} \hypersetup{ colorlinks=true, linkcolor=blue, filecolor=magenta, urlcolor=blue, citecolor=blue}

\usepackage{diagbox}

\newcommand{\SE}{\texttt{SourceXtractor++\ }}
\newcommand{\SEdot}{\texttt{SourceXtractor++}}
\newcommand{\metryka}{\texttt{Morfometryka\ }}
\newcommand{\metrykadot}{\texttt{Morfometryka}}
\newcommand{\profit}{\texttt{ProFit\ }}
\newcommand{\profitdot}{\texttt{ProFit}}
\newcommand{\gala}{\texttt{Galapagos-2\ }}
\newcommand{\galadot}{\texttt{Galapagos-2}}
\newcommand{\deepleg}{\texttt{DeepLeGATo\ }}
\newcommand{\deeplegdot}{\texttt{DeepLeGATo}}
\newcommand{\VIS}{I_{\scriptscriptstyle\rm E} }
\newcommand{\VISdot}{I_{\scriptscriptstyle\rm E}}
\newcommand{\sersic}{S\'ersic\ }
\newcommand{\sersicdot}{S\'ersic}
\newcommand{\partonedot}{EMC2022a}
\newcommand{\partone}{EMC2022a\ }
\newcommand{\orcid}[1]{\unskip\protect\href{https://orcid.org/#1}{\protect\includegraphics[width=8pt,clip]{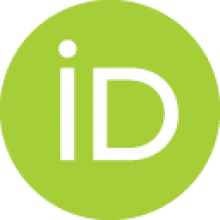}}}
\newcommand{\re}{$r_{\mathrm{e}}$\ }
\newcommand{\redot}{$r_{\mathrm{e}}$}
\newcommand{\ba}{$q$\ }

\newcommand{\badot}{$q$} 

\newcommand{\bt}{$\mathrm{b}/\mathrm{t}$\ }
\newcommand{\btdot}{$\mathrm{b}/\mathrm{t}$}

\begin{document}

\title{\Euclid preparation. XXVI. The \Euclid Morphology Challenge. Towards structural parameters for billions of galaxies}


\author{\normalsize{Euclid Collaboration: H.~Bretonni\`ere$^{1,2}\orcid{0000-0001-9935-9109}$\thanks{\email{hubert.bretonniere@ias.u-psud.fr}}, U.~Kuchner\orcid{0000-0002-0035-5202}$^{3}$, M.~Huertas-Company\orcid{0000-0002-1416-8483}$^{4,5,6,7}$, E.~Merlin\orcid{0000-0001-6870-8900}$^{8}$, M.~Castellano\orcid{0000-0001-9875-8263}$^{8}$, D.~Tuccillo$^{9}$, F.~Buitrago\orcid{0000-0002-2861-9812}$^{10,11}$, C.J.~Conselice$^{12}$, A.~Boucaud\orcid{0000-0001-7387-2633}$^{2}$, B.~H\"au\ss ler\orcid{0000-0002-1857-2088}$^{13}$
% ...
}}

\institute{$^{1}$ Universit\'e Paris-Saclay, CNRS, Institut d'astrophysique spatiale, 91405, Orsay, France\\
$^{2}$  Universit\'e Paris Cit\'e, CNRS, Astroparticule et Cosmologie, F-75013 Paris, France\\
$^{3}$ School of Physics and Astronomy, University of Nottingham, University Park, Nottingham NG7 2RD, UK\\
$^{4}$ Universit\'e Paris-Cit\'e, 5 Rue Thomas Mann, 75013, Paris, France\\
$^{5}$ Universit\'e PSL, Observatoire de Paris, Sorbonne Universit\'e, CNRS, LERMA, F-75014, Paris, France\\
$^{6}$ Instituto de Astrof\'isica de Canarias (IAC); Departamento de Astrof\'isica, Universidad de La Laguna (ULL), E-38200, La Laguna, Tenerife, Spain\\
$^{7}$ Instituto de Astrof\'isica de Canarias, Calle V\'ia L\'actea s/n, E-38204, San Crist\'obal de La Laguna, Tenerife, Spain\\
$^{8}$ INAF-Osservatorio Astronomico di Roma, Via Frascati 33, I-00078 Monteporzio Catone, Italy\\
$^{9}$ Instituto de F\'isica de Cantabria, Edificio Juan Jord\'a, Avenida de los Castros, E-39005 Santander, Spain\\
$^{10}$ Departamento de F\'{i}sica Te\'{o}rica, At\'{o}mica y \'{O}ptica, Universidad de Valladolid, 47011 Valladolid, Spain\\
$^{11}$ Instituto de Astrof\'isica e Ci\^encias do Espa\c{c}o, Faculdade de Ci\^encias, Universidade de Lisboa, Tapada da Ajuda, PT-1349-018 Lisboa, Portugal\\
$^{12}$ Jodrell Bank Centre for Astrophysics, Department of Physics and Astronomy, University of Manchester, Oxford Road, Manchester M13 9PL, UK\\
$^{13}$ European Southern Observatory, Alonso de Cordova 3107, Casilla 19001, Santiago, Chile\\
$^{14}$ Universit\"ats-Sternwarte M\"unchen, Fakult\"at f\"ur Physik, Ludwig-Maximilians-Universit\"at M\"unchen, Scheinerstrasse 1, 81679 M\"unchen, Germany\\
$^{15}$ Department of Astronomy, University of Geneva, ch. d\'Ecogia 16, CH-1290 Versoix, Switzerland\\
$^{16}$ Institut d'Astrophysique de Paris, UMR 7095, CNRS, and Sorbonne Universit\'e, 98 bis boulevard Arago, 75014 Paris, France\\
$^{17}$ Canada-France-Hawaii Telescope, 65-1238 Mamalahoa Hwy, Kamuela, HI 96743, USA\\
$^{18}$ Instituto de Matem\'{a}tica Estat\'{i}stica e F\'{i}sica, Universidade Federal do Rio Grande, 96203-900, Rio Grande, RS, Brazil\\
$^{19}$ University of Nottingham, University Park, Nottingham NG7 2RD, UK\\
$^{20}$ Aix-Marseille Univ, CNRS, CNES, LAM, Marseille, France\\
$^{21}$ ICRAR, M468, University of Western Australia, Crawley, WA 6009, Australia\\
$^{22}$ SRON Netherlands Institute for Space Research, Landleven 12, 9747 AD, Groningen, The Netherlands\\
$^{23}$ Kapteyn Astronomical Institute, University of Groningen, PO Box 800, 9700 AV Groningen, The Netherlands\\
$^{24}$ Institut de Recherche en Astrophysique et Plan\'etologie (IRAP), Universit\'e de Toulouse, CNRS, UPS, CNES, 14 Av. Edouard Belin, F-31400 Toulouse, France\\
$^{25}$ Centro de Estudios de F\'isica del Cosmos de Arag\'on (CEFCA), Plaza San Juan, 1, planta 2, E-44001, Teruel, Spain\\
$^{26}$ Universit\'{e} de Strasbourg, CNRS, Observatoire astronomique de Strasbourg, UMR 7550, F-67000 Strasbourg, France\\
$^{27}$ School of Physics, HH Wills Physics Laboratory, University of Bristol, Tyndall Avenue, Bristol, BS8 1TL, UK\\
$^{28}$ Max Planck Institute for Extraterrestrial Physics, Giessenbachstr. 1, D-85748 Garching, Germany\\
$^{29}$ Institut f\"ur Astro- und Teilchenphysik, Universit\"at Innsbruck, Technikerstr. 25/8, 6020 Innsbruck, Austria\\
$^{30}$ INAF-Osservatorio Astronomico di Capodimonte, Via Moiariello 16, I-80131 Napoli, Italy\\
$^{31}$ Institute of Cosmology and Gravitation, University of Portsmouth, Portsmouth PO1 3FX, UK\\
$^{32}$ INAF-Osservatorio di Astrofisica e Scienza dello Spazio di Bologna, Via Piero Gobetti 93/3, I-40129 Bologna, Italy\\
$^{33}$ Mullard Space Science Laboratory, University College London, Holmbury St Mary, Dorking, Surrey RH5 6NT, UK\\
$^{34}$ Dipartimento di Fisica e Astronomia "Augusto Righi" - Alma Mater Studiorum Universit\`{a} di Bologna, via Piero Gobetti 93/2, I-40129 Bologna, Italy\\
$^{35}$ INFN-Sezione di Bologna, Viale Berti Pichat 6/2, I-40127 Bologna, Italy\\
$^{36}$ Dipartimento di Fisica, Universit\`{a} di Genova, Via Dodecaneso 33, I-16146, Genova, Italy\\
$^{37}$ INFN-Sezione di Roma Tre, Via della Vasca Navale 84, I-00146, Roma, Italy\\
$^{38}$ Department of Physics "E. Pancini", University Federico II, Via Cinthia 6, I-80126, Napoli, Italy\\
$^{39}$ Instituto de Astrof\'isica e Ci\^encias do Espa\c{c}o, Universidade do Porto, CAUP, Rua das Estrelas, PT4150-762 Porto, Portugal\\
$^{40}$ Dipartimento di Fisica, Universit\'a degli Studi di Torino, Via P. Giuria 1, I-10125 Torino, Italy\\
$^{41}$ INFN-Sezione di Torino, Via P. Giuria 1, I-10125 Torino, Italy\\
$^{42}$ INAF-Osservatorio Astrofisico di Torino, Via Osservatorio 20, I-10025 Pino Torinese (TO), Italy\\
$^{43}$ INAF-IASF Milano, Via Alfonso Corti 12, I-20133 Milano, Italy\\
$^{44}$ Institut de F\'{i}sica d'Altes Energies (IFAE), The Barcelona Institute of Science and Technology, Campus UAB, 08193 Bellaterra (Barcelona), Spain\\
$^{45}$ Port d'Informaci\'{o} Cient\'{i}fica, Campus UAB, C. Albareda s/n, 08193 Bellaterra (Barcelona), Spain\\
$^{46}$ Institut d'Estudis Espacials de Catalunya (IEEC), Carrer Gran Capit\'a 2-4, 08034 Barcelona, Spain\\
$^{47}$ Institute of Space Sciences (ICE, CSIC), Campus UAB, Carrer de Can Magrans, s/n, 08193 Barcelona, Spain\\
$^{48}$ INFN section of Naples, Via Cinthia 6, I-80126, Napoli, Italy\\
$^{49}$ Dipartimento di Fisica e Astronomia "Augusto Righi" - Alma Mater Studiorum Universit\'a di Bologna, Viale Berti Pichat 6/2, I-40127 Bologna, Italy\\
$^{50}$ INAF-Osservatorio Astrofisico di Arcetri, Largo E. Fermi 5, I-50125, Firenze, Italy\\
$^{51}$ Centre National d'Etudes Spatiales, Toulouse, France\\
$^{52}$ Institut national de physique nucl\'eaire et de physique des particules, 3 rue Michel-Ange, 75794 Paris C\'edex 16, France\\
$^{53}$ Institute for Astronomy, University of Edinburgh, Royal Observatory, Blackford Hill, Edinburgh EH9 3HJ, UK\\
$^{54}$ ESAC/ESA, Camino Bajo del Castillo, s/n., Urb. Villafranca del Castillo, 28692 Villanueva de la Ca\~nada, Madrid, Spain\\
$^{55}$ European Space Agency/ESRIN, Largo Galileo Galilei 1, 00044 Frascati, Roma, Italy\\
$^{56}$ Univ Lyon, Univ Claude Bernard Lyon 1, CNRS/IN2P3, IP2I Lyon, UMR 5822, F-69622, Villeurbanne, France\\
$^{57}$ Institute of Physics, Laboratory of Astrophysics, Ecole Polytechnique F\'{e}d\'{e}rale de Lausanne (EPFL), Observatoire de Sauverny, 1290 Versoix, Switzerland\\
$^{58}$ Departamento de F\'isica, Faculdade de Ci\^encias, Universidade de Lisboa, Edif\'icio C8, Campo Grande, PT1749-016 Lisboa, Portugal\\
$^{59}$ Instituto de Astrof\'isica e Ci\^encias do Espa\c{c}o, Faculdade de Ci\^encias, Universidade de Lisboa, Campo Grande, PT-1749-016 Lisboa, Portugal\\
$^{60}$ Department of Physics, Oxford University, Keble Road, Oxford OX1 3RH, UK\\
$^{61}$ INFN-Padova, Via Marzolo 8, I-35131 Padova, Italy\\
$^{62}$ Universit\'e Paris-Saclay, Universit\'e Paris Cit\'e, CEA, CNRS, Astrophysique, Instrumentation et Mod\'elisation Paris-Saclay, 91191 Gif-sur-Yvette, France\\
$^{63}$ INAF-Osservatorio Astronomico di Trieste, Via G. B. Tiepolo 11, I-34143 Trieste, Italy\\
$^{64}$ Istituto Nazionale di Astrofisica (INAF) - Osservatorio di Astrofisica e Scienza dello Spazio (OAS), Via Gobetti 93/3, I-40127 Bologna, Italy\\
$^{65}$ Istituto Nazionale di Fisica Nucleare, Sezione di Bologna, Via Irnerio 46, I-40126 Bologna, Italy\\
$^{66}$ INAF-Osservatorio Astronomico di Padova, Via dell'Osservatorio 5, I-35122 Padova, Italy\\
$^{67}$ Institute of Theoretical Astrophysics, University of Oslo, P.O. Box 1029 Blindern, N-0315 Oslo, Norway\\
$^{68}$ Leiden Observatory, Leiden University, Niels Bohrweg 2, 2333 CA Leiden, The Netherlands\\
$^{69}$ Jet Propulsion Laboratory, California Institute of Technology, 4800 Oak Grove Drive, Pasadena, CA, 91109, USA\\
$^{70}$ von Hoerner \& Sulger GmbH, Schlo{\ss}Platz 8, D-68723 Schwetzingen, Germany\\
$^{71}$ Technical University of Denmark, Elektrovej 327, 2800 Kgs. Lyngby, Denmark\\
$^{72}$ Institut d'Astrophysique de Paris, 98bis Boulevard Arago, F-75014, Paris, France\\
$^{73}$ Max-Planck-Institut f\"ur Astronomie, K\"onigstuhl 17, D-69117 Heidelberg, Germany\\
$^{74}$ Aix-Marseille Univ, CNRS/IN2P3, CPPM, Marseille, France\\
$^{75}$ Universit\'e de Gen\`eve, D\'epartement de Physique Th\'eorique and Centre for Astroparticle Physics, 24 quai Ernest-Ansermet, CH-1211 Gen\`eve 4, Switzerland\\
$^{76}$ Department of Physics and Helsinki Institute of Physics, Gustaf H\"allstr\"omin katu 2, 00014 University of Helsinki, Finland\\
$^{77}$ NOVA optical infrared instrumentation group at ASTRON, Oude Hoogeveensedijk 4, 7991PD, Dwingeloo, The Netherlands\\
$^{78}$ Argelander-Institut f\"ur Astronomie, Universit\"at Bonn, Auf dem H\"ugel 71, 53121 Bonn, Germany\\
$^{79}$ Department of Physics, Institute for Computational Cosmology, Durham University, South Road, DH1 3LE, UK\\
$^{80}$ University of Applied Sciences and Arts of Northwestern Switzerland, School of Engineering, 5210 Windisch, Switzerland\\
$^{81}$ European Space Agency/ESTEC, Keplerlaan 1, 2201 AZ Noordwijk, The Netherlands\\
$^{82}$ Department of Physics and Astronomy, University of Aarhus, Ny Munkegade 120, DK-8000 Aarhus C, Denmark\\
$^{83}$ Centre for Astrophysics, University of Waterloo, Waterloo, Ontario N2L 3G1, Canada\\
$^{84}$ Department of Physics and Astronomy, University of Waterloo, Waterloo, Ontario N2L 3G1, Canada\\
$^{85}$ Perimeter Institute for Theoretical Physics, Waterloo, Ontario N2L 2Y5, Canada\\
$^{86}$ Space Science Data Center, Italian Space Agency, via del Politecnico snc, 00133 Roma, Italy\\
$^{87}$ Departamento de Astrof\'{i}sica, Universidad de La Laguna, E-38206, La Laguna, Tenerife, Spain\\
$^{88}$ Dipartimento di Fisica e Astronomia "G.Galilei", Universit\'a di Padova, Via Marzolo 8, I-35131 Padova, Italy\\
$^{89}$ Departamento de F\'isica, FCFM, Universidad de Chile, Blanco Encalada 2008, Santiago, Chile\\
$^{90}$ Centre for Electronic Imaging, Open University, Walton Hall, Milton Keynes, MK7~6AA, UK\\
$^{91}$ AIM, CEA, CNRS, Universit\'{e} Paris-Saclay, Universit\'{e} de Paris, F-91191 Gif-sur-Yvette, France\\
$^{92}$ Centro de Investigaciones Energ\'eticas, Medioambientales y Tecnol\'ogicas (CIEMAT), Avenida Complutense 40, 28040 Madrid, Spain\\
$^{93}$ Universidad Polit\'ecnica de Cartagena, Departamento de Electr\'onica y Tecnolog\'ia de Computadoras, 30202 Cartagena, Spain\\
$^{94}$ Infrared Processing and Analysis Center, California Institute of Technology, Pasadena, CA 91125, USA\\
$^{95}$ INAF-Osservatorio Astronomico di Brera, Via Brera 28, I-20122 Milano, Italy\\
$^{96}$ Junia, EPA department, F 59000 Lille, France\\
$^{97}$ SISSA, International School for Advanced Studies, Via Bonomea 265, I-34136 Trieste TS, Italy\\
$^{98}$ IFPU, Institute for Fundamental Physics of the Universe, via Beirut 2, 34151 Trieste, Italy\\
$^{99}$ INFN, Sezione di Trieste, Via Valerio 2, I-34127 Trieste TS, Italy\\
$^{100}$ Dipartimento di Fisica e Scienze della Terra, Universit\'a degli Studi di Ferrara, Via Giuseppe Saragat 1, I-44122 Ferrara, Italy\\
$^{101}$ Istituto Nazionale di Fisica Nucleare, Sezione di Ferrara, Via Giuseppe Saragat 1, I-44122 Ferrara, Italy\\
$^{102}$ Institut de Physique Th\'eorique, CEA, CNRS, Universit\'e Paris-Saclay F-91191 Gif-sur-Yvette Cedex, France\\
$^{103}$ Dipartimento di Fisica - Sezione di Astronomia, Universit\'a di Trieste, Via Tiepolo 11, I-34131 Trieste, Italy\\
$^{104}$ NASA Ames Research Center, Moffett Field, CA 94035, USA\\
$^{105}$ INAF, Istituto di Radioastronomia, Via Piero Gobetti 101, I-40129 Bologna, Italy\\
$^{106}$ INFN-Bologna, Via Irnerio 46, I-40126 Bologna, Italy\\
$^{107}$ Universit\'e C\^{o}te d'Azur, Observatoire de la C\^{o}te d'Azur, CNRS, Laboratoire Lagrange, Bd de l'Observatoire, CS 34229, 06304 Nice cedex 4, France\\
$^{108}$ Institute for Theoretical Particle Physics and Cosmology (TTK), RWTH Aachen University, D-52056 Aachen, Germany\\
$^{109}$ Department of Physics \& Astronomy, University of California Irvine, Irvine CA 92697, USA\\
$^{110}$ University of Lyon, UCB Lyon 1, CNRS/IN2P3, IUF, IP2I Lyon, France\\
$^{111}$ INFN-Sezione di Genova, Via Dodecaneso 33, I-16146, Genova, Italy\\
$^{112}$ INAF-Istituto di Astrofisica e Planetologia Spaziali, via del Fosso del Cavaliere, 100, I-00100 Roma, Italy\\
$^{113}$ Instituto de F\'isica Te\'orica UAM-CSIC, Campus de Cantoblanco, E-28049 Madrid, Spain\\
$^{114}$ Department of Physics, P.O. Box 64, 00014 University of Helsinki, Finland\\
$^{115}$ Ruhr University Bochum, Faculty of Physics and Astronomy, Astronomical Institute (AIRUB), German Centre for Cosmological Lensing (GCCL), 44780 Bochum, Germany\\
$^{116}$ Department of Physics, Lancaster University, Lancaster, LA1 4YB, UK\\
$^{117}$ Universit\'{e} Paris-Saclay, CNRS/IN2P3, IJCLab, 91405 Orsay, France\\
$^{118}$ Department of Physics and Astronomy, University College London, Gower Street, London WC1E 6BT, UK\\
$^{119}$ Astrophysics Group, Blackett Laboratory, Imperial College London, London SW7 2AZ, UK\\
$^{120}$ Univ. Grenoble Alpes, CNRS, Grenoble INP, LPSC-IN2P3, 53, Avenue des Martyrs, 38000, Grenoble, France\\
$^{121}$ Dipartimento di Fisica, Sapienza Universit\`a di Roma, Piazzale Aldo Moro 2, I-00185 Roma, Italy\\
$^{122}$ Institut f\"ur Theoretische Physik, University of Heidelberg, Philosophenweg 16, 69120 Heidelberg, Germany\\
$^{123}$ Zentrum f\"ur Astronomie, Universit\"at Heidelberg, Philosophenweg 12, D- 69120 Heidelberg, Germany\\
$^{124}$ Department of Mathematics and Physics E. De Giorgi, University of Salento, Via per Arnesano, CP-I93, I-73100, Lecce, Italy\\
$^{125}$ INAF-Sezione di Lecce, c/o Dipartimento Matematica e Fisica, Via per Arnesano, I-73100, Lecce, Italy\\
$^{126}$ INFN, Sezione di Lecce, Via per Arnesano, CP-193, I-73100, Lecce, Italy\\
$^{127}$ Institute of Space Science, Bucharest, Ro-077125, Romania\\
$^{128}$ Institute for Computational Science, University of Zurich, Winterthurerstrasse 190, 8057 Zurich, Switzerland\\
$^{129}$ Higgs Centre for Theoretical Physics, School of Physics and Astronomy, The University of Edinburgh, Edinburgh EH9 3FD, UK\\
$^{130}$ Universit\'e St Joseph; Faculty of Sciences, Beirut, Lebanon\\
$^{131}$ Department of Astrophysical Sciences, Peyton Hall, Princeton University, Princeton, NJ 08544, USA\\
$^{132}$ Helsinki Institute of Physics, Gustaf H{\"a}llstr{\"o}min katu 2, University of Helsinki, Helsinki, Finland\\
$^{133}$ Department of Mathematics and Physics, Roma Tre University, Via della Vasca Navale 84, I-00146 Rome, Italy\\
$^{134}$ Cosmic Dawn Center (DAWN)\\
$^{135}$ Niels Bohr Institute, University of Copenhagen, Jagtvej 128, 2200 Copenhagen, Denmark\\
$^{136}$ Departement of Physics and Astronomy, University of British Columbia, Vancouver, BC V6T 1Z1, Canada}

\date{}
\keywords{
    Galaxies: structure  --
    Galaxies: evolution --
    Cosmology: observations
}

\titlerunning{\Euclid Morphology Challenge - Structural parameters}
   \authorrunning{H. Bretonni\`ere, U. Kuchner, M. Huertas-Company}

 \abstract{
The various \Euclid imaging surveys will become a reference for studies of galaxy morphology by delivering imaging over an unprecedented area of $15\,000$ square degrees with high spatial resolution. In order to understand the capabilities of measuring morphologies from \textit{Euclid}-detected galaxies and to help implement measurements in the pipeline of the Organisational Unit MER of the Euclid Science Ground Segment, we have conducted the Euclid Morphology Challenge, which we present in two papers. While the companion paper focusses on the analysis of photometry, this paper  assesses the accuracy of the parametric galaxy morphology measurements in imaging predicted from within the Euclid Wide Survey.
We evaluate the performance of five state-of-the-art surface-brightness-fitting codes, \deeplegdot, \galadot, \metrykadot, \profit and \SEdot, on a sample of about $ 1.5~$ million simulated galaxies ($350\,000$ above $5\sigma$) resembling reduced observations with the \Euclid VIS and NIR instruments. The simulations include analytic \sersic profiles with one and two components, as well as more realistic galaxies generated with neural networks.
We find that, despite some code-specific differences, all methods tend to achieve reliable structural measurements ($<10\%$ scatter on ideal \sersic simulations) down to an apparent magnitude of about $\VIS=23$ in one component and $\VIS=21$ in two components, which correspond to a signal-to-noise ratio of approximately $1$ and $5$, respectively. We also show that when tested on non-analytic profiles, the results are typically degraded by a factor of $3$, driven by systematics.
We conclude that the official \Euclid Data Releases will deliver robust structural parameters for at least $400$ million galaxies in the Euclid Wide Survey by the end of the mission. We find that a key factor for explaining the different behaviour of the codes at the faint end is the set of adopted priors for the various structural parameters.}

\maketitle

\section{Introduction}
\label{sec:Intro}

Measurements of galaxy morphology offer easily accessible information for constraining physical processes that regulate galaxy growth and evolution. Galaxy morphologies are therefore among the most important observables available from extragalactic imaging campaigns and continues to be so throughout the era of big data astronomy. This is because the distribution of the stellar light emitted by a galaxy can be correlated to its stellar populations, angular momentum, and the star formation and merger histories (e.g. \citealp{cole2000, conselice2003, kormendy2004, schreiber2009, brennan2017}).

A fundamental goal of extragalactic astronomy is understanding how the diversity of galaxy morphologies is established across time. This is predicated on earlier observations, which already revealed that galaxies come in various types~(e.g. \citealp{1926ApJ....64..321H}). The most fundamental distinction differentiates disc-dominated structures that often appear with bright spiral arms and bulge-dominated galaxies with smooth light distributions. Most galaxies are in fact a combination of both shapes, featuring both a bulge and a disc with varying weights. This simple scheme describes the essential building blocks of nearby galaxies. However, a descriptive classification for grouping galaxies into two rough classes is a simplification, and in reality the visible part of most galaxies result from a combination of multiple components. 

Characterising and classifying galaxies based on their optical morphologies is not straightforward. A number of different approaches for quantifying galaxy structure and morphology have been developed, documented, and tested in the last few decades, each designed with specific applications in mind. The general goal of all of these methods is to obtain a quantitative measurement -- and an error budget -- of the morphological properties of galaxies that are easy to understand, use, quantify, and replicate. Contemporary examples include visual classifications (e.g. \citealp{lintott2008, mortlock2013, bait2017}), non-parametric morphologies (\citealp{conselice2003_CAS, lotz2004, pawlik2016}),  1D intensity profile fitting of a galaxy's light distribution, either treating each galaxy as a whole (e.g. \citealp{sersic1968, peng2002, buitrago2008, buitrago2013}) or decomposing them into two separable components (2D surface brightness fitting, e.g. \citealp{simard2011, lang2014}), machine learning techniques (e.g. \citealp{huertas2008, huertas2011, vega2021}), and structural kinematics (\citealp{schreiber2009, falcon2017, vandesande2017}). The increasingly challenging nature of observations of fainter and more distant galaxies makes defining and distinguishing between different structures a non-trivial task.  Traditional visual classifications also become ambiguous for many objects, especially for early-type galaxies. In addition, techniques need to be able to efficiently deal with the ever increasing sample sizes of galaxies in contemporary and future all-sky surveys, with an increased statistical accuracy. Light profile fitting is a quantitative, generally automatic, or semi-automatic, and often a faster approach, compared to the qualitative visual classification process. This is especially important for statistical approaches using the very large datasets we are expecting with missions such as \Euclid in the near future. 

\Euclid is a European Space Agency $1.2\,\mathrm{m}$ space-based telescope mission, primarily designed to investigate dark energy and dark matter by mapping a large fraction of the visible sky \citep{laureijs2011}. In order to achieve this goal, \Euclid will conduct a Wide Survey of around $1.5$ billion galaxies out to $z\sim3$ with relatively high spatial resolution wide-field optical and near-infrared (NIR) imaging, as well as low-resolution grism spectroscopy ($R \sim 250$). These data will be provided by the VIS instrument, which features one broad optical band called $\VIS$, covering approximately $\SI{540}{\nano\meter}$ to $\SI{900}{\nano\meter}$ (i.e. covering most of the usual $r$, $i$, and $z$ bands), and a mean image quality of $\ang{;;0.17}\,\mathrm{FWHM}$ \citep{cropper2010}. The Euclid Wide Survey will therefore provide a unique combination of high spatial resolution and wide area coverage, enabling studies of galaxy morphology and structure with unprecedented statistics. The uncommonly large wavelength range of the VIS  filter provides unknown effects for determining galaxy morphologies with \Euclid since no previous large studies have used such a wide filter. While this filter was especially designed with \Euclid core cosmological science in mind, it is essential to fully characterise the use of this filter for the measurement of galaxy morphologies. \Euclid's other instrument is the Near Infrared Spectrometer and Photometer (NISP), which will observe in three IR bands, $Y_{\scriptscriptstyle\rm E}$, $J_{\scriptscriptstyle\rm E}$, and $H_{\scriptscriptstyle\rm E}$, covering approximately $950$ to $2020\,\mathrm{nm}$ \citep{schirmer_2022}. 

\Euclid's nominal requirements are to image $15\,000\,\mathrm{deg}^2$ or $35\,\%$ of the accessible sky down to at least a $10\sigma$ depth of magnitude $\VIS=24.5$ in the optical and down to a 5$\sigma$ depth of magnitude $24.3$ at NIR wavelengths ($Y_{\scriptscriptstyle\rm E} = 24.3$, $J_{\scriptscriptstyle\rm E} = 24.5$, and $H_{\scriptscriptstyle\rm E} = 24.4$). Observing strategies and initial tests of the instrument forecast higher sensitivity than the nominal requirements. In addition, the Euclid Deep Survey will provide images two magnitudes deeper in a smaller area of $40\,\mathrm{deg}^{2}$, as part of the deep fields. \Euclid will thus provide an unprecedented number of high spatial resolution images for morphological measurements, which will be an extraordinary database for a range of legacy science questions including galaxy formation and evolution, as well as a plethora of follow-up projects.

The \sersic law \citep{sersic1968} is a commonly used parametric model to describe galaxy radial profiles, which can describe a variety of shapes, from a disc or underlying smooth component of spiral galaxies (\citealp{freeman1970, kormendy1977}), to elliptical galaxies and bulges \citep{devaucouleurs1948}. The practice of fitting the \sersic law to astronomical images of objects has become widely used. Its aim is to measure and quantify the shapes of galaxy profiles (i.e. the surface brightness profile).
The success of \sersic profiling for morphology measurements has been repeatedly shown. For example, massive elliptical galaxies are well described by one-component \sersic profiles (\citealp{graham2003, trujillo2001}) out to around eight effective radii \citep{tal2011}. Deep imaging of large samples of face-on late-type galaxies confirm that this type is well represented by an exponential profile (\sersic profile of $n=1$) down to faint limits of $\mu = 27\,\mathrm{mag}\,\mathrm{arcsec^{-2}}$ \citep{pohlen2006} out to at least $ 17$ effective radii \citep{bland2005}.

Given the large number of galaxies that will be observed by \Euclid, it is essential to obtain a fast and reliable way of measuring morphological parameters of galaxies from images. In order to understand the capabilities of measuring morphologies and structures from \Euclid-detected galaxies, we have created the Euclid Morphology Challenge to test, quantify, and evaluate the performance of galaxy morphology measurements by existing parametric fitting codes on simulated \Euclid data. The structural measurements evaluated in this work are not tailored to a specific science case. Rather, we provide a comparison of the measurements of parameters (\Euclid data products) that will enable astronomers to investigate a range of research questions related to galaxy evolution and morphologies or structures with \Euclid. For example, as \sersic indices are an approximation to statistically distinguish early- from late-type galaxies, probing those indices in a large range of redshift can help us understand morphology evolution. They will also be combined with other parameters of interest such as colour or stellar mass to scrutinise current models. The depth and volume of Euclid will constrain these relations and open a variety of investigations needed to make progress in galaxy evolution science.

The challenge comprises a simulated dataset of five fields, each realised with single-\sersicdot, double-\sersicdot, and neural-network-generated galaxies in the $\VIS$ band. In addition, one of the fields has been simulated in the  NIR ($Y_{\scriptscriptstyle\rm E}$, $J_{\scriptscriptstyle\rm E}$, and $H_{\scriptscriptstyle\rm E}$) bands, and in the five \emph{u, g, r, i,} and \emph{z} Rubin Observatory bands to test the accuracy of multi-band-based model fitting with ancillary data. While Rubin will only cover the southern hemisphere, other facilities such as CFHT (MegaCam) or DES will also cover the northern hemisphere in similar bands. The companion paper (Merlin et al. 2022; hereafter \partonedot) provides a visualisation of the bandwidth and wavelengths (see their Fig.~1).

In this work, we focus on quantifying galaxy structures through analytic functions that describe the shape of the surface brightness profile of each galaxy. The outcome is a set of parameters that allow the reconstruction of the 2D photometric shape of a galaxy, and thus provides important information for the statistical study of galaxy evolution. To carry out this challenge, we have invited a number of developers of widely used software packages to retrieve morphologies and structures from our large dataset of simulated galaxies. Five teams participated in the challenge. Each team tested the performance of their codes on a common set of simulated \Euclid galaxies that was provided to them. The codes are (in alphabetical order) \deepleg \citep{tuccillo2018}, \galadot\footnote{\url{https://github.com/MegaMorph/galapagos}} \citep{haeussler2022}, \metryka \citep{ferrari2015}, \profitdot\footnote{\url{https://github.com/asgr/ProFound}} \citep{robotham2017}, and \SEdot\footnote{\url{https://github.com/astrorama/SourceXtractorPlusPlus}}
\citep{SE1, SE2}. At their cores, all of the software packages describe the morphology or structure of each galaxy from its surface brightness distribution. The five participating model-fitting software packages are described in detail in \partonedot\  and in the individual software publications referenced in each section. All but one (\deeplegdot) make use of parametric methods, which use functional forms to fit the light distributions from imaging data. \deepleg bases its photometric galaxy profile modelling on convolutional neural networks. All of them fitted at least a single profile to each galaxy in the $\VIS$ band, and some teams and codes have extended the challenge to include the simultaneous fitting of multiple images at different wavelengths.

\medskip 
We present the comparison analysis based on the Euclid Morphology Challenge in this paper.  We investigate  the outcomes from the five participating codes on simulated \Euclid galaxies. Each software package incorporates its own preferred scheme for dealing with the data and was run by the developers or developing teams themselves. Each participant was free to choose setup parameters and criteria according to their best practice and experience, with the hope that this would ensure the best possible outcomes. This could include independent tests or cross-checks from comparing their software to a subset of the `true' parameters of the simulated data, which we made available to the developers. Therefore, we can expect that each code developer's knowledge contributes to the best possible performance of each code. No further specifics, for example in relation to the way of preparing or handling the data, was given to the participants. Each code has different ways of identifying unreliable fits, and we refer the reader to the publications describing each code for additional information. Our goal in this paper is to probe the robustness and accuracy of the most optimal outcome of each software package, examine the code-to-code scatter, and investigate the known bias towards over-estimating the fitting accuracy. This paper presents a tabulated score of the performance of each code with the ultimate goal of using the optimal code for future \Euclid observations. Ultimately, one such code will be implemented in the official \Euclid pipeline to retrieve galaxy morphology parameters for Euclid legacy science. 

In the rest of this paper, we first describe the data that formed the base of the challenge (Sect.~\ref{sec:data}): these are single-\sersic simulations, double-\sersic simulations, and what we call `realistic' simulations that use a variational auto-encoder (VAE) trained on observed COSMOS galaxies. We then describe the metric we designed to quantify the comparison between codes (Sect.~\ref{sec:metrics}). In our results section (Sect.~\ref{sec:results}), we discuss each parameter separately and include a comparison of the recovery statistics, for both single-\sersic and double-\sersic runs. In Sect.~\ref{sec:multiband}, we briefly summarise multi-band fits for the four codes that provided multi-band results. This is an in-depth investigation that was briefly introduced in the companion paper using the same challenge data, but devoted to comparing results for photometry. The first sub-sections of each `result section' detail an in-depth analysis. Readers interested in the summary only will find overview comparisons in the summary figures (Figures \ref{fig:summary_ss}, \ref{fig:summary_ds}, and \ref{fig:real_summary}) and in the `global score' sub-sections (Sections \ref{sec:ss_score}, \ref{sec:ds_score}, and \ref{sec:real_scores}). Section \ref{sec:uncertainty} focusses on quantifying the uncertainty predictions that were requested as part of the challenge. We conclude our analysis with a global score in Sect.~\ref{sec:conclusion}. One goal of this challenge is to find elements that will help to make an appropriate choice for the task of measuring morphological parameters for galaxies observed with \Euclid. The score we developed here is, however, not able to represent all science objectives, for which individual choices will be required. Information about the reproducibility of the results can be found in the appendix. 

\section{Data}\label{sec:data}

The Euclid Morphology Challenge addressed the robustness of structural measurements by comparing `True' input parameters of simulated \Euclid galaxies to outcomes (fitted) `predicted' values that are output from  the software packages (often referred to as `codes' for simplicity) we test. Simulated galaxies with known input parameters provide full control over measurement errors while minimising systematic errors. In this section, we briefly introduce the data used in the challenge. For more information, we refer the reader to the companion paper, \partonedot.

We created five fields of $25000 \times 25000\,\mathrm{pixel}$ each, at $\ang{;;0.1}\,\mathrm{pixel}^{-1}$ scale, corresponding to a field of view (FoV) of about $0.482\,\mathrm{deg}^2$. The fields were made available to the Challenge participants through an online repository, which included a description, lists of source positions and true values of one field that included single-\sersic and double-\sersic information for internal consistency checks and for training purposes. Each field was realised in three versions that are described in more detail below: single-\sersicdot profiles, double-\sersic profiles; and simulations with realistic morphologies for the $\VIS$ band. In one of the five fields we also provide double-\sersic simulations in eight different imaging bands, simulating the three NIR $Y_{\scriptscriptstyle\rm E}$, $J_{\scriptscriptstyle\rm E}$, $H_{\scriptscriptstyle\rm E}$ filters and ancillary data from the five optical Rubin bands \emph{u, g, r, i}, and \emph{z} to test multi-band capabilities. 

\begin{figure}
\centering
   \includegraphics[width=1\linewidth]{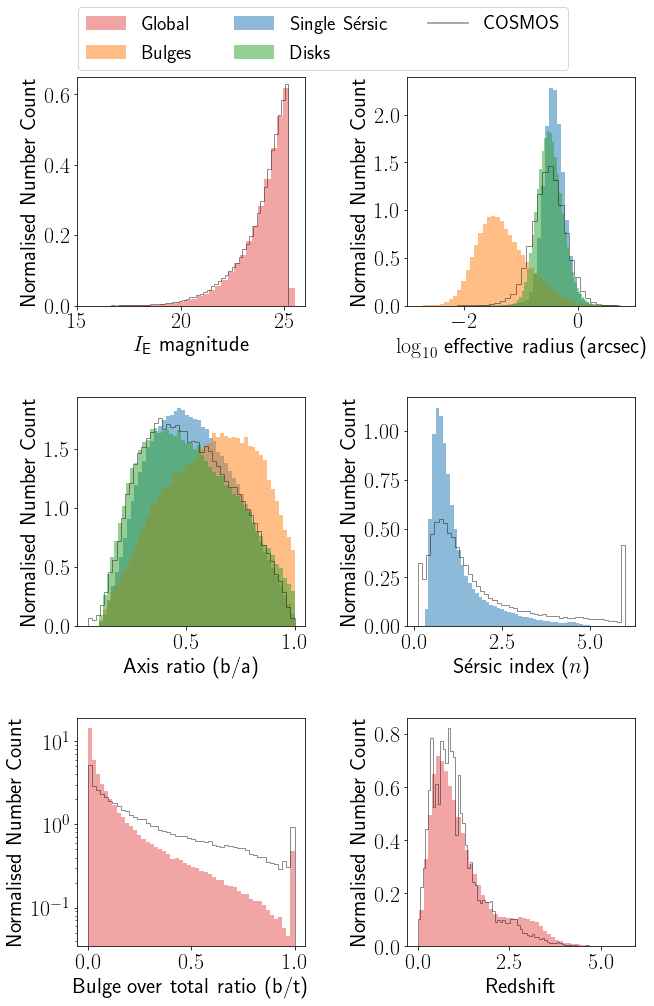}
   \caption{Distributions of the simulated `true' galaxy parameters measured in the Euclid Morphology Challenge. Top left: $\VIS$ distribution down to $5\sigma$ detections. Top right: effective radii for the single component galaxy (blue), and for bulges (orange), and discs (green) separately. Middle Left: Axis ratio distributions. Middle right: \sersic index distributions for single-component galaxies. We note that \sersic indices of the bulges are fixed to $n=4$, and the discs to $n=1$. Bottom left shows the bulge-to-total ratio distribution. The black solid line shows the COSMOS distribution. We also note that for \btdot, the $y$-axis is on a logarithmic scale. The distributions are normalised such that the area is equal to $1$. This figure is replicated from \partonedot.}
   \label{fig:param_distribution} 
\end{figure}

We simulated roughly $314\,000$ galaxies in each field, ranging from $\VIS\simeq15$ to $\VIS\simeq30$ magnitudes. For each field we provided five lists of objects in the format: ID, $x$, $y$ (pixel space) to the participants. Four lists were created which included the simulated objects brighter than a given VIS nominal signal-to-noise ratio (S/N) thresholds for $100\sigma$ ($\VIS \simeq\,22$), $10\sigma$ ($\VIS\simeq\,24.6$), $5\sigma$ ($\VIS\simeq\,25.25$), and $1\sigma$ ($\VIS\simeq\,27.1$). The fifth list contains all simulated sources, including objects below S/N$=1$. We asked the participants to fit those galaxies with at least an S/N over $5\sigma$, where we defined the S/N of a source as the S/N of a point-source in a circular aperture with a diameter of $2\arcsec$, and thus this value corresponds to galaxies brighter than $\VIS = 25.3$. It is important to note that this definition of the S/N does not consider each galaxy's relative profile, and could impact the completeness in less concentrated profiles (lower \sersic index or larger effective radius). The vast majority (more than $99\%$) of galaxies have a magnitude $\VIS$ fainter than $20$ (Fig.~\ref{fig:param_distribution}), which should be kept in mind when examining the results.

The input catalogues were created using the \texttt{EGG} simulator \citep[version v1.3.1,][]{EGG}, which outputs a double-\sersic components catalogue. The single-\sersic catalogues are derived from the double-\sersic with empirical formulae to match observations such as the one by the \textit{Hubble Space Telescope} (HST).
Figure~\ref{fig:param_distribution} gives an overview of the distributions of the parameters we analyse in this paper for all galaxies with an S/N greater than $5\sigma$: $\VIS$, effective radius $r_{\mathrm{e}}$ (plotted as logarithmic, $\log_{\rm{10}}r_{\mathrm{e}}$); axis length ratio \badot; \sersic index $n$ for all simulated single component galaxies; and bulge-to-total ratio \bt, which is also shown for double component galaxies. The $5\sigma$ limit is defined based on the total flux of the galaxy, and roughly corresponds to $\VIS=25.3$ (see \partone for more details). We describe in more detail the generation of these galaxies in the following sections. We note that the fitted \sersic indices only range from $0.3$ to $6$, which are \texttt{Galsim}-related limitations. The same is true for \badot, where restrictions prevent the simulation of galaxies with an ellipticity larger than $0.9$ (\ba smaller than $0.1$).

The galaxy images were then created using the \texttt{Galsim} software. This challenge was designed to mimic the observational depth and conditions of the Euclid Wide Survey \citep{scaramella2021}. The point spread function (PSF) models the  expected behaviour of the telescope and the VIS instrument. It is  more complex than a Gaussian PSF, but has a full width at half maximum (FWHM) equivalent to $\ang{;;0.17}$. To convolve the images, the PSF was over-sampled to different degrees: $6$ times in VIS; $6$ times in NIR at $\ang{;;0.3}$ pixel scale; and no oversampling in the external bands. Participants received a version of the Euclid PSF before oversampling to use for their measurements. There are no reported temporal or spatial variations in the models, which were taken from Euclid's Scientific Challenge $8$\footnote{Euclid's Scientific Challenges are benchmark tests organised inside the Euclid Consortium in preparation for the launch of the satellite.}. Thus, the PSF is assumed to be constant over the FoV. Rubin's PSFs were simulated with \texttt{PhoSim} \citep{phosim2015}. We also added noise that matches the Euclid Wide Survey depth, with the noise a sum of two sources, a Gaussian and a Poisson component. The fact that we did not include correlated noise could be a limitation of the simulation. Detailed information about the simulation procedure can be found in \partonedot.

Our analyses are performed on a common catalogue that consists of $212\,000$ objects for the single-\sersic simulations, $207\,064$ for the double-\sersic simulations, and $204\,229$ for the realistic morphologies. Due to a technical issue with one of the contributing software packages that occurred during the measurements of the mono-band single- and double-\sersic simulations of one of the fields, only four of the five fields were completed by all the participants. As a consequence, we only used the four completed fields for our analysis, and only three fields for the double-\sersic case because one of the fields was used for the multi-band analysis only. Several codes provide a number of individual quality flags with further information on their fits, including details in relation to reliability. While it is out of the scope of this paper to analyse all the different flags of each code, we test and discuss some important flags in Appendix \ref{sec:flags}. We explain our decisions and production steps for the common catalogues in more detail in \partonedot.

\subsection{Single-\sersic simulations} \label{sec:single_sersic_simulations}

Single-\sersic profile simulations were created using the \texttt{Galsim} software \citep[version v2.2.1][]{galsim} following a \sersic profile, which is a characterisation of the intensity $I(r)$ of the galaxy as a function of radius. The flux varies with the distance to the centre according to the following relation:

\begin{equation}
    I(r) \propto \exp\left[-b_n\left(\dfrac{r}{r_{\mathrm{e}}}\right)^{1/n}\right]\;,
\end{equation}
where \re is the effective or half-light radius, the radius in which half of the galaxy's flux is contained.
This is usually considered as a proxy for the size of the galaxy and is sometimes abbreviated to `radius' in this work. The \sersic index is denoted $n$, which is a shape parameter describing the curvature of the function. It drives the steepness of the light profile, and thus describes its shape or concentration. Typically, a profile with $n = 4$ fits well to elliptical galaxies, and for $n = 1$, the \sersic law forms an exponential function, which is often used to describe a disc. We note the presence of $b_n$, which can be approximated by $b_n=2n-1/3$, which links $n$ and $r_{\mathrm{e}}$ \citep{ciotti1991}.  \texttt{Galsim} simulates the surface brightness profiles at high spatial resolution, which we then sample at the image pixel scale. This is important to do in order to avoid aliasing effects, especially when the \sersic index is large.

The galaxy model is then sheared to match the desired ellipticity, or \badot, which is the semi-minor over semi-major axis of the ellipse shape. The normalisation factor is fixed afterwards to match the total magnitude of the object.

\subsection{Double-\sersic simulations}
\label{sec:double_sersic_simulations}

Galaxy formation and evolution studies gain essential knowledge from tracing the individual galaxy components, that is to say bulges and discs, by fitting two-component models. At the simplest level, light profile decompositions enables the classification of galaxies according to their bulge-dominance. 
Double-component galaxies are each simulated with \texttt{Galsim} as a pixel-wise sum of two profiles, one profile for a bulge and one for a disc. The disc is simulated with a \sersic profile with $n=1$, which thus simplifies to an exponential profile:

\begin{equation}
I_{\mathrm{disc}}(r) \propto \exp\left[-b_1 \left(\dfrac{r}{r_{\mathrm{e}}}\right)\right]\, .
\end{equation}
The bulge profile is fixed with a \sersic index of $n=4$, so that the total profile combines to:

\begin{equation}
    I(r) \propto \, (1-\mathrm{b}/\mathrm{t}) \exp\left[-b_1\left(\dfrac{r}{r_{\mathrm{e,b}}}\right)\right] + \mathrm{b}/\mathrm{t}\,\exp\left[-b_{4} \left(\dfrac{r}{r_{\mathrm{e,d}}}\right)^{1/4}\right]\;.
\end{equation}
The two profiles are then sheared to fit the desired ellipticity, $q_{\rm{b}}$ and $q_{\rm{d}}$. The flux is first scaled to generate galaxies with suitable \btdot, and then the global flux is re-scaled to match the global flux of the galaxy. The two components are always aligned to the same position angle, and the PSF is applied to the global profile. iven the overall aim of the challenge to probe the capacity of software packages that attempt galaxy model fitting, we chose to test the codes on ideal galaxy simulations with known and fixed \sersic indices to control for variations across the software packages.

In addition, we created one field with double-\sersic galaxies that includes images in nine bands, which will be relevant for tests of multi-band fitting routines (Sect.~\ref{sec:multiband}). The structural properties in all bands are kept constant, and therefore our simulations do not model wavelength dependent structural changes.

\subsection{Realistic simulation}\label{sec:data_real}
\label{sec:realistic_simulations}
Simulated galaxy images are inherently difficult to produce realistically, which is why most tests for morphology measurements focus on simulating and fitting smooth analytic profiles. The Euclid Morphology Challenge also provides a dataset with more realistic galaxies learned following a data-driven approach using deep neural networks. This is described in detail in \citeauthor{bretonniere2022} (2022(a); referred to as B22 from here onwards). Very briefly, we use a deep generative model called the variational auto-encoder \citep{Kingma2019},  that compresses and decompresses images to learn a probabilistic latent representation of the training set distribution. Using HST images the model learns how to simulate real 2D noiseless galaxy profiles at a VIS-like resolution. A second generative model, called Normalising Flow \citep{papamakarios2021} is then used to condition the latent distribution with the structural parameters. The resulting architecture, called the Flow-Variational AutoEncoder (FVAE), can therefore simulate galaxies directly from a catalogue of parameters, provided that the training set properly covers the range of values. The advantage of the FVAE compared to a classical VAE or other generative network is the ability to constrain the physical parameters of the emulated galaxies. 

Given the lack of very large and bright galaxies in the HST data used for training,  this dataset does not include galaxies larger than $0.2$ arcminutes or brighter than $20.5\,\mathrm{mag}$. This only represents around $1\%$ of the $314\,000$ simulated galaxies per field. Although this dataset should allow us to quantify the performance of the different codes in more realistic conditions, it is important to emphasise that these simulations are not perfect. Indeed, the conditioning of the latent space with galaxy morphology is not always exact, which can introduce a systematic bias in what we call the `true' values for these realistic fields; we refer the reader to the discussions in B22. We also note that the model slightly differs from the one used in B22, in the sense that the magnitude is also a parameter conditioned by the Flow, which is then also re-calibrated using $\texttt{Galsim}$. This dependence on the Flow allows us to keep the correlation between morphology and magnitude. The post-processing steps (PSF and noise) are the same as described in the previous sections.

\section{Metrics} \label{sec:metrics}
As in the companion paper \partonedot, we use four main indicators to evaluate and compare the different codes: completeness ($\mathcal{C}$)\footnote{The completeness $\mathcal{C}$ measures the fraction of objects for which there is a successful fit, see \partone for details.}; bias ($\mathcal{B}$); dispersion ($\mathcal{D}$); and outlier fraction ($\mathcal{O}$). We also combine these values into a global score, $\mathcal{S}$, to ease the comparison of the different codes. Each of these parameters is computed for each galaxy structural parameter ($p$), and is plotted in bins of apparent magnitude to quantify the impact of signal-to-noise. In the following, we provide a definition of each of these accuracy estimators, which slightly differs from the ones used in \partonedot. These differences were necessary to better capture the specifics of our parameter distribution, in particular the large impact of outliers in the dispersion values.
\subsection{Bias}\label{sect:bias}

The individual bias $b_{p}$ on a structural parameter $p$ of a galaxy is defined as the difference between the predicted value, $\mathrm{Pred}_{p}$, and the true simulated value, $\mathrm{True}_{p}$:
\begin{equation}
b_{p} = (\mathrm{Pred}_{p} - \mathrm{True}_{p})\;,
\label{eq:bias}
\end{equation}
where \mbox{$p=\{r_{\mathrm{e}}, q, n\}$} for single-\sersic fits and \mbox{$p=\{\mathrm{b}/\mathrm{t},r_{\mathrm{e,b}},r_{\mathrm{e,d}},q_{\mathrm{d}}, q_{\mathrm{b}}\}$} for double component fits. Sometimes it is appropriate to calculate the relative bias, $\tilde{b}_{p}$, which is defined as
\begin{equation}
\tilde{b}_{p} = \dfrac{\mathrm{Pred}_{p} - \mathrm{True}_{p}}{\mathrm{True}_{p}}\;.
\label{eq:bias_rel}
\end{equation}

\noindent The use of either the absolute or relative bias depends on the parameter. For example, the same absolute bias has a different meaning in a small galaxy than in a large galaxy: a measurement error of $\ang{;;0.1}$ for a galaxy of $r_{\mathrm{e}}=\ang{;;0.2}$ is more problematic than the same error on a galaxy with $r_{\mathrm{e}}=\ang{;;3.0}$. This is not the case for other parameters, such as \ba and \btdot,  which have a constrained dynamical range between $0$ and $1$. 
We also chose to use the absolute bias for the \sersic index, even though this is less straightforward to measure, since the dependence of the profile on $n$ is not linear. For galaxies with $n>4$, the impact of increasing $n$ on the surface brightness profile is small, which implies that errors on large \sersic indices are generally less severe than on small values of $n$.  However, since this dependence is not linear, the relative bias does not properly encapsulate this behaviour. In order to make the interpretation easier, we simply use the same absolute definition of $b$. The choice is also motivated by the fact that the majority of galaxies in our simulations have a low \sersic index, for which the absolute bias is well suited (see Fig.~\ref{fig:param_distribution}).

We also define the global bias $\mathcal{B}_{p}$ of a population as the median of all individual biases of the population, $\boldsymbol{b}_{p}$:  
\begin{equation}
\mathcal{B}_{p} = Q_{0.5}(\boldsymbol{b}_{p}) \;,    
\end{equation}
or if we take the relative bias,
\begin{equation}
\mathcal{\widetilde{B}}_{p} = Q_{0.5}(\boldsymbol{\tilde{b}}_{p}) \;,
\end{equation} which is the value reported in all subsequent sections. 
A statistically unbiased measurement thus corresponds to $\mathcal{B}_{p}=0$. Notice that $\mathcal{B}_{p}$ can have positive and negative values if a given parameter is over- or under-estimated, respectively. This metric is computed on all the objects of the common catalogue, without removing the outliers, which are discussed in Sect.~\ref{sect:badfit}.

\subsection{Dispersion}

\begin{figure}[h!]
\centering
\includegraphics[width=1\linewidth]{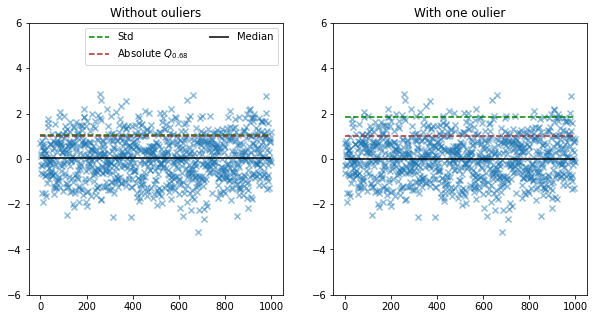}
\caption{Illustration of our dispersion metric choice. In both plot, we plot the median, the standard deviation and our definition of the dispersion, defined Eq.~\ref{eq:disp} for a Normal Gaussian distribution. In the right figure, we add an outlier at $y = 100$. We can see that our definition is not sensible of the presence of an outlier, compared to the standard deviation.}
\label{fig: std_quantile} 
\end{figure}

The dispersion of a population, $\mathcal{D}_{p}$ on a parameter $p$ is defined as the $0.68$ quantile ($Q_{0.68}$) of the absolute population biases from which we subtract the median bias:
\begin{equation}
    \mathcal{D}_{p} = Q_{0.68}\left( \left|\boldsymbol{b}_{p}\right| - Q_{0.5}\left(\boldsymbol{b}_{p}\right)\right) \;.
    \label{eq:disp}
\end{equation}
Here again, the absolute bias $b$ is used for \badot, $n$, and \btdot, while the relative bias $\tilde{b}$ is used for the effective radii.
The median bias is removed to recentre the distribution around zero, so that the quantile matches the significance of a standard deviation. We use the $0.68$ quantile because it is less sensitive to outliers than the standard deviation. Outliers are quantified independently (see Sect.~\ref{sect:badfit}). We note, however, that for Gaussian distributions both $Q_{0.68}$ and the standard deviation correspond to the same measurement. Figure~\ref{fig: std_quantile} illustrates the advantage of our dispersion metric compared to a simple standard deviation, comparing the classic standard deviation with our definition in presence of a single outlier. Whenever we use the absolute error $\tilde{b}$, we define the dispersion as $\widetilde{\mathcal{D}}_{p}$.

\subsection{Outlier fit fraction}\label{sect:badfit}
In addition to bias and dispersion, we also quantify the fraction of `outliers', which could equally be called `fraction of bad fits'. We define an outlier on a given structural parameter $p$ when its bias $b_{p}$ is larger than a given threshold ($t_b$), which we fix to be $t_b = 0.5$ for all parameters $p$. The fraction of outliers ($\mathcal{O}$) is thus the number of objects above the threshold divided by the total number of objects in the considered bin. 
Since the bias $b$ is not always defined in the same way for all parameters (see Sect.~\ref{sect:bias}), the meaning of $\mathcal{O}$ also differs in the following three cases. Firstly, for the effective radius: because we use the relative bias $\tilde{b}$, $t_b=0.5$ means that we consider an outlier if the relative error is larger than $50\%$.
Secondly, for the axis ratio and bulge-to-total ratio: because the bias is absolute, but the range of possible values is limited to $[0,1]$,  $t_b=0.5$ means that an outlier is defined when the error is larger than $50\%$ of the maximum possible error.
Finally, for the \sersic index: since the bias is not relative and the range is not bounded, the outlier definition cannot be seen as a percentage in this case; see the discussion in Sect.~\ref{sect:bias}. We emphasise here that the bias and dispersion metrics are computed including the outliers.

\subsection{Global score}
Finally, in order to summarise the overall performance of a given code and to compare more easily the codes to one another, we define a global score $\mathcal{S}_{p}$ on a given parameter $p$, which encapsulates the four previous measurements $\mathcal{C}$, $\mathcal{B}_{p}$, $\mathcal{D}_{p}$, $\mathcal{O}_{p}$:
\begin{equation}
    \mathcal{S}_{p} = (1-\mathcal{C})+ \sum_{i} w_i \left(k_{\mathcal{B}}\mathcal{B}_{p, i} + k_{\mathcal{D}}\mathcal{D}_{p, i} + k_{\mathcal{O}}\mathcal{O}_{p, i} \right)\;.
    \label{eq:score}
\end{equation}
We note that our three metrics, $k_{\mathcal{B}}, k_{\mathcal{D}}$, and $k_{\mathcal{O}}$ are weights applied to each of the different precision indicators. In our case, we set the same relative weight that has been calibrated empirically, so that the order of magnitude of the score, and thus its interpretation, is consistent with the companion paper \partonedot:
\begin{equation}
    k_{\mathcal{B}} = k_{\mathcal{D}} = k_{\mathcal{O}} = 2.1 \;.
\end{equation}

\noindent With this calibration, scores generally range from $0.2$ to $2$, the lower the better. The sum is performed over bins of apparent magnitude. The different $w_i$ are therefore factors that weight the score with regard to the S/N of the bin and the fraction of objects in the bin (fewer objects and lower S/N will lead to a smaller weight, and thus smaller impact on $\mathcal{S}$); see \partone for more details, where the definitions of the diagnostics are similar, but not identical, due to different use cases. We emphasise that the score is intended to provide a first-order estimation of the performance of the different codes using a single number, but should not be used on its own to chose a `best code' appropriate for every scenario. This is due to a number of additional important considerations, like the execution time or user-friendliness, which are left out. We therefore acknowledge that our global score is a simplification and  point out that alternative metrics, which could be adapted for specific science goals, might result in different conclusions. In order to support the user in tailoring the diagnostics to their individual science case, we have created an interactive plotting tool, which is published alongside this paper. It enables the recreation and adaptation of most figures shown in this paper. We describe this tool in Appendix \ref{app:app1}.


\section{Results} 
\label{sec:results}
Summarising the results in a reasonable number of figures is difficult, since the problem is multi-dimensional with several degeneracies between the different structural parameters. For simplicity, we only show the metrics as a function of apparent $\VIS$ magnitude in the main text as taken from the `true' input values, which is a proxy for S/N. This is a limited representation of the complexity of the problem, but it is a reasonable trade-off between readability and information provided. We also provide an online interactive plotting tool\footnote{\url{https://share.streamlit.io/hbretonniere/euclid_morphology_challenge}}  for full exploration of the data. Using this tool it is possible to investigate independently how the fits trend with other parameters, such as \sersic index or size. In Figs.~\ref{fig:trumpet_re_z} and \ref{fig:summary_redshift} of the appendix, we show and comment on an example of morphological parameters as a function of the true redshift. 

The results are presented as follows. For each type of simulation -- single-\sersicdot, double-\sersic and realistic -- we measure our three metrics ${\mathcal{B}}$, ${\mathcal{D}}$, and ${\mathcal{O}}$ for each structural parameter and every code on a common dataset containing only galaxies for which all codes provide a valid fit (see also the companion paper \partonedot). In this way, we ensure a fair comparison between the different codes. These values are summarised in Tables \ref{table: single sersic} (single-\sersic and realistic) and \ref{table: double sersic} (double-\sersicdot).
Throughout the next sections, we step through our metrics analysis for each of the datasets by discussing two main types of figure. The first figure type is a scatter plot of magnitude versus $b$ or $\tilde{b}$ for individual objects.  Because the dispersion increases towards fainter fluxes (high magnitudes), the scatter plots produce a trumpet-like shape, and are therefore referred to as `trumpet plots'. The two metrics, $\mathcal{B}$ and $\mathcal{D}$ are represented with a running orange line ($\mathcal{D}$ represented as error bars centred on $\mathcal{B}$). In this first type of figure, we also show the distribution of the bias $b$ on the right inset plot, with the reference $0$ bias in thick blue lines, and the overall bias in dashed white lines. The outlier threshold $t_b$ is represented by dashed red lines. The second type of plot, which we call the `summary figure', shows our three metrics $\mathcal{B}$, $\mathcal{D}$, and $\mathcal{O}$ values in $11$ bins of magnitude, from magnitude $14$ to $26$. This allows us to plot in the same figure the five different codes for a direct comparison.


\subsection{Single-\sersic results}
\label{sec:ss}
In this section, we analyse results from the fitting of single-component \sersic functions that describe the radial surface brightness profile, fitted on the $\VISdot$-band images only. Figure~\ref{fig:summary_ss} summarises the results, along with Table~\ref{table: single sersic} and Sect.\ref{sec:ss_score}. In addition, Figure~\ref{fig:residuals} shows residuals between the simulation and the modelled galaxies.
Naturally, single-\sersic fits are less sensitive to small scale features, since they essentially smooth over the individual components of a galaxy. Despite this drawback, they are generally the fastest and most straightforward measure of the sizes (via the half-light radius, Sect.~\ref{sec:ss_re}), axis ratios (Sect.~\ref{sec:ss_ba}), and shapes (via the \sersic index Sect.~\ref{sec:ss_n}) of galaxies. All participants returned results for this analysis, which is why figures in this section have five individual results for comparison.

\subsubsection{Half-light radius}\label{sec:ss_re}
\begin{figure}
\centering
    \includegraphics[width=\linewidth]{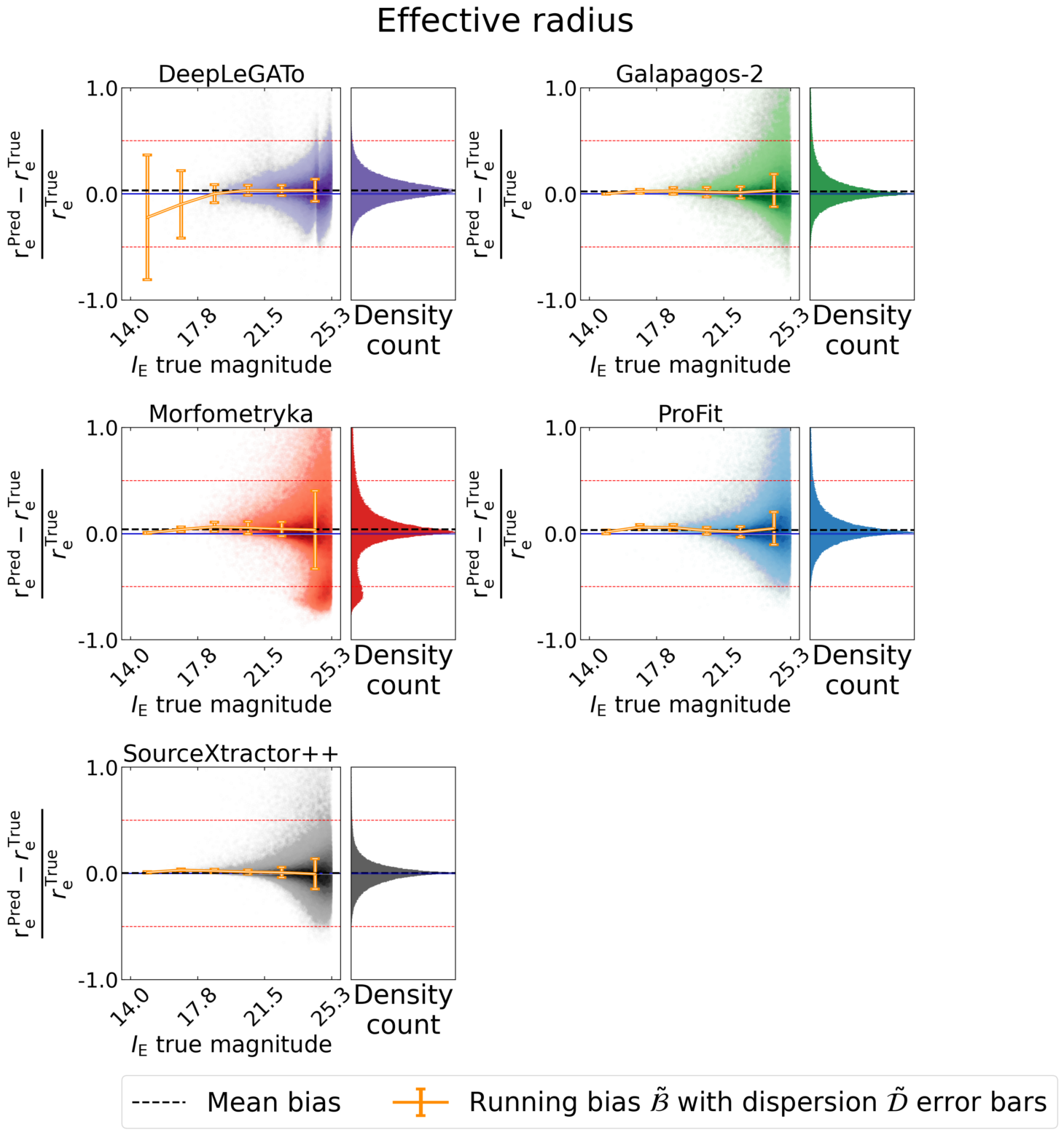}
    \caption{Scatter plots showing the recovery of the half-light radius measured from the single-\sersic simulation. Each panel shows a different code. The main plot of each panel shows the relative bias per galaxy as a function of apparent $\VIS$ magnitude, while we summarise the bias distribution as a histogram on the right. The opacity is proportional to the density; the darker colours mean more points. The blue solid line highlights a zero bias for reference, and the grey dashed line represents the mean value of the bias for all magnitude bins together. The orange points indicate the running mean bias $\mathcal{B}$ in bins of magnitude, with error bars representing the dispersion $\mathcal{D}$ (see Sect.~\ref{sec:metrics}).}
    \label{fig:trumpet_ss_re}
\end{figure}

Figure~\ref{fig:trumpet_ss_re} shows that the global behaviour of all five software packages is similar, with the expected trumpet shape visible in all plots: the scatter increases for faint objects. Moreover, the scatter plots generally do not show a significant bias (with the exception of \deepleg for bright objects). Another commonality of all codes is that the trumpet plot is skewed towards positive values, that is the majority of outliers (points outside the two red dashed lines) are due to an overestimation of the size.

Beyond this common general behaviour, some peculiarities are notable. This includes the bias in \metrykadot's plot (in red), indicating a bi-modality at the faint end, with around $13\%$ of objects consistently fitted with a lower radius than expected (the relative bias is around $-0.5$). This is due to convergence problems for objects close to the lower limit, when the fits do not update beyond the first guesses that the software uses, so outputs stall at \sersic indices between $0.1$ and $0.2$. \metryka recognises the unreliability of these fits with an internal flag that is given to objects with sizes smaller than the PSF's FWHM. Generally, these objects also have low \sersic indices. This flag, `\texttt{TARGETISSTAR}', is designed to flag stars, which these are not, but their small sizes and low \sersic indices are recognised internally as such. Such flags were not provided to the authoring team as part of the challenge. They represent around $14\%$ of the common catalogue. We decided to keep these objects in the overlapping catalogue even after the flags were provided. The reason for this is that removing them would bias codes that were generally able to fit these objects, and because of the non-negligible fraction of the catalogue they represent. Nevertheless, even if \metryka is not able to fit these objects, they are able to recognise the problem and flag them. We show in appendix Fig.~\ref{fig: trumpet_re_no_stars_flag} a version of the trumpet plot without those particular objects.

\deepleg (in purple) also shows a characteristic behaviour, with a strong negative bias and dispersion for very bright objects ($\VIS <18$), and an apparent discontinuity around $24.5$ mag. The first can be explained by the fact that the dataset used to train the model lacks bright objects which are rare in the observations. This is a well known effect of machine learning models, which are sensitive to the distribution of properties apparent in the training dataset. The second distinctive observation of all \deepleg plots, the discontinuity around $24.5$, is a direct consequence of the training strategy of the neural networks in bins of S/N. The abrupt change corresponds to a change of the deep learning model. Indeed, in an attempt to improve performance on both bright and faint objects, the \deepleg algorithm was trained separately for two sets of objects, objects fainter and brighter than magnitude $24.5$ (which corresponds to an S/N of $10$). This leads to two sets of weights and thus to two models, which can and do behave differently. This behaviour is seen in all structural parameters for which \deepleg produced results.

Looking ahead to the `summary plot' in Fig.~\ref{fig:summary_ss}, the first row of the plot compares the effective radius measurements that we are discussing here. Each column shows one of the three accuracy indicators: bias ($\mathcal{B}$); dispersion ($\mathcal{D}$); and outliers fraction ($\mathcal{O}$). We note that to better highlight the small differences between the codes, the $y$-axis range has been reduced. 

The first column, $\mathcal{B}$, reveals that in general all codes slightly overestimate galaxy sizes, which confirms the trend seen in the trumpet plots. Only \deepleg dramatically under-estimate the radius of the very bright galaxies, with a decreasing bias from $-0.4$ (outside the plotted area) at $\VIS=14.5$ to $-0.05$ at $\VIS=17.5$. In addition to the lack of bright objects in the training set, this can be explained by the fact that \deepleg works with a fixed stamp size of $64\times 64\,\mathrm{pixel}$, which can cut the edges of the galaxy profile and thus lead to an under-estimation of its radius. We can also see that \profit very slightly under-estimates the radius for the first bin (very bright galaxies). However, given that this bin has less than ten galaxies, the statistics may not be large enough to point to a particular trend. We again note that the first four bins only hold around $100$ galaxies, which represent less than $1\%$ of the entire catalogue. Importantly though, the absolute value of the bias remains smaller than $7\%$ for all magnitudes and all codes (and for $\VIS > 17$ for \deeplegdot, as discussed), which means that despite their different approaches, there are no major differences between the $\mathcal{B}$ values of the different codes. We can see that for the three brightest bins, \galadot, \metrykadot, and \SE perform very similarly, with \gala reaching a slightly smaller bias. \profitdot's bias is less stable; tt first has a slightly higher bias, which decrease between $\VIS=17$ and $\VIS=23.5$. For those intermediate magnitudes, \gala and \SE perform very similarly, while \deepleg and \metryka have a higher positive bias. Finally, for the very faint galaxies ($\VIS > 24$), \SE has a bias close to zero, followed by \metrykadot, \deeplegdot, \gala and \profitdot. 

The second column of the summary figure compares the dispersion $\mathcal{D}$ of all codes. The trends are generally comparable, staying below $0.1$ at $\VIS <24$ for all codes except for \deepleg for bright objects. Here again, and for the same reasons explained in the previous paragraph, \deepleg shows a high dispersion, decreasing from about $0.8$ (off the displayed plotting area) at $\VIS=14.5$ to $0.2$ at $\VIS=17.5$. We can also see the higher dispersion for \profit in the first magnitude bin. The four codes behave similarly with differences of only a few percent for $\VIS < 23.5$, with \SE having the smaller dispersion, followed by \profit and \galadot, \deepleg and \metrykadot. For fainter objects, \deeplegdot's dispersion stays below $0.10$, while \SEdot, \gala and \profit increase to $0.15$. \metryka shows the largest dispersion, up to $0.45$ (again, off the plotting area) for the lowest S/N bin. As seen in the trumpet plot, the dispersion at the faint end is dominated by a long tail in the distribution, with a large fraction of objects being estimated to be too large.

Regarding the fraction of outliers (third column), we see that at the bright end, all codes except \deepleg have no bad fits (the only bin with a non-zero outlier fraction is \profit and that concerns only one galaxy).
For $\VIS < 23$, all the codes have less than $10\%$ outliers, with \profit and \gala showing the smallest numbers of bad fits, followed by \SE and \metrykadot.
For fainter objects, all measurements except for \deeplegdot, and to some extend \SEdot, increase significantly, up to approximately $30\%$ for \metryka and $20\%$ for \profit and \galadot. On the contrary, \deepleg has close to zero outliers for $23\leq \VIS \leq 26$ and \SE also keeps a relatively small fraction of bad fits, with up to $5\%$ for the fainter objects.
\metrykadot's outlier fraction for faint objects is due to the accumulation of galaxies around $b=~-0.5$, which we have commented on before and are flagged during a regular output catalogue with the flag `\texttt{TARGETISSTAR}' (see also Fig.~\ref{fig: trumpet_re_no_stars_flag}). We remind the reader that even if the individual three metrics in Fig.~\ref{fig:summary_ss} seem unfavourable  for \deepleg measurements of bright galaxies, this has little impact on the global score $\mathcal{S}$, affecting only $93$ galaxies, less than $1\%$ of the fitted catalogue.

\subsubsection{Axis ratio}\label{sec:ss_ba}
\begin{figure}
\centering
    \includegraphics[width=\linewidth]{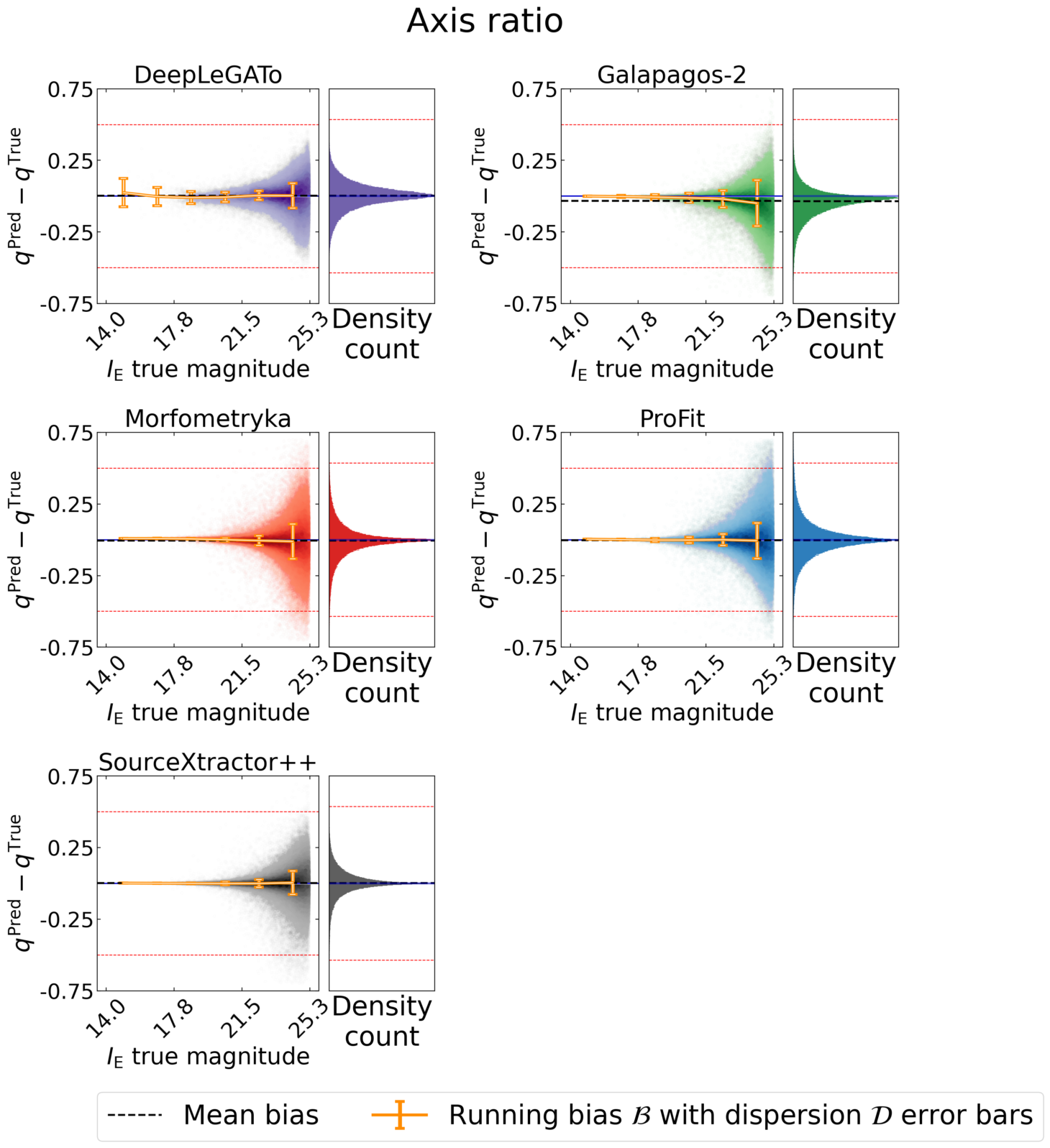}
    \caption{Fitting results for the axis ratio \ba of the single-\sersic simulation. See caption of Fig.~\ref{fig:trumpet_ss_re} and Sect.~\ref{sec:results} for further information.}
    \label{fig:trumpet_ss_q}
\end{figure}
We now move on to the axis ratio \badot. Recall that \ba has the opposite interpretation compared to ellipticity, a high \ba describing a circular galaxy. We see in the trumpet plot of Fig.~\ref{fig:trumpet_ss_q} an overall good recovery from all codes, with almost zero bias and a reasonably low dispersion. The discontinuities between S/N bins for \deepleg is much less noticeable, and the bias for bright objects is also lower. Evidently, also \metrykadot's buildup of unreliable size measurements for small objects (and \sersic indices as we subsequently see in the next section) are not a problem for providing accurate axis ratios.

The second row of Fig.~\ref{fig:summary_ss} shows the summary of the three metrics for \badot. Axis ratios are measured remarkably well, with a bias smaller than $3\%$ for $\VIS < 26$ for all codes, and for $\VIS < 23$ for \galadot. \gala has a slightly larger bias than the other codes for the faint objects, with a tendency to estimate more elongated galaxies. However, it remains smaller than $0.07$ even in the faintest object bin. We still see a large bias for \deeplegdot, which oscillates between around $-0.09$ and $0.07$ from $\VIS=14$ to $\VIS=17$ (cut by the $y$-axis range in the graph for visualisation purposes). For $\VIS < 24$, \SE and \profit behave similarly well (nearly no bias), followed by a fraction of percent for \metryka and \galadot. \metryka over-estimates \ba for faint objects and under-estimates it for bright objects. In the last (faintest) magnitude bin, we can see that \SE and \deepleg slightly over-estimate \badot, while the other three under-estimate it, which could suggest that the problem comes from the difficulty of the task at very low S/N, rather than a problem linked to the estimation of the PSF.

Regarding the dispersion, all codes except \deepleg have a smooth increase with magnitude, from zero up to respectively $0.10$ for \SE and \deeplegdot, $0.15$ for \metryka and \profitdot, and $0.20$ for \galadot, and it remains smaller than $0.1$ for all codes at $\VIS < 24$. For $\VIS < 22$, \metryka and \SE achieve the smallest dispersion. \deeplegdot's high dispersion at the bright end relates to issues already expanded on previously.
 
The outlier fraction (third column in Fig.~\ref{fig:summary_ss}) is overall below $1\%$ for all codes and magnitudes. This is another sign that the ellipticity is one of the parameters which is generally recovered reliably by all software packages, even though an outlier threshold of $0.5$ is quite permissive. Indeed, galaxies with a true value of $0.5$ cannot be fitted as outliers, but we chose to keep this definition for simplicity of the metric. Furthermore because the metric is the same for all codes, we believe this comparison to be fair. We can see that even though \deepleg has the strongest bias and dispersion for bright objects, they are still well below the outlier threshold, and stay very close to zero even for the faintest galaxies. For the other software packages, the fraction of outliers starts to be non-zero for $19\leq \VIS \leq 21$. 
The interested reader is invited to use the interactive plotting tool released together with this work to investigate the result on the fraction of outliers. It allows one change (and therefore to decrease) the outlier threshold.

We highlight that the error in the axis ratio measurement is the sum of at least two procedures: the prediction of the two semi-axis lengths (impacted by the S/N and the PSF) but also of the position angle (PA) of the galaxy, necessary to define the two semi-axis. We note that the PA is not part of our current comparison.

\subsubsection{\sersic index} \label{sec:ss_n}
\begin{figure}
\centering
    \includegraphics[width=\linewidth]{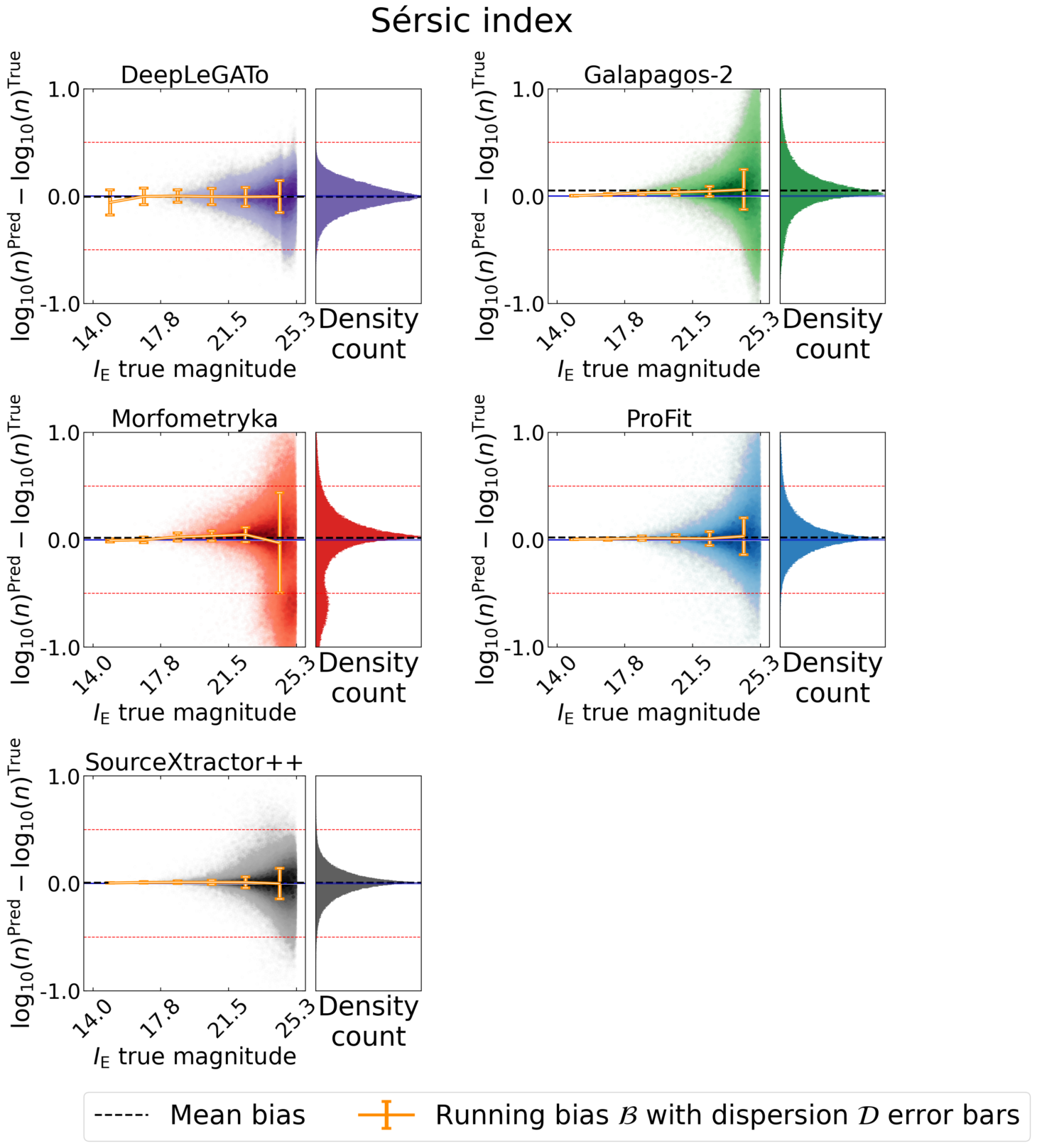}
    \caption{Fitting results for the \sersic index of the single-\sersic simulation. See caption of Fig.~\ref{fig:trumpet_ss_re} and Sect.~\ref{sec:results} for further information.}
    \label{fig:trumpet_ss_n}
\end{figure}

In this section we inspect the estimation of the \sersic index of galaxies (Fig.~\ref{fig:trumpet_ss_n}). As a reminder, the \sersic function is a simplified model that does not capture the entire galaxy, but gives important information about how the intensity varies with radius. Compared to other morphological parameters retrieved from single-\sersic model fitting, the \sersic index is regarded as the most challenging parameter to recover \citep{buitrago2013,dosreis2020}.
Because the dependence of light profiles on the \sersic index is exponential, we always analyse $\log_{10}(n)$ instead of $n$ in the following (see e.g. \citealp{kelvin2012} for an extended discussion).

All codes display the familiar trumpet shapes with the known caveats in \deepleg and \metrykadot. Beyond that, we observe that \deeplegdot, \metryka and \SE tend to be skewed towards negative values for faint objects (indicating the prediction of smaller $\log_{10}(n)$ compared to the truth), while \gala and \profit show the opposite trend.

\begin{figure}
\centering
    \includegraphics[width=\linewidth]{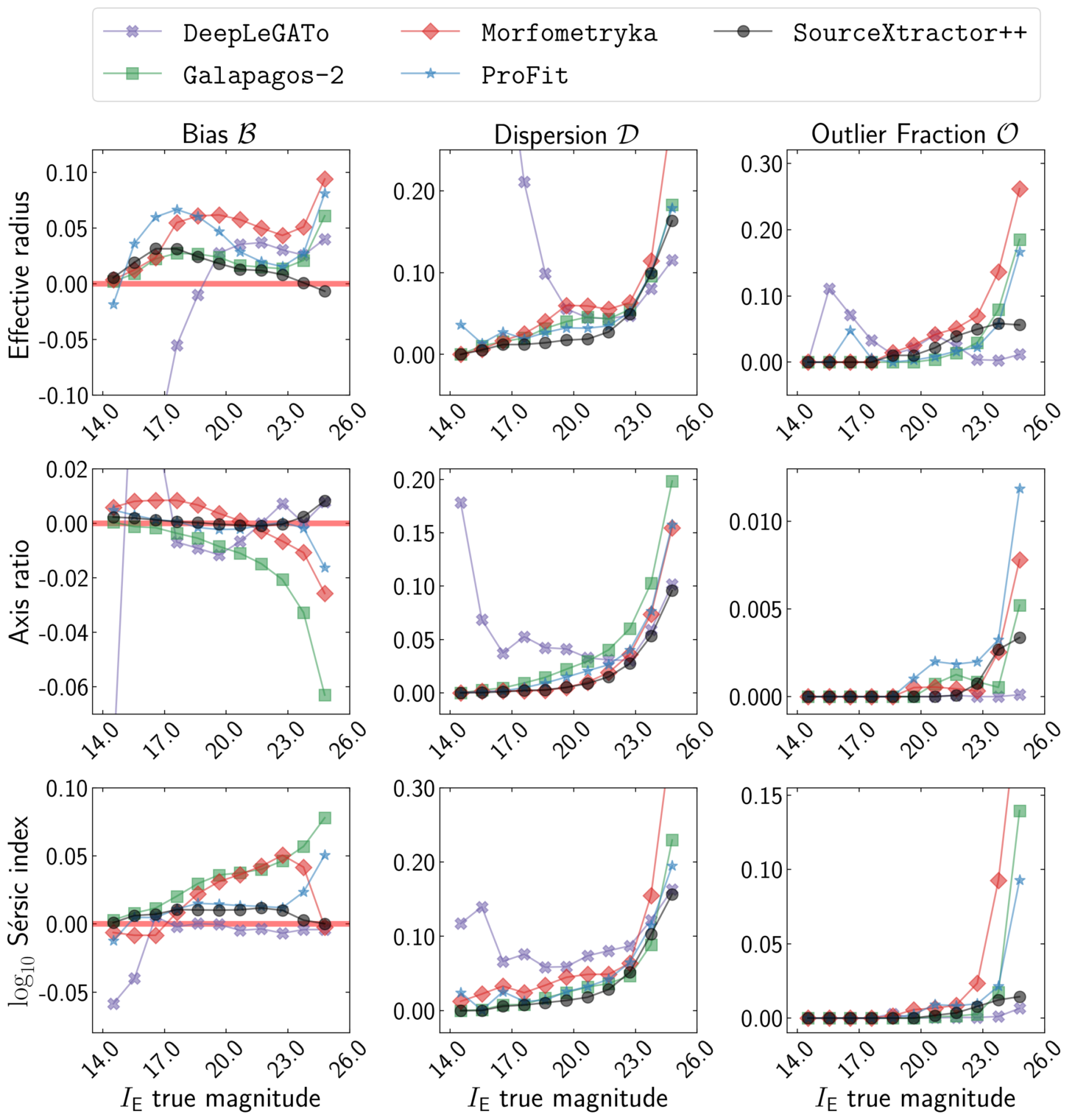}
    \caption{Summary plot for the single-\sersic simulation. The different rows show the results for the three different structural parameters: half-light radius \re (top), axis ratio \ba (middle) and \sersic index $n$ (bottom). Columns represent (1) the mean bias $\mathcal{B}$, (2) the dispersion $\mathcal{D}$, and (3) the fraction of outliers $\mathcal{O}$, per bin of $\VIS$ magnitude (see text for details). We note that the $y$-axis is sometimes cut at low values to highlight the small differences between the software packages. Each code is plotted with a different colour as labelled.}
    \label{fig:summary_ss}
\end{figure}

The third row of Fig~\ref{fig:summary_ss} presents the metrics for the logarithm of the \sersic index. While \deeplegdot's performance for fitting bright objects is less biased compared to the previous parameters, it still has the largest negative bias for the smallest magnitude bins, which means it predicts bright galaxies without steep cores (i.e., bulges).
Beyond this bright end, \deepleg is the only code that does not over-estimate the \sersic index, which means it does not predict steeper galaxy profiles in their cores. For fainter galaxies, from $\VIS=17$ to $\VIS=26$, \deepleg achieves the most robust bias calibration, mitigated by the fact that it has the highest dispersion. \SE and \profit have a similarly small bias (around $0.01$) for $\VIS < 23$, which then decrease close to zero for \SE and increases to around $0.5$ for \profitdot. \metrykadot's and \galadot's bias steadily increase for ever fainter galaxies. \gala increase up to $0.07$, while \metryka abruptly falls to $-0.1$ due to the known accumulation of objects that were not successfully modelled.

The behaviour of the dispersion (second column) is similar for all codes except for \deepleg for $\VIS < 23$, with a dispersion lower than $0.10$. \SE has the lowest dispersion, followed by \gala and \profitdot, \metrykadot, and \deeplegdot. Here again, the difference between the four first codes is very marginal. The dispersion $\mathcal{D}$ then increases for every code, up to $0.16$ for \SE and \deeplegdot, $0.2$ for \profitdot, $0.25$ for \galadot, and $0.65$ for \metrykadot, which can once again be explained by the cluster of points around $0.5$.
None of the codes suffer from bad fits (third column, $\mathcal{O}$) for $\VIS < 19$, and just up to few percents for $\VIS < 22$. The fraction then increases steeply at faint magnitudes. The increase is highest for \metrykadot, from about $1\%$ at $\VIS=20$ up to $34\%$ at the faintest bin, again related to the discussed failed fits. \gala increases to $15\%$ only in the faintest bin. \deepleg achieves the lowest number for all magnitudes, followed by \SE also at the faint end.

\subsubsection{Global scores}\label{sec:ss_score}
The blue numbers in Table~\ref{table: single sersic} summarise the global scores (see Eq.~\ref{eq:score}) for the three parameters of the single-\sersic simulations and for the five codes. An average global score $\mu_{\mathcal{S}}$ is also provided. They are also plotted in the first part of Fig.~\ref{fig:global scores}. The best score is obtained for \SEdot, which achieves a value of $\mathcal{S}=0.28$. In addition, the table also shows that some codes behave better than others for some specific structural parameters. For example, \metryka is better for the axis ratio than for the effective radius, where it is highly penalised by the large dispersion for faint objects that we discussed. We emphasise again that this score is very sensitive to the different weights on the number of objects, the S/N, and the weights of the metrics. In particular, the weights of the smallest magnitude bins(from $\VIS=14$ to $\VIS=19$) have close to no impact on the score, because of the very small number of objects in those bins. It explains why \deepleg has a good global score while performing worse than the other codes for bright objects, while other codes like \gala or \metryka perform best for certain parameters. By the nature of how we set up the metric, the order of the global score ranking can therefore change if we adjust the different weights to reflect a specific emphasis. We encourage the reader to explore the interactive tool released with this work, to tune this score to their particular science case.

\begin{table}
\caption{Comparison of the scores $\mathcal{S}$ obtained by the different software packages in all structural parameters for the single \sersic simulations. Numbers in blue are results from $\textsc{Galsim}$ simulations (discussed in Sec. \ref{sec:ss}), and numbers in black quote results from measurements of simulations with the deep generative model FVAE (discussed in Sec. \ref{sec:real}). The last column is the mean of the parameters. A smaller $\mathcal{S}$ means a better fit.}
\scalebox{0.75}{
\begin{tabular}{|c|c|c|c|c|c} 
\cline{1-5}
\diagbox{FVAE}{\textcolor{blue}{\texttt{Galsim}}} & $\mathcal{S}_{r_{\rm{e}}}$                    & $\mathcal{S}_{\rm{b/a}}$                  & $\mathcal{S}_{\rm{n}}$                    & $\mu_{\mathcal{S}}$                           &   \\ 
\cline{1-5}
DeepLeGATo                               & \diagbox{$\emptyset$}{\textcolor{blue}{0.37}} & \diagbox{$\emptyset$}{\textcolor{blue}{0.25}} & \diagbox{$\emptyset$}{\textcolor{blue}{0.38}} & \diagbox{$\emptyset$}{\textcolor{blue}{0.33}}  &   \\ 
\cline{1-5}
Galapagos-2                              & \diagbox{2.05}{\textcolor{blue}{0.58}}         & \diagbox{0.79}{\textcolor{blue}{0.43}}         & \diagbox{1.29}{\textcolor{blue}{0.60}}        & \diagbox{1.38}{\textcolor{blue}{0.54}}         &   \\ 
\cline{1-5}
Morfometryka                             & \diagbox{$\emptyset$}{\textcolor{blue}{1.10}} & \diagbox{$\emptyset$}{\textcolor{blue}{0.37}} & \diagbox{$\emptyset$}{\textcolor{blue}{1.20}} & \diagbox{$\emptyset$}{\textcolor{blue}{0.89}} &   \\ 
\cline{1-5}
ProFit                                   & \diagbox{1.82}{\textcolor{blue}{0.47}}        & \diagbox{0.65}{\textcolor{blue}{0.21}}        & \diagbox{0.78}{\textcolor{blue}{0.40}}        & \diagbox{1.09}{\textcolor{blue}{0.36}}        &   \\ 
\cline{1-5}
SourceXtractor++                         & \diagbox{1.84}{\textcolor{blue}{0.38}}        & \diagbox{0.60}{\textcolor{blue}{0.18}}        & \diagbox{0.75}{\textcolor{blue}{0.29}}         & \diagbox{1.06}{\textcolor{blue}{0.28}}        &   \\
\cline{1-5}
\end{tabular}}
\label{table: single sersic}
\end{table}

\subsection{Double-\sersic results}\label{sec:ds}
We now analyse the measurements from the double-\sersic simulations. Figure~\ref{fig:summary_ds} summarises the results, along with Table~\ref{table: double sersic} and Sect.\ref{sec:ds_score}.

As expected, separating the galaxy light into two components is a more degenerate problem than the single-\sersic model fitting. This is enhanced by the fact that bulges and \bt in our sample are generally small, that is the bulge component has a low S/N compared to the disc (see Fig.~\ref{fig:param_distribution}). We also note that \metryka did not provide results for the bulge-disc decomposition. It is therefore excluded from the comparison in the following sections. Another difference compared to the single-\sersic dataset is that one of the fields contained multiple bands including \Euclid NIR and Rubin filters. In the following we only show results for $3/5$ fields with VIS-only data. The multi-band dataset is analysed separately in Sect.~\ref{sec:multiband}. Finally, we note that while the simulations were made with a bulge \sersic index fixed to $n=4$, and a disc with a fixed $n=1$, we asked the participants to also model the galaxies with a free bulge \sersic index. We compare the results for free and fixed bulge fittings in Sect.~\ref{sec:fix_free}. Here, we concentrate on the model using a fixed value of $n$. Notice that because \deepleg does not fit a model, it does not have those two different versions.

\subsubsection{Bulge-to-total flux ratio}\label{sec:single_bt}
\begin{figure}
\centering
    \includegraphics[width=\linewidth]{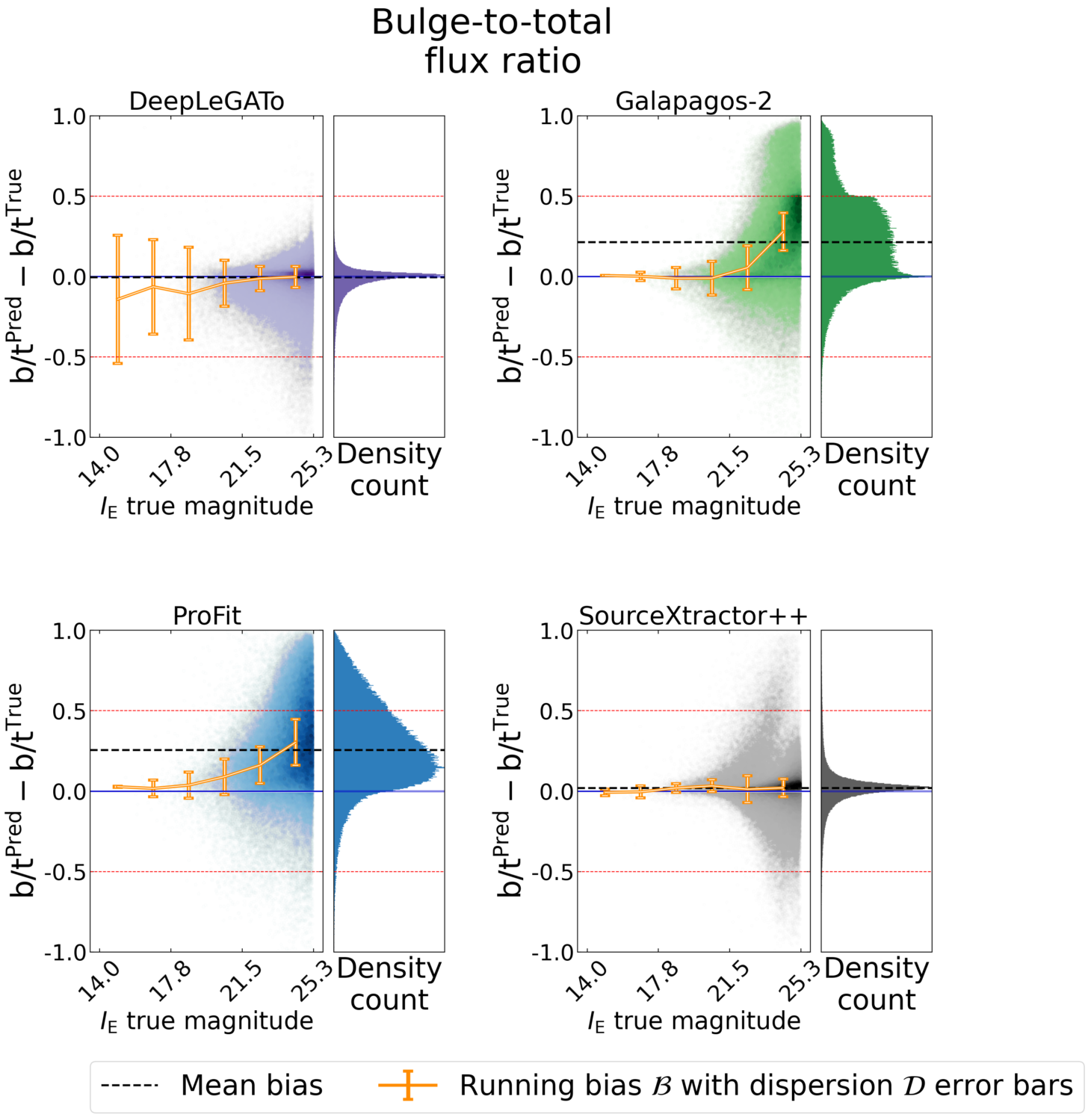}
    \caption{Fitting results for the bulge-to-total flux ratio using the double-\sersic simulation. See caption of Fig.~\ref{fig:trumpet_ss_re} for further information.}
    \label{fig:trumpet_ds_bt}
\end{figure}

We first inspect how accurately the bulge-to-total flux ratio \bt is recovered.
The results are shown in Fig.~\ref{fig:trumpet_ds_bt}. First, we see that \SE and \deepleg are less impacted by the low S/N at the faint end of the plot than the other two codes, with the trumpet shape highly concentrated towards zero bias (peaked Gaussian distribution in the histograms). \gala and \profit have highly non-Gaussian distributions of biases, with a tendency of over-estimating \btdot~for faint objects. This is obvious both in the distributions of \bt and of the bulge radius (Fig.~\ref{fig:trumpet_ds_reb}). This suggests that in cases where the bulges are small and faint, these codes tend to fail to properly disentangle the flux of the bulge from the flux of the disc. As a consequence, a part of the disc's flux gets attributed to the bulge. A possible explanation for the \SE and \deepleg ability to avoid this effect could be the use of favourable priors. Surprisingly, the figure shows that the metrics are better for faint objects, where the constraining power of the data is theoretically the lowest, and therefore the estimation is mostly driven by the prior. \SE uses an explicit prior of $0.022$ for \btdot, which matches the average \bt in the simulation. It was calibrated by the participants on a sub-sample of the dataset with known ground truth.
\deepleg also implicitly learns the prior from the data during training, by maximising the likelihood.
\gala uses arbitrary priors and initially places half the light in the bulge and half in the disc. \profit starts with reasonable initial guesses for the profile solution based on runs of the \texttt{ProFound} software on the cutouts \citep{robotham2018}, but these initial conditions remain less accurate than the ones used by \SEdot. These trends seem to confirm that the information contained in the images at the faint end is limited and therefore the final results are in most cases driven by the priors.

The summary of the metrics is provided in Fig.~\ref{fig:summary_ds}; the first row detailing \btdot. For $\VIS < 23$, \gala achieves the lowest bias, followed by \SEdot. \profit has a tendency to over-estimate \btdot, even for the brighter objects with increasing bias up to $0.37$ for the faintest objects. \gala has a similar bias in the faint end, but starts rising at fainter magnitudes ($\VIS\simeq23$ versus $\VIS\simeq19$ for \profitdot). \deepleg starts to be competitive around $\VIS=20$, and achieves the lowest bias at the faint end, followed by \SEdot. \deepleg generally under-estimates \btdot, which is the opposite trend than the one seen in the other codes. This may be due to \deeplegdot's learning being driven by the implicitly learned prior rather than by a disentangling of light based on profile fitting.
\gala has the smallest dispersion (second column) for the brightest objects, but then $\mathcal{D}$ increases to $0.14$ for fainter objects. This is comparable to \profit from $\VIS\simeq17.5$ onwards. \deepleg has a high dispersion up to $\VIS\simeq21$, which decreases from $0.5$ to $0.05$ at the faint end -- a similar dispersion to \SEdot. \SE stays relatively stable at all magnitudes, with dispersion between $\sim0.05$ (bright) and $0.10$ (faint). 
The trends for the outlier fractions are similar in all codes, with \profitdot's outliers starting to increase from $\VIS\sim19$ onwards and up to a fraction of $30\%$ for the faintest galaxies. \gala has close to no outliers up to $\VIS\simeq22$ and a fraction of $0.28$ for the faintest galaxies. Compared to \galadot, \SE has a slightly larger outlier fraction, but then keeps outliers to under $5\%$ in the faintest bins.
\deepleg retains the lowest number of outliers for $20<\VIS<26$, but reports some bad fits among the brightest objects. 

\subsubsection{Bulge half-light radius}\label{sec:ds_reb}
\begin{figure}
\centering
    \includegraphics[width=\linewidth]{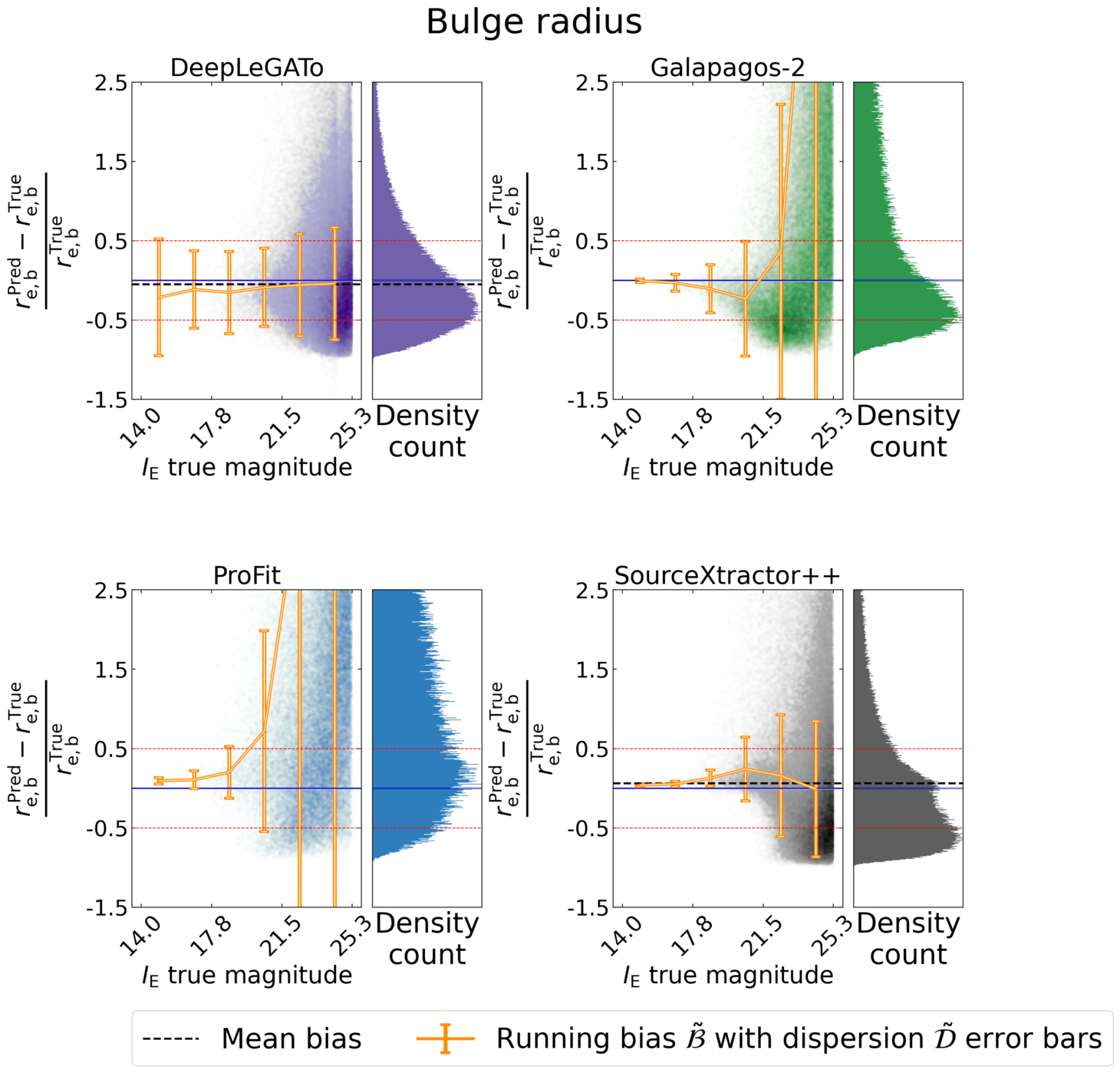}
    \caption{Fitting results for the bulge radius using the double-\sersic simulation. Notice that only four codes provided results for the double-\sersic simulation. From top to bottom and from left to right: \deeplegdot; \galadot; \profitdot; and \SEdot. See caption of Fig.~\ref{fig:trumpet_ss_re} for further information.}
    \label{fig:trumpet_ds_reb}
\end{figure}

We now inspect the estimation of the effective radius of the bulge component. Figure~\ref{fig:trumpet_ds_reb} clearly reflects the difficulty in obtaining reliable structural measurements of bulges. First, for all codes, the bias distributions are skewed towards positive values, that is an over-estimation of the true size. This can be directly linked to the fact that the bulge-to-total flux ratios are generally over-estimated. Figure~\ref{fig:param_distribution} shows that bulge radii are small, and because the bulges are generally smaller than the discs, they are submerged inside the disc profiles, making it increasingly challenging to accurately estimate their radii.

The second row of the summary figure (Fig.~\ref{fig:summary_ds}) details this observation. We note that the scale is logarithmic for the bias and the dispersion, to help appreciate the differences for faint objects. We can see that the four codes (except \deeplegdot) for the first two bins of very bright objects) have a similar value absolute for $\VIS<20$. \SE and \profit slightly over-estimate the radius while \deepleg and \gala slightly under-estimate it. For fainter objects, \deepleg keeps the lowest bias, followed by \gala and \SE and then \profit for $\VIS < 22.5$. For the challenging faint galaxies, \SE decreases to close to zero bias, which could be explained by the correct choice of priors, as discussed in the previous subsection. \gala and \profitdot's $\mathcal{B}$ rise up to approximately $10$ and $60$, respectively, at the faint end.
A similar behaviour is visible in the dispersion: \profit and \gala increase in similar ways up to $4$ at $\VIS=23$, which rises up to $80$ for \profitdot, while \deepleg and \SE keep their dispersions below $1$.
The challenge of fitting bulges becomes even more obvious when we look at the outlier fraction. Indeed, we can see that for the faintest bins -- and always according to our arbitrary definition of outlier -- more than half of the galaxies are poorly fit, close to $100\%$ for \profitdot. For brighter objects ($\VIS < 23.5$), \gala maintains the lowest number of outliers, from close to zero to around $10\%$, while \SE goes up to $\sim30\%$, and \profit $50\%$. Again, this seems to reflect the fact that when the fit can be robustly constrained by the data because it has high S/N, \gala performs well since the prior is not that relevant. 

\begin{figure}
\centering
    \includegraphics[width=\linewidth]{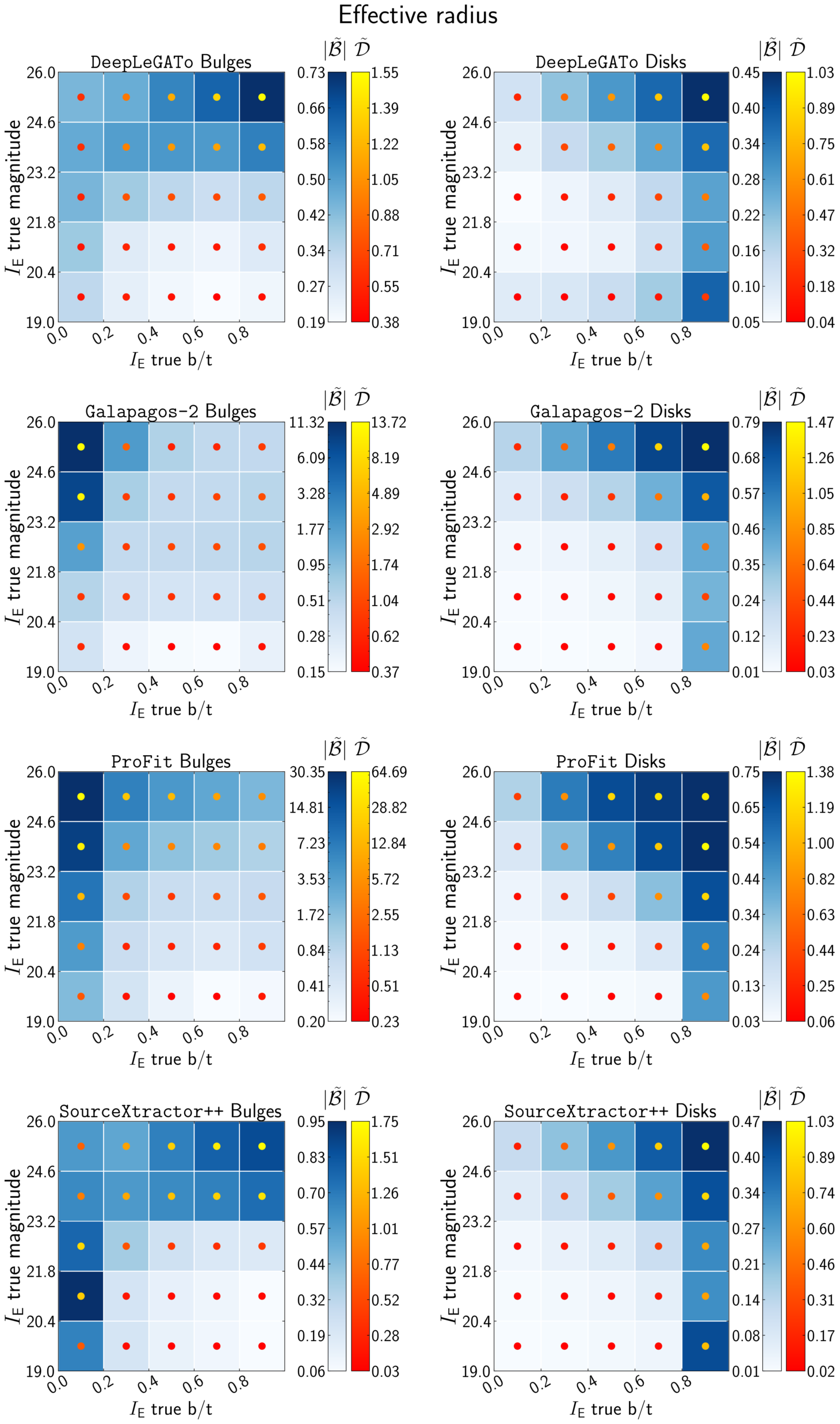}
    \caption{Absolute bias $|\widetilde{\mathcal{B}}|$ and dispersion $\widetilde{\mathcal{D}}$ for the effective radius of bulge (left column) and disc (right column) components in the double-\sersic simulation, as a function of bulge-to-total ratio ($x$-axis) and apparent $\VIS$ magnitude ($y$-axis). Each row shows a different code. For \profit and \galadot, the colour scale is logarithmic. In each panel, the colour of the squares is proportional to the mean bias $\mathcal{D}$ (lighter being smaller), and the colour of the dot inside each square indicates the dispersion $\mathcal{D}$ (redder being lower). For most of the codes, we find the expected behaviour: both the bias and the dispersion increase for faint objects, as well as at small \bt for bulges and large \bt for discs.}
    \label{fig:re_2D}
\end{figure}

In order to better understand this large bias and fraction of outliers, we show in Fig.~\ref{fig:re_2D} the different metrics as a function of the bulge-to-total fraction ($x$-axis) in addition to magnitude ($y$-axis). It is well known that the accuracy of bulge-disc decompositions are correlated with magnitude and bulge-to-total ratios.
Understanding the metrics in relation to the true value of a galaxy's \bt can help to disentangled those two effects. In this figure, we want to highlight the absolute magnitude of the bias and dispersion,
independent of their sign. The plot therefore shows for which types of objects measurement errors are large versus where they are small. For this, we compute the absolute mean bias per bin of magnitude and \bt, $$|\widetilde{\mathcal{B}}_{r_{\mathrm{e}}}| = \overline{|\boldsymbol{\tilde{b}_{r_{\mathrm{e}}}}|} \;,$$ while the dispersion is the same as for the other cases (see Eq.~\ref{eq:disp}).
In this figure, the colour of the square shows the bias $|\widetilde{\mathcal{B}_{p}}|$ (lighter colours indicate smaller bias), and the coloured discs indicate the dispersion (the redder the point the smaller the dispersion). The first column plots results for the bulge radius, the second for the disc radius, and each line is a different software code. We note that we limit the magnitudes to faint galaxies ($\VIS > 18.5$) and that for \profit and \galadot, the colour-bars are on a logarithmic scale to accommodate the large values. The expected behaviour is particularly clear for \profit (third row), which we use here to for demonstration. The bias of the bulge radius $\mathcal{B}$ becomes smaller for brighter and more bulge-dominated galaxies (lower right corner of the plot) and the dispersion is low. On the contrary, a faint galaxy with small \bt has a high bias and high dispersion. The opposite is seen for disc radii: biases are highest in faint bulge-dominated galaxies. The figure therefore confirms that most of the catastrophic fits for \gala and \profit correspond to faint galaxies with low \btdot. When the bulge component is dominant, the overall accuracy improves significantly. For example, for \galadot, the dispersion stays below $1.5$ if we remove the extreme bin of \bt (\bt $< 0.2$), and the bias remains under $2$. We can see the same behaviour for \profit if we remove the low \bt (first column), and faint objects (top row), with a dispersion and bias staying lower than $3$. 
The plot also uncovers some unexpected behaviour: \SE struggles to measure bulge and disc radii for faint bulge dominated objects, but also for brighter objects (disc radius) and bright objects with small bulges.
We encourage the reader to go to the online platform and adapt those graphs according to their interests, for example removing the extreme cases, for a better visualisation.

\subsubsection{Disc half-light radius}
\begin{figure}
\centering
    \includegraphics[width=\linewidth]{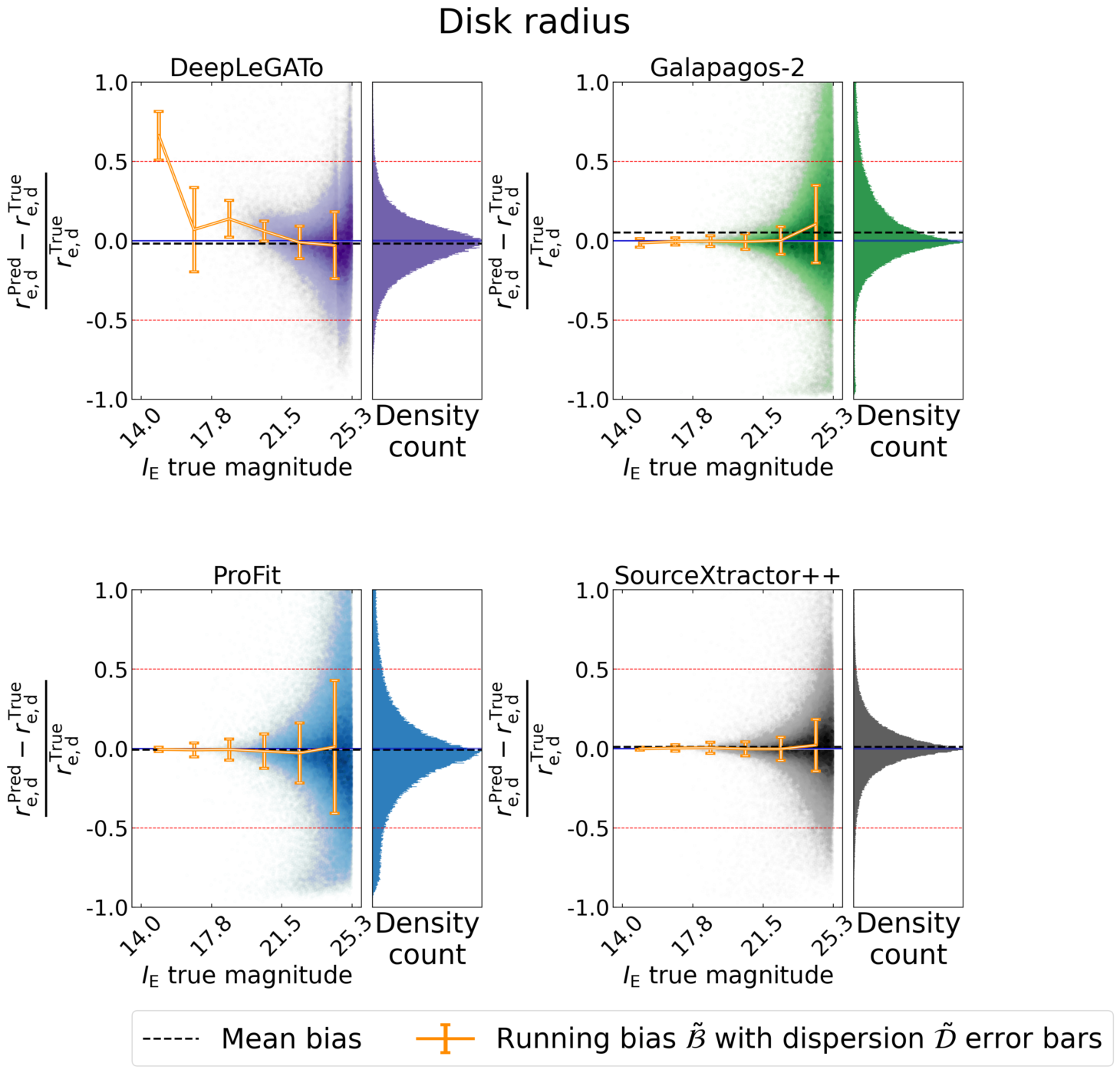}
    \caption{Fitting results for the disc radius using the double-\sersic simulation. See caption of Fig.~\ref{fig:trumpet_ss_re} for further information.}
    \label{fig:trumpet_ds_red}
\end{figure}

Figure~\ref{fig:trumpet_ds_red} shows the trumpet plots for the half-light radius measurements of the disc component. Results are noticeably more symmetric than for the bulge component and in fact are similar to the results reported for the single-\sersic case. One noticeable difference is the bias of \deeplegdot, which is inverted; bright galaxies are estimated with larger discs compared to the truth. As previously discussed this is related to discs generally being larger than bulges and the small bulges contained in the simulations. While being symmetric, the `trumpets' (and thus the bias distributions) are significantly wider, with prominent wings in the histograms.

The third row of Fig.~\ref{fig:summary_ds}, confirms that the overall reliability of the estimation of the disc structural parameters is comparable to the single-\sersic \re fit, with a slightly higher bias and dispersion. Beyond that global view, trends for bright galaxies are opposite to these for the single-\sersic radius estimation: an over-estimation of the radius for \deepleg and an under-estimation for the three others. We can see that all codes maintain absolute biases smaller than $0.04$ -- apart from \deepleg, for galaxies brighter than $21$, and the fainter bin of \gala (with a bias of $0.15$). \SE retains its disc radius bias close to $0$ over all magnitudes except for the last one, where it goes up to $0.04$. \profit has a slightly larger negative bias at intermediate magnitudes. Similarly to bulges, the dependence on magnitude is not as obvious as for the single-\sersic case, because of the additional dependence on \btdot. However, the impact is less obvious for discs, given the \bt distribution skewed towards small bulges (Fig.~\ref{fig:param_distribution}). The second column of Fig.~\ref{fig:re_2D} again explores bias and dispersion for \bt and magnitude trends. It shows that accuracy increases for bright objects and low \btdot.
Regarding the dispersion (Fig.~\ref{fig:summary_ds}), we can see a steady increase with magnitude, peaking at $0.19$, (\SEdot), $0.21$ (\deeplegdot), $0.30$ (\galadot), and $0.47$ (\profitdot).
The outlier fraction is less linear but with similar ranking. \profit has the lowest fraction of bright outliers, but is the highest in the faint bins. For the faint bin, \deepleg has the smallest fraction, followed by \SEdot. The fractions in the last bins are nevertheless higher than for the single-\sersic fit, with respectively, $5\%$, $10\%$, $29\%$, and $30\%$ for \deeplegdot, \SEdot, \galadot, and \profitdot.

\subsubsection{Bulge axis ratio}
\begin{figure}
\centering
    \includegraphics[width=\linewidth]{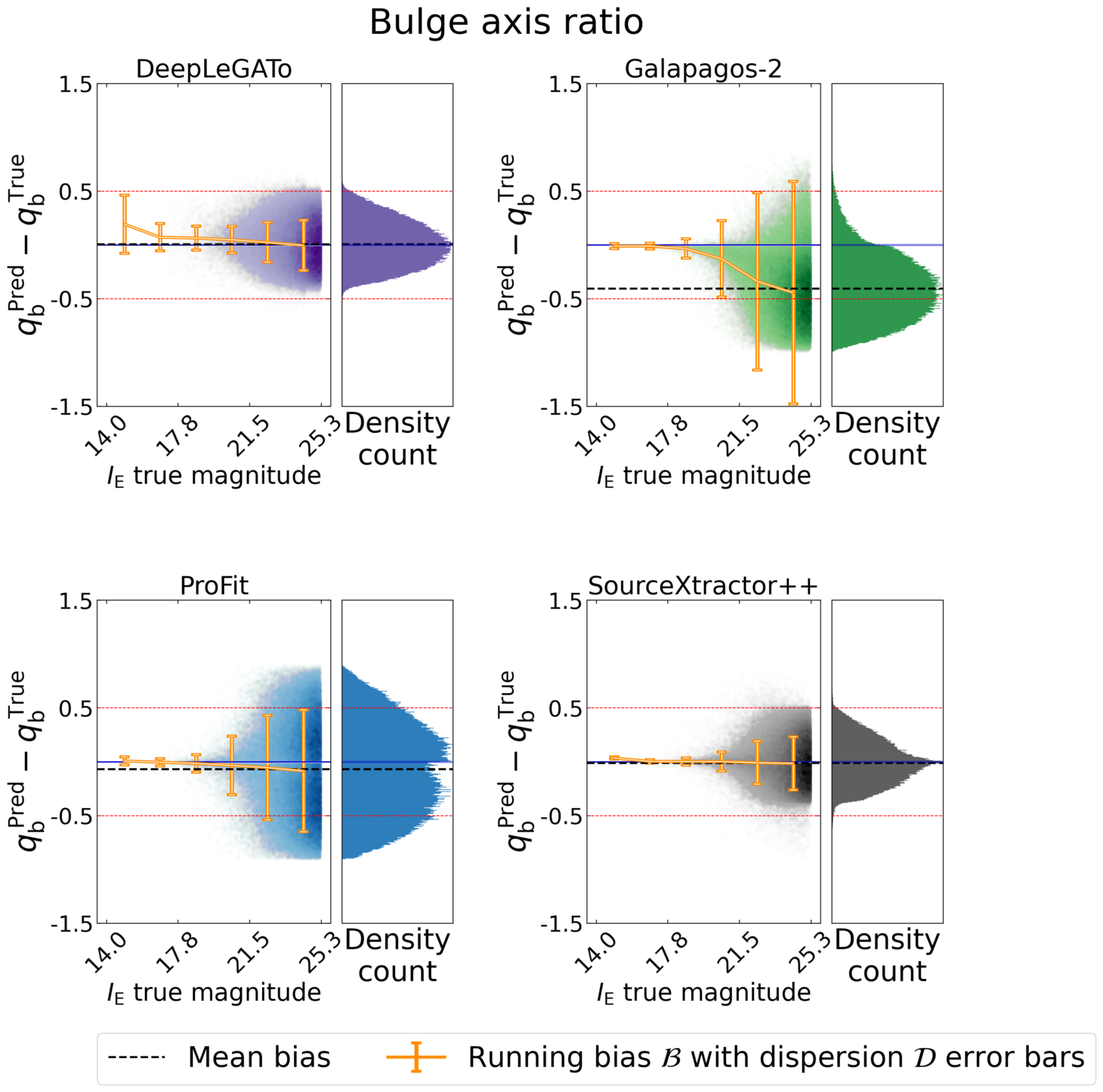}
    \caption{Fitting results for the bulge axis ratio using the double-\sersic simulation. See caption of Fig.~\ref{fig:trumpet_ss_re} for further information.}
    \label{fig:trumpet_ds_qb}
\end{figure}

Figure~\ref{fig:trumpet_ds_qb} presents the accuracy in the estimation of the axis ratio \ba of the bulge components. The characteristic trumpet shape is no longer preserved, and distributions tend to be flatter, especially for the faint objects.
These results are quantified in the fourth row of Fig.~\ref{fig:summary_ds}. \SEdot, \profitdot, and \deepleg maintain an absolute bias smaller than $0.1$ for $17 < \VIS < 26$. \SE has close to no bias, while \profit has a tendency to under-estimate the bulges \badot, that is predicting galaxies that are too elongated. It is the opposite for \deeplegdot, which over-estimates \badot, especially for the brightest galaxies. \gala is well calibrated for $\VIS < 19$, and then starts to under-estimate \badot, with a negative bias down to $\mathcal{B}=-0.42$ on the faintest galaxies.
For the dispersion $\mathcal{D}$, \deepleg and \SE achieve the lowest values for faint objects, around $0.25$. \profit and \gala have a strong increase for $\VIS > 20$, up to $0.5$ and $1$, respectively. For brighter objects, all codes except \deepleg achieve comparable results. Finally, \deepleg and \SE achieve a very low outlier fraction only with few percent.
For $\VIS < 20$, the three codes (excluding \deeplegdot) behave similarly, but then $\mathcal{O}$ rises for \profit and \gala for $\VIS < 22.5$, and ends at $0.3$ for \profit and $0.42$ for \galadot. \deeplegdot's fraction of outliers ranges from approximately $100\%$ (bright) to $1\%$ (faint).

We also investigated the 2D distributions of the metrics as a function of magnitude and \btdot, in the same way as we did for the radius in Fig.~\ref{fig:re_2D}. We found that removing cases with extreme \bt significantly improves the results at all magnitudes. We let the interested readers explore this behaviour with the online tool.

\subsubsection{Disc axis ratio}

\begin{figure}
\centering
    \includegraphics[width=\linewidth]{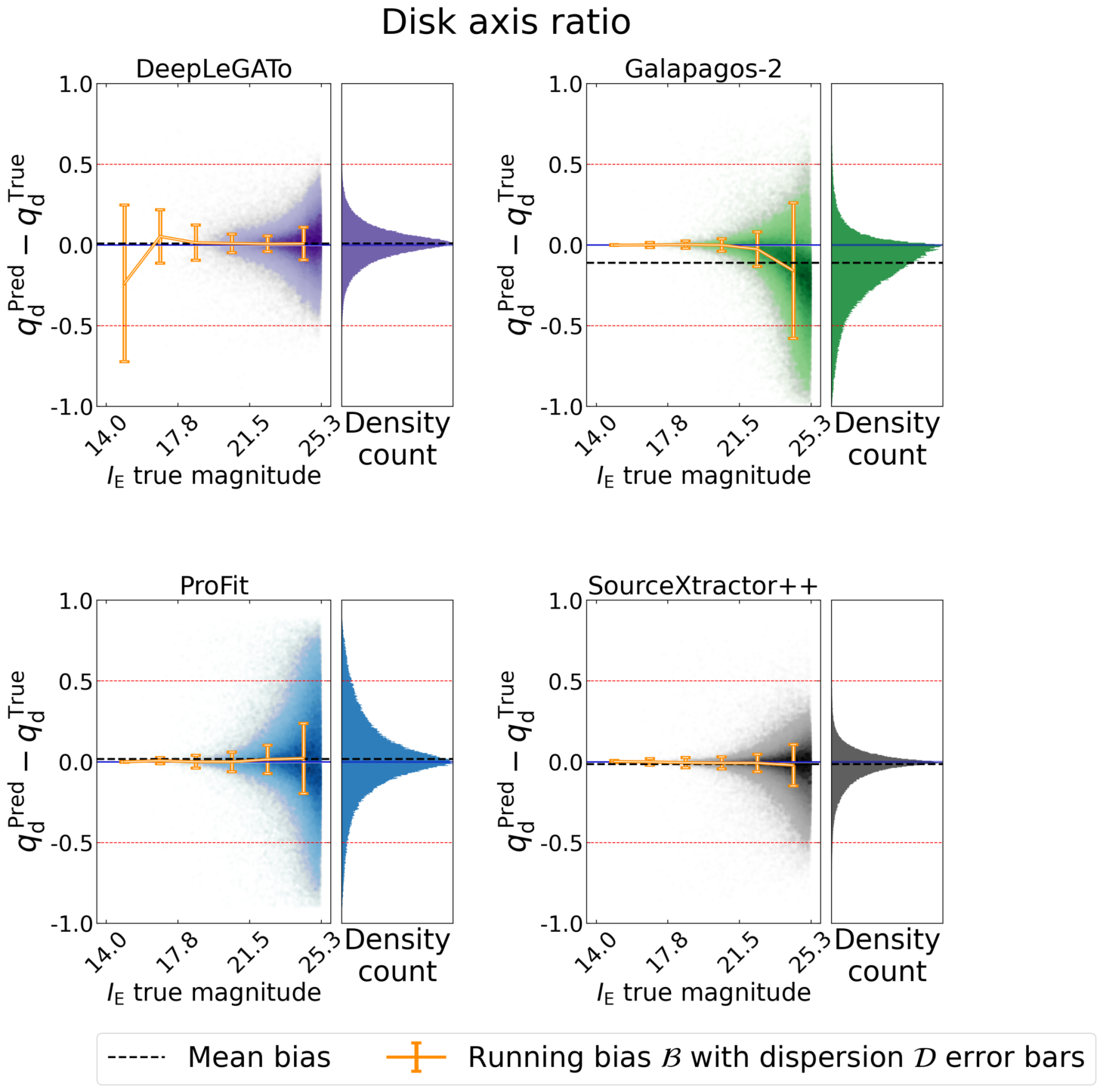}
    \caption{Fitting results for the disc axis ratio using the double-\sersic simulation. See caption of Fig.~\ref{fig:trumpet_ss_re} for further information.}
    \label{fig:trumpet_ds_qd}
\end{figure}

In general, software packages were able to measure the axis ratio \ba of the disc components (Fig.~ \ref{fig:trumpet_ds_qd}) more accurately than for the bulges. They are comparable to results from the single-\sersic case, albeit with a higher dispersion and a larger negative bias for faint objects for \gala which tends to under-estimate \badot.

\begin{figure}
\centering
    \includegraphics[width=\linewidth]{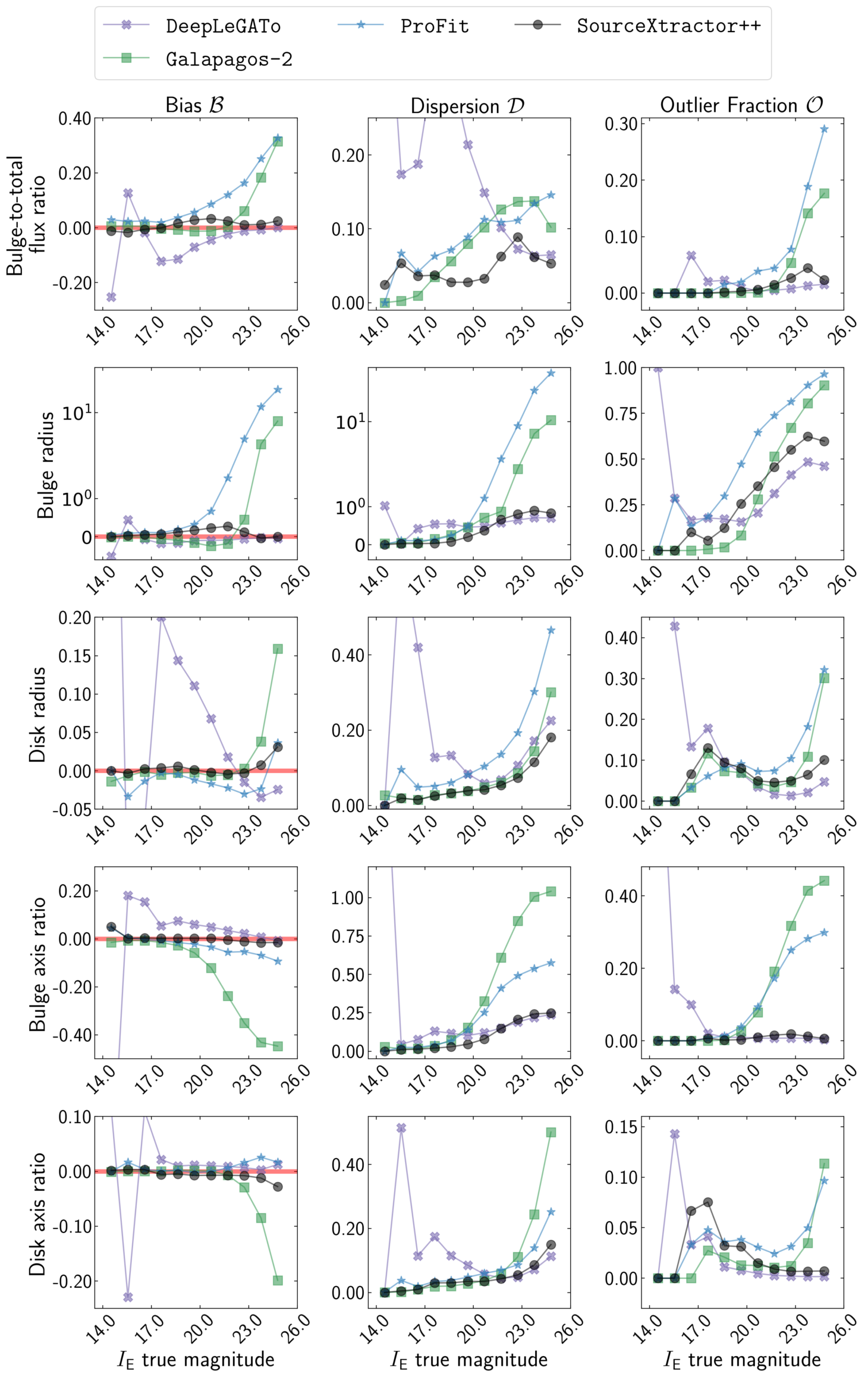}
    \caption{Summary plot for the double-\sersic simulations. From top to bottom: bulge effective radius, disc effective radius, bulge axis ratio, disc axis ratio, and bulge over total flux ratio. See caption of Fig.~\ref{fig:summary_ss} for further information.}
    \label{fig:summary_ds}
\end{figure}

We can make a more in-depth comparison of the metrics by looking at the last row of Fig.~\ref{fig:summary_ds}. Their general behaviour is comparable to the bulge axis ratio, but with smaller values. The absolute bias remains smaller than $0.3$ for all codes for $\VIS < 23$ (apart from \deepleg which is again unreliable for bright objects). \galadot's and \profitdot's biases are well calibrated for $\VIS < 22.5$, but then decline to a value of $-0.2$. For the faintest bins, \SE has a slight tendency to under-estimate the axis ratio, while \profit and \deepleg over-estimate it.
For the dispersion, \gala is also the best calibrated for $\VIS < 21$, but increases up to $0.45$ for the faintest bins, while \profit increases to $0.23$, \SE to $0.18$, and \deepleg to $0.15$. \deepleg again starts to be comparable to other codes for $\VIS > 20$, and improves to achieve the smallest dispersion for faint objects. 
\gala is also the best calibrated for $\VIS \lesssim18$ for the fraction of outliers, with less than $4\%$ of outliers, and up to $12\%$ in the faintest bins. It is still the second lowest for intermediate bins, followed by \SE and \profit by a few percent. From magnitudes$18$ to $26$, \deepleg achieves the lowest number of bad fits, below $3\%$, followed by a fraction of percent by \SE in the faintest bins. At intermediate magnitudes ($\VIS \simeq 17$), \deepleg and \profit increase to around $5\%$, and \SE to $7\%$.

\subsubsection{Multi-band fits}\label{sec:multiband}
Galaxies change appearance with varying wavelengths \citep{kelvin2012, vulcani2014, kennedy2015}.
As a result, the chosen waveband may influence the classification and the determination of a galaxy structural parameters (see e.g., \citealp{haeussler2022} for a detailed discussion). As discussed in the introduction, in addition to the VIS images, which deliver the highest spatial resolution, \Euclid will also provide NIR images in three filters. In addition, a variety of ground-based surveys such at the LSST will overlap with the \Euclid footprint.
While the main focus of the Euclid Morphology Challenge is on VIS, we included the option to test the capability of software packages to fit images in multiple wavelength ranges. 
The multi-wavelength simulations we provided are rather simplistic, with only the total magnitude and the bulge-to-total ratio \bt changing with wavelength. While the first is extensively analysed in \partonedot, we focus here on the results for \btdot. We expect that \bt is best recovered in VIS and challenging in other bands due to their lower S/N, lower resolution, noise correlation, and artefacts related to the re-sampling. However, it is interesting to check whether the constraints provided by VIS help to improve the morphology estimated in the lower resolution images.

\begin{figure}
\centering
    \includegraphics[width=\linewidth]{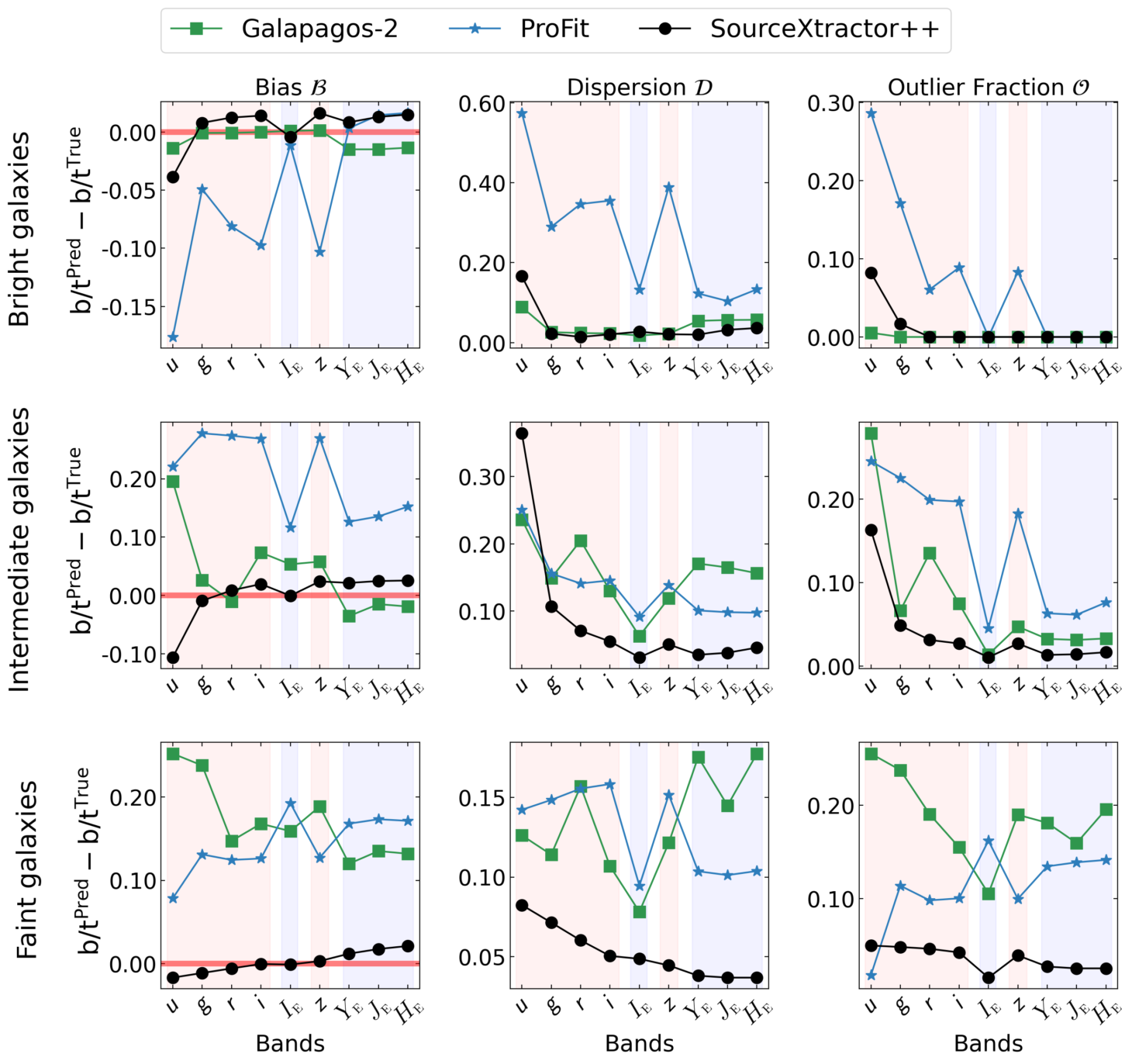}
    \caption{Results for the fitting of \bt in double-\sersic multi-band data. The three lines represent the results for three different selections of galaxies. From top to bottom: bright galaxies ($14 < \VIS < 20$); intermediate galaxies ($20 < \VIS < 23$); and faint galaxies ($23 < \VIS < 26$). The three columns represents our three metrics, the bias $\mathcal{B}$, the dispersion $\mathcal{D}$, and the outlier fraction $\mathcal{O}$, which are plotted on the $y$-axis of the corresponding columns. The $x-$axis represents the different bands, ordered by increasing wavelength ($\VIS$ overlaps with $g$, $r$, and $i$). The background colours show the \Euclid bands in blue and the Rubin bands in red.}
    \label{fig:bt_multiband}
\end{figure}

We received multi-band fitting measurements from \galadot, \profitdot, and \SEdot. Not every team interpreted the task to provide multi-band fitting in the same way and thus methods and decisions vary from code to code.
The \gala team ran all the bands simultaneously to produce the different parameters. In their bulge-disc decompositions, they fixed all the parameters apart from the magnitude, for which complete freedom to vary with wavelength was ensured. The \bt we compare in Fig.~\ref{fig:bt_multiband} is constructed from these magnitude outputs.
We note that the results are only shown as a function of $\VIS$ magnitudes, which is the deepest image by far. The strength of codes like \gala lies in improvements for shallower data,  like the NIR images. These can be explored in the online tool.
\SE also fitted all the bands in a joint analysis, with the exception of \btdot, which \SE provides directly. This means that the \bt parameter was fit independently in each band and the overall model amplitude could scale freely.
\profit fitted all bands independently, and thus galaxies can have different structural parameters in the different bands. This choice disadvantages the fitting process in the faint or low S/N bands (filters with narrow pass-bands). It did however give us a good indication that $\mathcal{B}$, $\mathcal{D}$, and $\mathcal{O}$ increase for all morphological parameters that we probe, from $\VIS$ to NIR y band, typically from a few percent in bright galaxies to $10$ and more percent in faint galaxies. We note that \profit has the option for a multi-band joint analysis, but this mode was not used for the challenge.

Figure~\ref{fig:bt_multiband} summarises the results of fitting \bt across the nine bands, roughly arranged by wavelength\footnote{{Recall that $\VIS$ is very wide, $\SI{550}- \SI{900}{\nano\meter}$, essentially combining Rubin's \emph{r, i}, and \emph{z} bands. See Fig.~1 of \partone for more information.}} (four \Euclid filters highlighted by blue shading and five Rubin filters highlighted by red shading), and for three classes of galaxies: bright ($14\leq\VIS\leq20$, top row); intermediate ($20\leq\VIS\leq23$, middle raw); and faint ($23\leq\VIS\leq26$, bottom row). The figure uses successful results from galaxies in the one multi-band field, combined in an overlapping catalogue of around $70\,700$ objects.
Bias $\mathcal{B}$, dispersion $\mathcal{D}$ and outlier fraction $\mathcal{O}$ increase in the five Rubin compared to the \Euclid bands. This is expected, since they are narrower and thus have less throughput, and being a ground-based telescope, Rubin's PSF is larger than \Euclid's PSF ($\ang{;;0.7}$ vs. $\ang{;;0.17}$), which leads to a lower resolution. Second, looking at the brighter galaxies, we see that the difference between \Euclid and Rubin bands is larger for \profit than for the two other software packages. This is a direct indication of the benefit of fitting images simultaneously.
A multi-band approach increases the S/N of measurements in faint bands \citep[e.g.,][]{haeussler2013}.
On the contrary, for \gala and \SEdot, we can see that there is close to no effect of the band width on the three metrics of the bright objects. The effect stays low even for intermediate and faint galaxies (second and third rows). An interesting exception is the extremely faint Rubin $u$ band, where results are especially unreliable. This result relates back to the dominance of $\VIS$, which is by far the widest, and thus offers the most information from its deeper image.

In general, results are comparable to the analysis of single-band $\VIS$ \bt measurements (see Sect.\,\ref{sec:single_bt} and Fig.~\ref{fig:summary_ds}). Differences between \gala and \SE become apparent for faint galaxies (bottom row), where \SE results are consistently more stable across all bands. The difference again highlights the importance of the prior or initial guess, especially for faint galaxies.
These results suggest that the simultaneous analysis of \Euclid and Rubin will improve the accuracy of measurements in the narrow bands, mainly by exploiting information from the deep $\VIS$ band. It also highlights the important synergies between the two experiments~\citep{2022zndo...5836022G}.

\subsubsection{Free versus fixed bulge component}\label{sec:fix_free}
\begin{figure}
\centering
    \includegraphics[width=\linewidth]{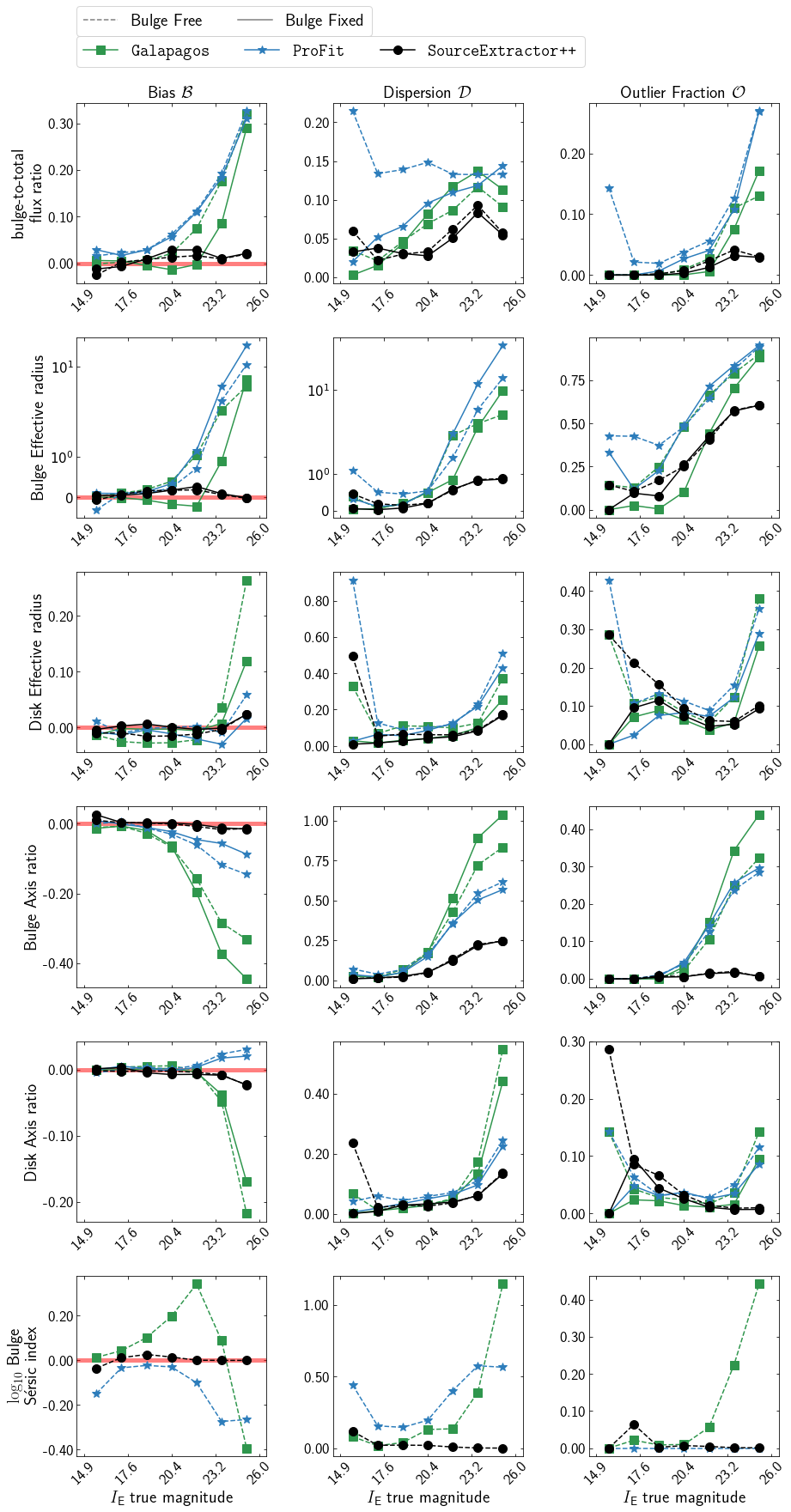}
    \caption{Comparison of the fitting between the fixed (solid line) and free (dashed line) bulge \sersic index models. We can see that overall, letting $n$ as a free parameter worsens the results according to our metrics for the disc component parameters and \btdot, while it is less clear for the bulge parameters. In addition to the usual double-\sersic parameters, we show in the last row the fitting of the bulge \sersic index for the ``bulge free'' model.}
    \label{fig: comparison_nbfree}
\end{figure}

We additionally asked participants to perform the bulge-disc decomposition with two different settings. In the first, participants kept the \sersic index of the bulge fixed to $n=4$, which is also the value that was used in the simulations. In the second setting, the \sersic index was set as a free parameter.
A comparison is provided in Fig.~\ref{fig: comparison_nbfree}, where dashed lines indicate the case of a free bulge \sersic index and solid lines when the bulge is fixed. The disc effective radius and axis ratio measurements (right column) deteriorate in all codes when the bulge \sersic index is left free, from a few percent to $10\%$ in extreme cases (fainter objects). 
The effect is weakest for \SEdot, which seems to be less sensitive to the change of model. As discussed before, the favourable results could come from their choice of prior for $n$. We notice the most dramatic effect of switching from free to fixed bulge \sersic index in \profitdot, which increases the dispersion of \bt measurements by more than $10\%$ for bright objects. 
Changes for the bulge component are less clear, with some codes achieving better results with $n$ free and others worse. For instance, \gala has a higher bias when inferring the bulge effective radius with a free bulge \sersic index, while \profit obtains a better result for bias and dispersion, but much higher outlier fraction for bright objects. Finally, the change in the bulge components for \SE is very small, which could confirm the interpretation about priors, for which we saw in the previous section that it is stronger for bulge components.
In some cases, we observe a significant difference between the free and fixed \sersic index models especially for the very bright objects. We cannot find a straightforward explanation for this behaviour, but we highlight again that this bin represents only a handful of objects.
Finally, the last row of Fig.~\ref{fig: comparison_nbfree} illustrates how well a fixed \sersic index $n=4$ is recovered when the parameter is left free in the model. We observe the same type of behaviour as for the other bulge-component parameters: \SE achieves the lowest bias and dispersion, driven by an appropriate prior selection, which is in this case perfectly known. \gala achieves comparable results for the very bright objects, but then tends to over-estimate $n$ for faint galaxies, that is giving too steep bulge profiles. The dispersion also gets higher, as for \profitdot. The effect on the bias is the opposite for \profitdot, which always under-estimates $n$.

\subsubsection{Global scores}\label{sec:ds_score}
Finally, Table \ref{table: double sersic} presents the scores $\mathcal{S}$ for the double-\sersic simulations, fitted with a fixed (in red) and a free (in black) bulge \sersic index.
The score reflects and summarises the analysis previously discussed. First, the average score for the bulge component parameters (first number of the last column) is much larger than the values obtained for the disc parameters. It is indeed penalised by the large values of bulge radius fitting, which range from $2.42$ to $89.37$. For the disc components, we see that the average score is much lower than for bulges, with values smaller than $1.2$ and hence comparable to the single-\sersic simulation. \deepleg again scores lowest for the two components, since it maintains accuracy for faint objects, which are the majority of the dataset. This outweighs its poor performance at the bright end. On the contrary, software packages that achieve better results for the bright objects are penalised by their lower performance at the faint end. 
The table also summarises differences between free and fixed \sersic index, as discussed in the previous section. 

\begin{table*}
\centering
\caption{Comparison of the scores $\mathcal{S}$ obtained by the different codes in all structural parameters for the double \sersic simulation (with a fixed bulge \sersic index fit in red, and with with a free bulge S\'ersic index fit in black). The last column is the mean of the parameters, first for the bulge components and $\mathrm{b/t}$, second for the disc components and $\mathrm{b/t}$. $\texttt{DeepLeGATo}$ is not a model-fitting algorithm, and thus has no fixed or free bulge \sersic index modes. A smaller $\mathcal{S}$ means a better fit.}
\scalebox{0.8}{
\begin{tabular}{|c|c|c|c|c|c|c|} 
\hline
\diagbox{Free bulge fit}{\textcolor{red}{Fix bulge fit}} & $\mathcal{S}_{r_{\rm{e, b}}}$ & $\mathcal{S}_{r_{\rm{e, d}}}$ & $\mathcal{S}_{q_{\mathrm{b}}}$  & $\mathcal{S}_{q_{\mathrm{d}}}$ & $\mathcal{S}_{\rm{b/t}}$ & $\mu_\mathrm{b}, \mu_\mathrm{d}$ \\ 
\hline
~\texttt{DeepLeGATo} & \diagbox{$\emptyset$}{\textcolor{red}{2.42}} & \diagbox{$\emptyset$}{\textcolor{red}{0.49}} & 
\diagbox{$\emptyset$}{\textcolor{red}{0.50}} &
\diagbox{$\emptyset$}{\textcolor{red}{0.21}} &
\diagbox{$\emptyset$}{\textcolor{red}{0.23}} &
\diagbox{$\emptyset$}{\textcolor{red}{1.05 ; 0.31}} \\ 
\hline
\texttt{Galapagos-2} & \diagbox{17.45}{\textcolor{red}{23.13}} &
\diagbox{1.46}{\textcolor{red}{0.91}} &
\diagbox{2.74}{\textcolor{red}{3.50}} &
\diagbox{1.29}{\textcolor{red}{0.51}} &
\diagbox{0.99}{\textcolor{red}{0.95}} &
\diagbox{7.06 ; 1.26}{\textcolor{red}{9.19 ; 0.96}} \\ 
\hline
\texttt{Profit} & \diagbox{37.65}{\textcolor{red}{89.37}} & 
\diagbox{1.37}{\textcolor{red}{1.14}} &
\diagbox{1.92}{\textcolor{red}{1.76}} &
\diagbox{0.60}{\textcolor{red}{0.51}} &
\diagbox{1.22}{\textcolor{red}{1.17}} &
\diagbox{13.6 ; 1.06}{\textcolor{red}{30.76 ; 0.93}} \\ 
\hline
\texttt{SourceXtractor++} & \diagbox{3.00}{\textcolor{red}{3.00}} &
\diagbox{0.51}{\textcolor{red}{0.48}} & 
\diagbox{0.54}{\textcolor{red}{0.53}} &
\diagbox{0.29}{\textcolor{red}{0.29}} &
\diagbox{0.27}{\textcolor{red}{0.26}} &
\diagbox{1.26 ; 0.36}{\textcolor{red}{1.26 ; 0.34}} \\
\hline
\end{tabular}}
\label{table: double sersic}
\end{table*}

\subsection{Realistic simulation results}
\label{sec:real}
We call `realistic' simulations those based on an approach using deep neural networks (Sect.~\ref{sec:realistic_simulations}). Figure~\ref{fig:real_summary} summarises the results, along with Table~\ref{table: single sersic} and Sect.\ref{sec:real_scores}.

These are therefore inherently different, and by design more closely resemble real galaxies. In our analysis, each galaxy is characterised by the three parameters effective radius \redot, axis ratio \badot, and \sersic index $n$, comparable to the description of single-\sersic simulations, but with more complex morphologies. We also notice that, as explained earlier in Sect.~\ref{sec:data_real}, the limited control of the simulation input parameter can induce a bias in the true parameters. A comparison with the results of the \sersic simulations is therefore not straightforward, but it offers valuable insights into what impact complex structures have on galaxy profile fitting. Notice that only \galadot, \profitdot, and \SE provided predictions for the realistic simulation, which is why figures are limited to three panels. To the best of our knowledge, a comparison study on the same set of realistic galaxy morphologies has not been conducted before.

\subsubsection{Half-light radius}

\begin{figure}
\centering
    \includegraphics[width=\linewidth]{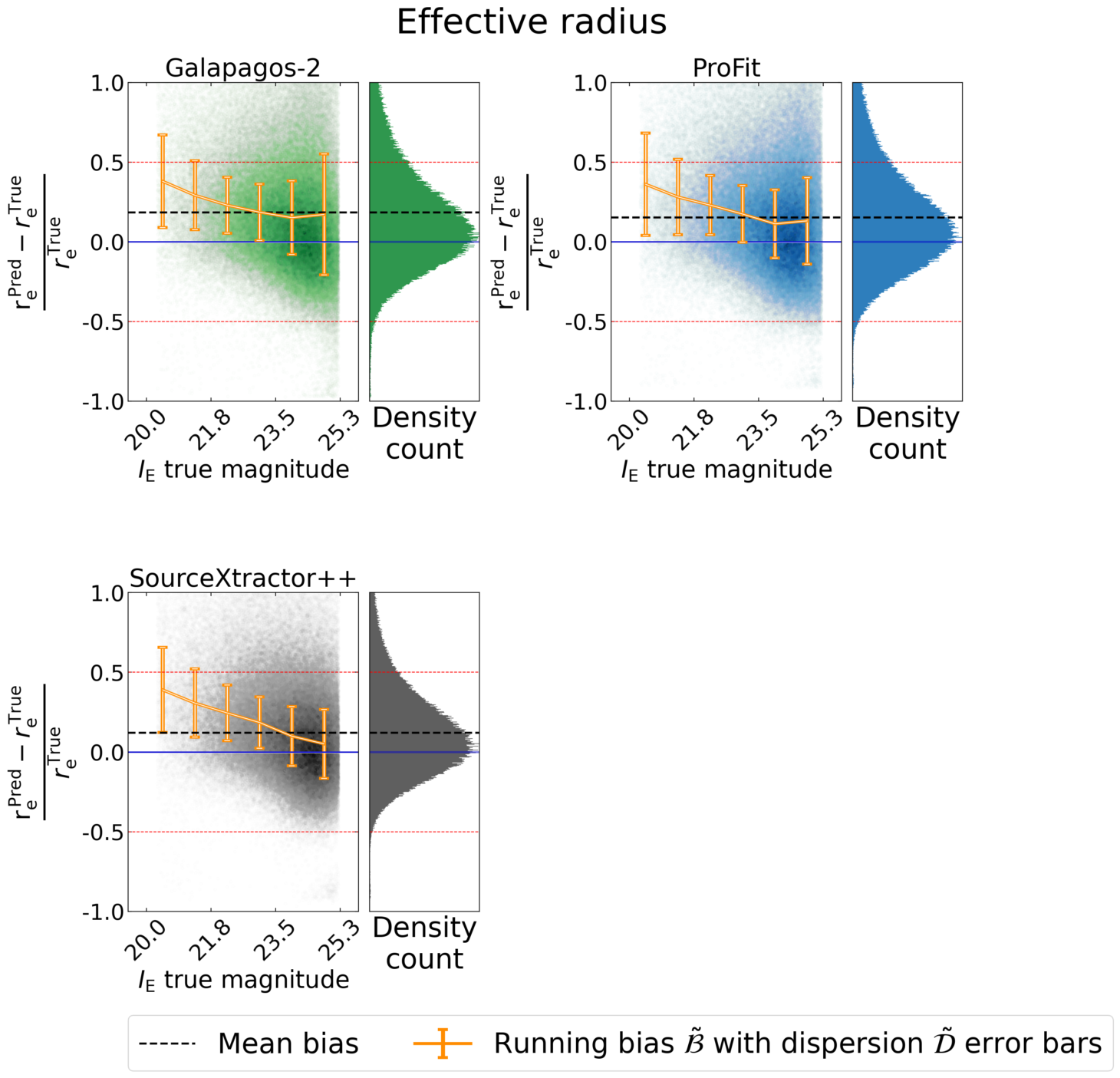}
    \caption{Fitting results for the effective radius of the realistic simulation. Notice that only three codes provided results for the double-\sersic simulation. From top to bottom and from left to right: \galadot, \profitdot, \SEdot. See caption of Fig.~\ref{fig:trumpet_ss_re} for further information.}
    \label{fig:real_trumpet_r}
\end{figure}

Figure~\ref{fig:real_trumpet_r} shows the scatter plot for the half-light radius based on the deep learning simulations. The overall performance is degraded compared to the single-\sersic simulation, meaning that the distributions are wider and the measurements are significantly biased, especially at the faint end. However, these factors are difficult to disentangle. A decreasing accuracy is expected because the \sersic model is less suited for complex surface-brightness profiles compared to the analytic simulations. However, the process of generating these simulations is not free of biases either. As explained in Sect.~\ref{sec:data}, galaxies in this case have been generated following a data-driven approach trained on HST observations. The mapping between structural parameters and galaxy images is also learned empirically and it is therefore not perfect. Interestingly, all codes behave similarly, with an under-estimation of the radius for faint objects. This might be an indication that the degradation could be partly explained by the fact that the input is also biased. If this is the case, then we can expect performance for all metrics on real \Euclid observations to be between the analytic and the realistic results.

The top row of Fig.~\ref{fig:real_summary} (we note that because for the realistic galaxies, bright objects were not simulated for this type of simulation, we reduce the $x$-axis range from $\VIS=20$ to $\VIS=26$) shows that the bias is similar for the three codes for $\VIS <23$, but remains more stable for \gala ($\mathcal{B}=14$), followed by \profit ($\mathcal{B}=19$) and \SE ($\mathcal{B}=24$).
The dispersion ranges from around $0.3$ ($0.39$ for \profitdot) at the bright end to $0.6$ at the faint end, with few percent of differences. This is a factor of approximately$3$ increase compared to the analytic \sersic simulation.
The fraction of outliers also increases to $15\%$ (\SEdot), $23\%$ (\profitdot), and $34\%$ (\galadot), which is slightly higher compared to the single-\sersic case. For the three codes, $\mathcal{O}$ follows a U-shape, increasing at the bright and faint ends. The increase for bright objects might again reflect a simulation bias since the model was trained with a small number of very bright galaxies, but also the fact that brighter galaxies have more structures, and thus a larger departure from a \sersic profile.

\subsubsection{Axis ratio}

\begin{figure}
\centering
    \includegraphics[width=\linewidth]{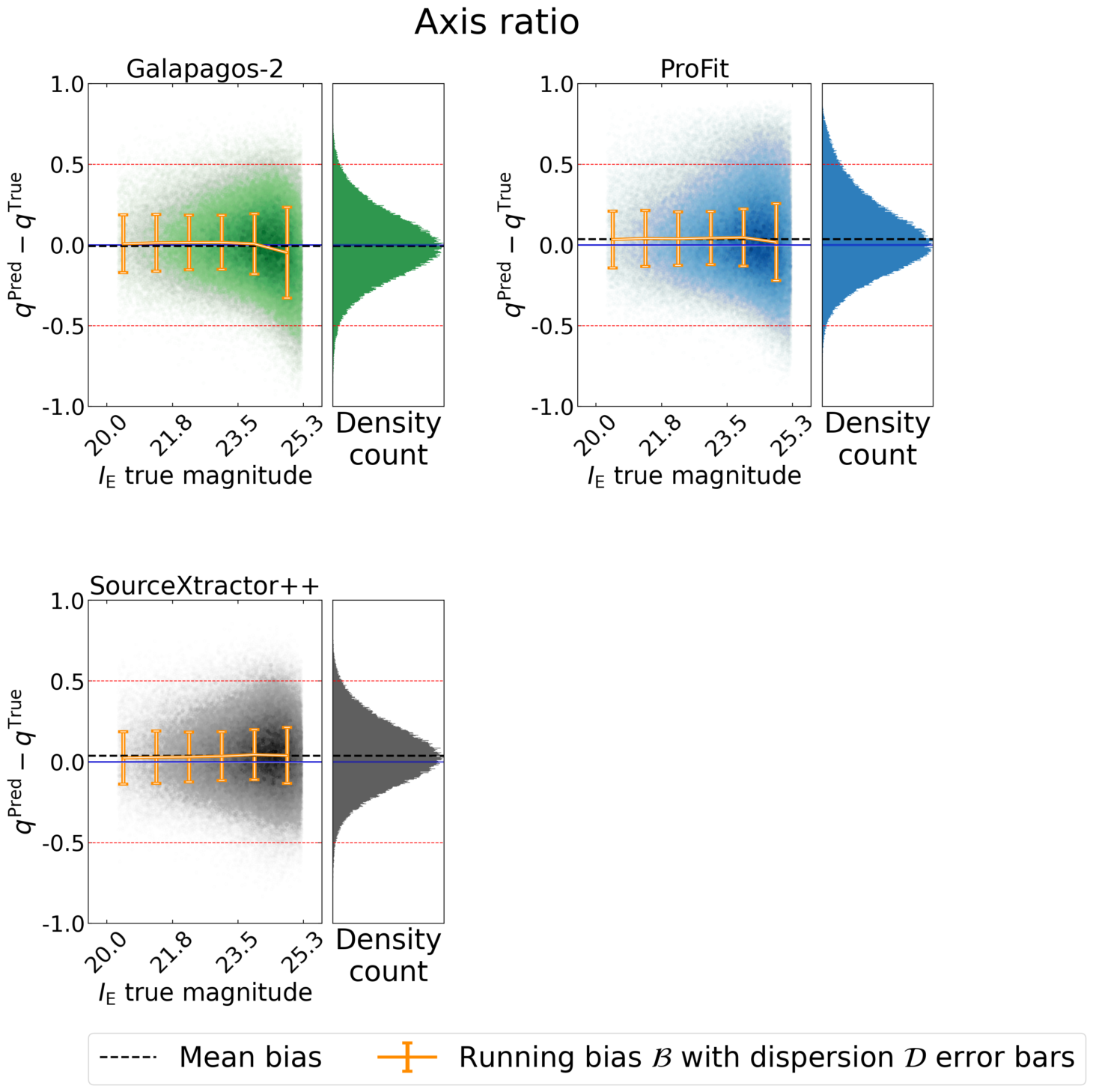}
    \caption{Fitting results for the axis ratio of the
Realistic simulation. See caption of Fig.~\ref{fig:trumpet_ss_re} for further information.}
    \label{fig:real_trumpet_q}
\end{figure}

Figure~\ref{fig:real_trumpet_q} shows the results for the galaxy axis ratio \badot. 
The first impression is similar to that for the half-light radius: the wings of the trumpet shape are wider compared to the single-\sersic case; however, the bias is no longer an issue as it was for the radius. The three codes are overall very consistent with each other.

Looking at the second row of Fig.~\ref{fig:real_summary}, we see a small bias for all codes, from less than a $0.01$ for \gala ($0.02$ for \SE and $0.035$ for \profitdot) up to $0.04$ for \profit and \SE for faint objects. As for the single-\sersic case, the bias for \gala and \profit decreases in the fainter magnitude bins, down to $-0.04$ for \galadot, but are low in general ($|\mathcal{B}|<0.04$). This might be an indication that the input simulations are less affected by systematics for this structural parameter in particular. The overall ellipticity is indeed less dependent on the galaxy surface-brightness profile.
The dispersions are somewhat higher than for the single-\sersic case, but stable for \SE around $0.16$, and from $0.17$ up to $ 0.28$ for the other two software packages. For the outliers, \SE and \gala achieve similar results, with a fraction around $2\%$ ($4\%$ for \gala at the fainter end), while \profit seems more affected by the features in the simulation, with $4.5\%$ to $7\%$ of outliers, which is approximately $10$ times higher than for the single-\sersic case (compared to $5$ times larger for the other software codes).

\subsubsection{\sersic index}

\begin{figure}
\centering
    \includegraphics[width=\linewidth]{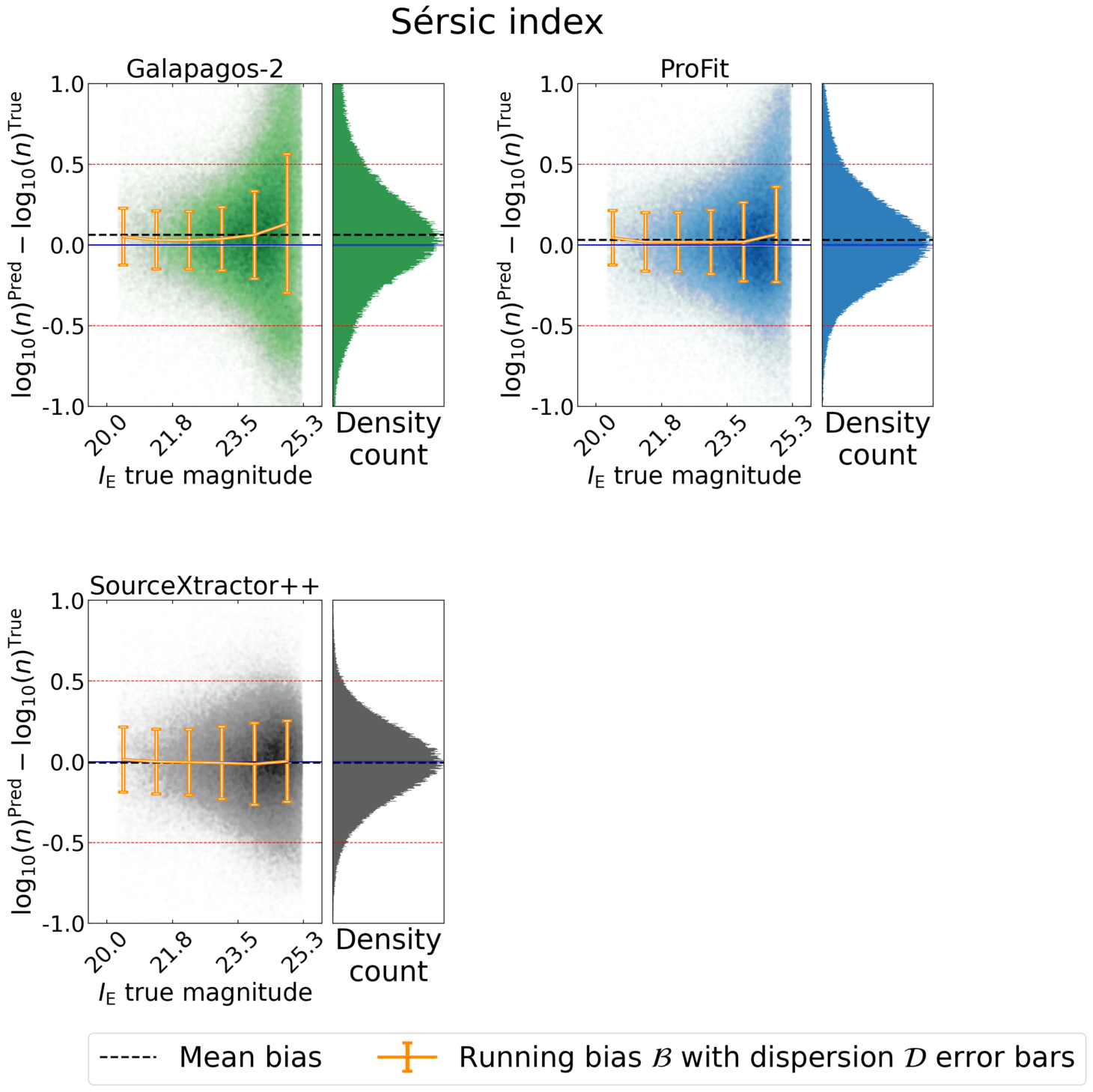}
    \caption{Fitting results for the \sersic index of the
realistic simulation. See caption of Fig.~\ref{fig:trumpet_ss_re} for further information.}
    \label{fig:real_trumpet_n}
\end{figure}

\begin{figure}
\centering
    \includegraphics[width=\linewidth]{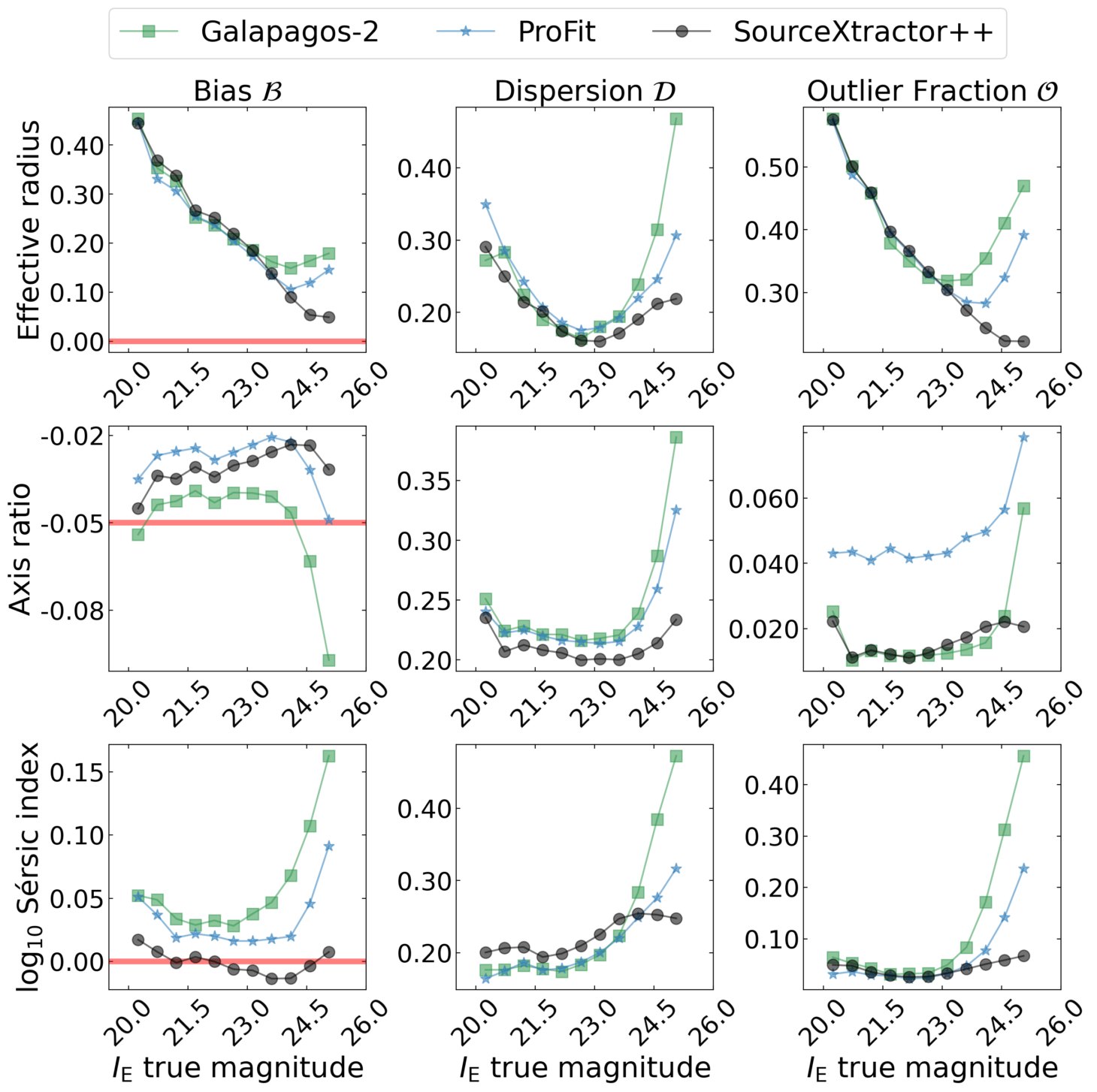}
    \caption{Summary plot for the realistic simulation. From top to bottom: effective radius; axis ratio; and \sersic index. See caption of Fig.~\ref{fig:summary_ss} for further information.}
    \label{fig:real_summary}
\end{figure}

We finally focus on the estimation of the \sersic index $n$. As explained in Sect.~\ref{sec:ss_n}, we analyse $\log_{10}(n)$ instead of $n$. Figure~\ref{fig:real_trumpet_n} is very similar to that of the axis ratio: the trumpet shape is wider than for the single-\sersic case, but with no major bias.
In the last row of Fig.~\ref{fig:real_summary}, we see that indeed, \SE has close to no bias for the all ranges of magnitude, while \profit and \gala retain a low bias ($<0.05$), with an increase up to $0.14$ for \gala in the faintest magnitude bin. 
The dispersion of \gala and \profit are very similar (and the lowest, compared to \SEdot) around $0.18$, here again until the faintest magnitudes. In the last bin, \SE achieves a reasonable bias ($0.25$), while \profit goes up to $0.30$ and \gala to $0.43$. The dispersion and biases are roughly $2$ times larger than in the single-\sersic case.
The same behaviour can be seen in the outlier fraction, with a small fraction ($5\%$) for $\VIS < 23$. The outlier fraction of \SE does not increase significantly for fainter objects, but $\mathcal{O}$ reaches $20\%$ in \profitdot, and $40\%$ in \galadot, which is, for all codes, much larger than the outlier fractions we reported from simulations created with analytical profiles.

\subsubsection{Global scores}\label{sec:real_scores}
Table~\ref{table: single sersic}, which we already discussed in Sect.~\ref{sec:ss_score}, also shows in black the results for the more realistic simulations. Overall, measurements are less accurate in the deep learning generated images, with a degradation between $2$ times to $5$ times, based on our metrics. The most stable parameter compared to the analytic simulation is \badot, followed by $n$. The hardest to capture in realistic simulations is the radius \redot. Overall, \SE and \profit achieve similar results, where \profit benefits from its high completeness, and \galadot, though less complete, achieves robust results in more complex morphologies. Average scores $\mu_\mathcal{S}$, range from $1.8$ (\galadot), to $2.5$ (\profitdot), and $3.8$ (\SEdot). 

\subsection{Uncertainty quantification}\label{sec:uncertainty}
Another goal of this challenge is to review the ability of software packages to predict the uncertainty related to each measure. Each participant was asked to provide the $1\sigma$ uncertainty for every parameter and every galaxy, called $\sigma_{p}$. To quantify the quality of this prediction, we tested whether each parameter prediction $P_{p}$ of each galaxy falls inside the predicted error interval,
\begin{equation}
T_{p} \in [P_{p}-\sigma_{p}, P_{p}+\sigma_{p}].
\end{equation}

\noindent If it does, then we attach a flag `$1$'. If it does not fall within this interval, then the prediction is flagged `$0$'. Following a frequentist approach\footnote{We only have one prediction for each object, thus we are not able to perform a proper Bayesian analysis.}, if $\sigma_{p}$ is indeed the $1\sigma$ uncertainty, then $68\%$ of the galaxies should be flagged $1$. \deepleg does not feature in this section because the network only provides point estimates and no reliable uncertainty estimation was available in the version that was used to obtain results for this challenge.

\subsubsection{Single-\sersic simulation}

\begin{figure}
\centering
    \includegraphics[width=\linewidth]{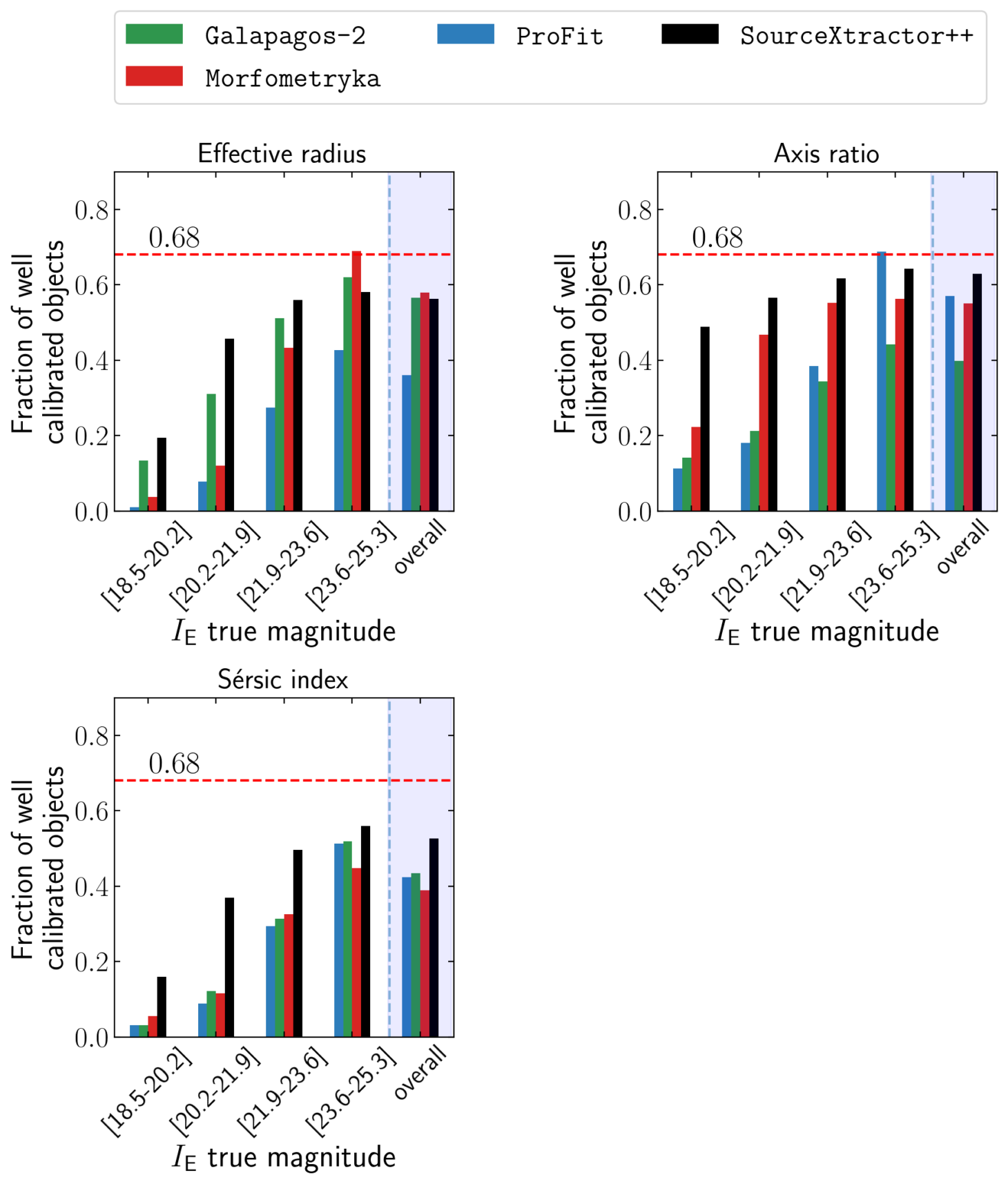}
    \caption{Uncertainty calibration for the single-\sersic simulation. The $x$-axis represents four bins of $\VIS$ magnitude. The $y$-axis shows the fraction of object per bin of magnitude for which the True value of a parameter falls in an interval of the predicted $1\sigma$ uncertainty centred on the predicted value. The final bin is for all objects, regardless of their magnitude.}
    \label{fig:ss_uncertainty}
\end{figure}

Figure~\ref{fig:ss_uncertainty} assesses how well uncertainties are predicted in each code for the single-\sersic-simulations per bin of $\VIS$ magnitude. The red dashed line indicates the fraction of objects that should be obtained if the uncertainty was well calibrated. We also show the overall calibration of the entire sample, which we present on the same plot on the right, highlighted by blue shading.

It becomes immediately obvious that the uncertainty is always under-estimated for all codes and parameters. \galadot, \metrykadot, and \SE under-estimate the uncertainties of their effective radius measurements in similar ways: around $58\%$ of object are well-calibrated (instead of $68\%$). Size uncertainty estimations are worse for bright objects, which is common for all codes. This suggests that the uncertainty is mostly related to the flux of the objects. This under-estimation is most substantial in \metrykadot, which has almost no well-calibrated bright objects, but over-estimates uncertainties in the faint end. This behaviour is similar, but less striking for \SEdot.
The middle and right panels show uncertainty calibrations for axis ratio and \sersic index measurements. The performance we just described is roughly comparable in all parameters. Nevertheless, reported uncertainties for the axis ratio seem to be best calibrated by \SE (the bar is close to $68\%$). They are also less affected by the magnitude. \profitdot's uncertainty estimation of axis ratios is also much better calibrated than for the radius, with a global score of $60\%$, similar to \metrykadot. Overall, \gala has only $45\%$ of its axis ratio measurements well calibrated. This is also the average estimation for uncertainties of the \sersic index (between $45\%$ and $50\%$ are correctly estimated), worse in bright objects than in faint objects.

\begin{figure}
\centering
    \includegraphics[width=\linewidth]{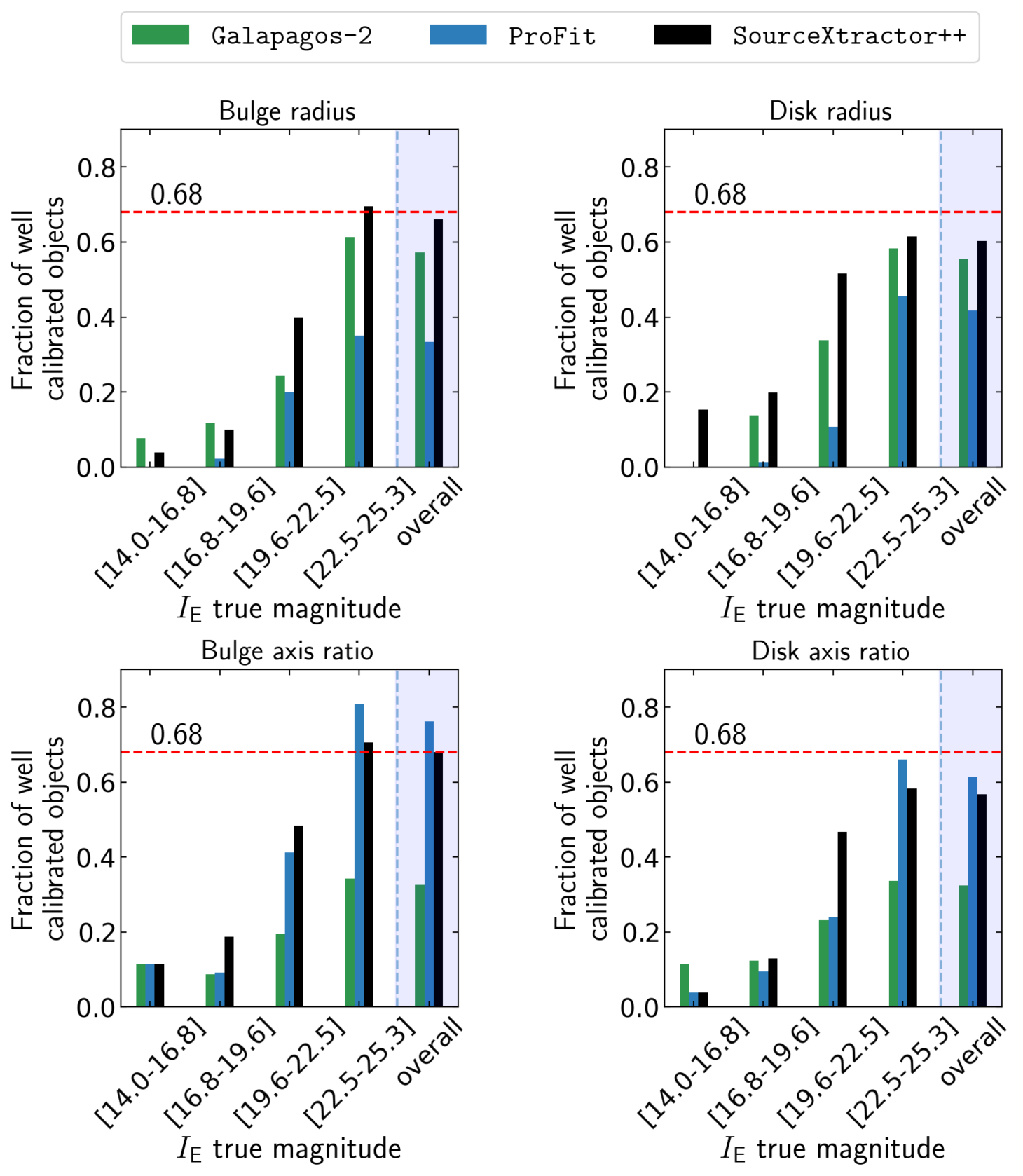}
    \caption{Uncertainty calibration for the double-\sersic simulation. See caption of Fig.~\ref{fig:ss_uncertainty} for further information.}
    \label{fig:ds_uncertainty}
\end{figure}

\subsubsection{Double-\sersic simulation}
In Fig.~\ref{fig:ds_uncertainty} we evaluate uncertainty estimations of measurements from the double-\sersic simulations. We collect and compare uncertainty estimates for three codes \SEdot, \galadot, and \profitdot.
While we see similar trends to single-\sersic estimates--better calibrations for faint than for bright objects, we also notice some improvements. For example, \SE is overall well calibrated for the bulge radius and for the bulge \badot. Importantly though, the all codes still under-estimate their uncertainties, especially in bright objects. For the disc \badot, \profit is better calibrated, with a score of $60\%$, while \gala only scores $30\%$, and the same for the bulge \badot. Some of these may be alleviated by employing \galadot-specific flags, which were not used in this analysis but introduced in \partonedot, and investigated in detail in \cite{haeussler2022}. For the bulge \badot, \profit over-estimates the uncertainty. For the disc radius, all the codes are under calibrated, with a score of about $60\%$ for \SE and \galadot, and $40\%$ for \profitdot.

\section{Summary and conclusions}
\label{sec:conclusion}

The Euclid surveys will become a reference dataset for studies involving galaxy structure and morphology due to the unique combination of a wide area and high spatial resolution. In this paper, we have described the results of the Euclid Morphology Challenge to quantify the performance of five state-of-the-art surface-brightness-fitting software packages on simulated \Euclid imaging data. 
This paper is the second part of the two papers discussing the EMC. While the companion paper (Euclid Collaboration: Merlin et al. 2022) focusses on the results related to photometry, we have focussed on the results concerning the morphological parameters only.
We compare the results after measuring structural properties of simulated galaxies in fields that mimic \Euclid observations with the VIS instrument from \deeplegdot, \galadot, \metrykadot,
\profitdot, and \SEdot. The simulations we use include single- and double-component \sersic profiles, as well as more realistic data-driven generated galaxies. In addition, one field was provided in multiple bands that included \Euclid NIR $Y_{\scriptscriptstyle\rm E}$, $J_{\scriptscriptstyle\rm E}$, and $H_{\scriptscriptstyle\rm E}$ filters and ancillary data from the five optical Rubin bands, \emph{u, g, r, i}, and \emph{z}, to test possible multi-band fitting routines.

\begin{sidewaysfigure*}
\centering
    \includegraphics[width=\linewidth]{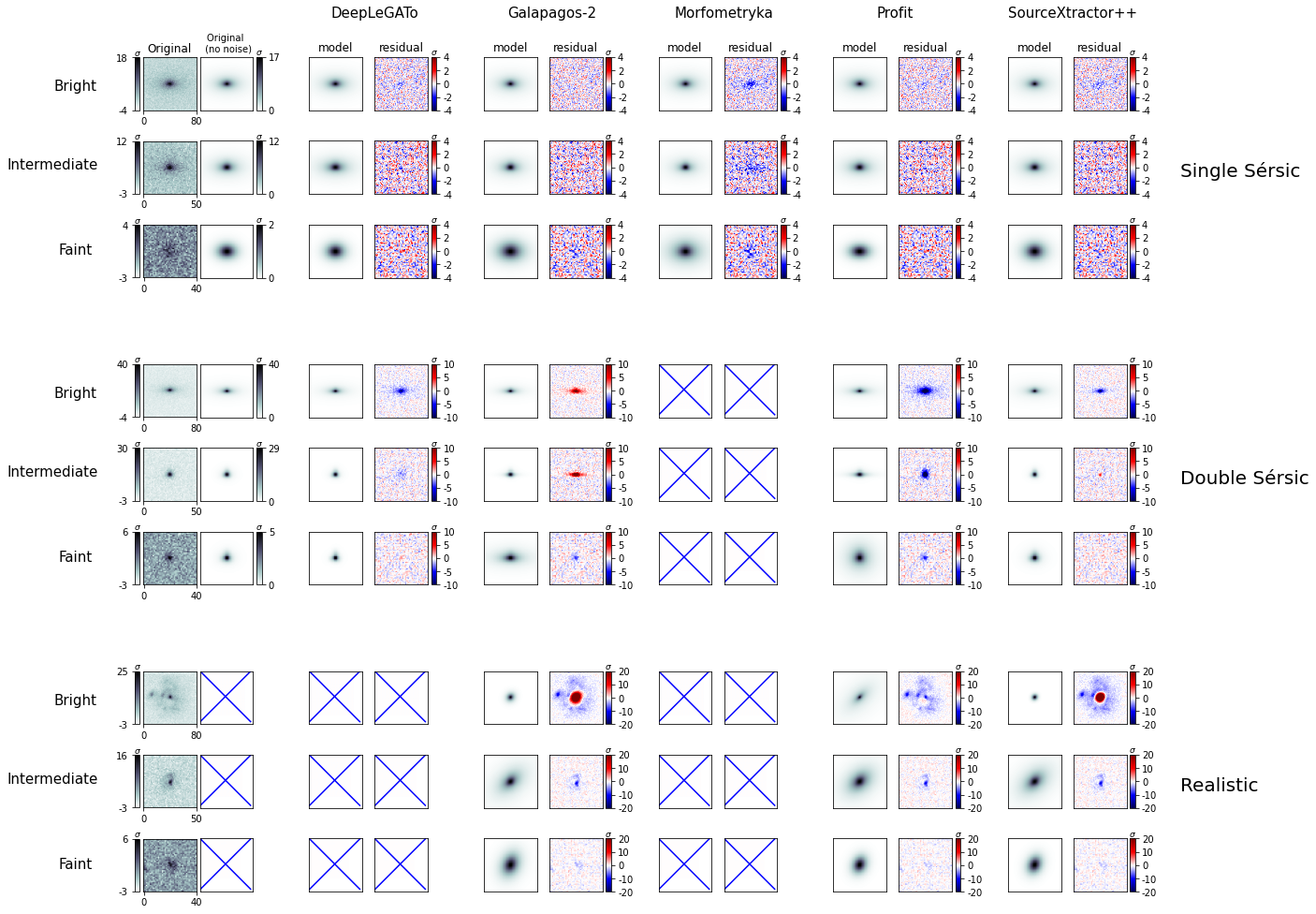}
    \caption{Illustration of residuals from the galaxy modelled with the predicted parameters by the different software packages. The plot has the same structure for the three types of simulations (single-\sersic in the first three rows, double-\sersic in the three middle rows, and realistic galaxies in the last three rows). For each type, the first row presents a bright galaxy, the second an intermediate galaxy, and the third a faint one). In the two first columns, we show the actual image of the galaxy, with and without noise. Then, each pair of columns presents: (1) the noiseless model constructed from the predicted parameters of each code, and (2) the residual between the noisy true galaxy and the noise-less model. All the colour bars are related to the Euclid Wide Survey noise level. For this reason, a perfect model would lead to a colour bar ranging from $-3\sigma$ and $3\sigma$. Finally, to better capture the details of the residuals, we clipped the value to a certain $\sigma$ value, which is fixed for all the codes and magnitudes in a certain type of simulation. For the residuals, the colour bar is symmetric, centred on zero.}
    \label{fig:residuals}
\end{sidewaysfigure*}

Figure~\ref{fig:residuals} visually summarises some of the main results of this work. This figure shows one example of a bright, intermediate, and a faint galaxy as fitted by each software package and for each type of simulation. We also show the residual images obtained after subtracting the best-fit model from the original image. The figure illustrates several of the key trends raised in this work.

Single-\sersic simulations (first three rows) yield the best results -- with the lowest residuals -- always remaining below $4\sigma$ of the noise level. All codes show very similar results, confirming that one-component fits are robust down to an S/N of $5$ for simple simulations. 

Bulge-disc decompositions (middle three rows) are more challenging to perform, as evident from the residuals of subtraction, with clear features remaining. However, the results vary for each code. \deepleg and \SE, in particular, appear to show smaller residuals at the faint end. We argue that this result is mainly due to assuming a prior distribution that is close to the true distribution, as opposed to the other codes. The fit at low S/N is therefore highly dependent on the assumed prior, reflecting the limited constraining power of the data.

The fitting of realistic simulations (last three rows) results in systematic uncertainties that arise from using simple \sersic models to describe the surface brightness distribution of galaxies at high spatial resolution. As a consequence, the residuals disclose the more complex structures that the basic model is unable to capture. For example, the brightest example galaxy (first row of the block) reveals a complex morphology with spiral arms and clumps. \gala and \SE tend to fit the bright bulge of the galaxy, but not the disc. However, in order to capture the global flux distribution of the galaxy, the bulge flux is over-estimated, leading to a very strong positive residual in the bulge, and a negative residual in the disc. \profit fits a galaxy with a lower \sersic index, which leads to a lower residual, where only the non-linear complex features of the galaxy remains. For the intermediate and faint galaxies (last two rows), we see here again some complex structures in the residual, which are very similar in all the codes. To our knowledge, this is the first time that several fitting codes have been tested on the same non-analytic simulations. 

\begin{figure*}
\centering
    \includegraphics[width=\linewidth]{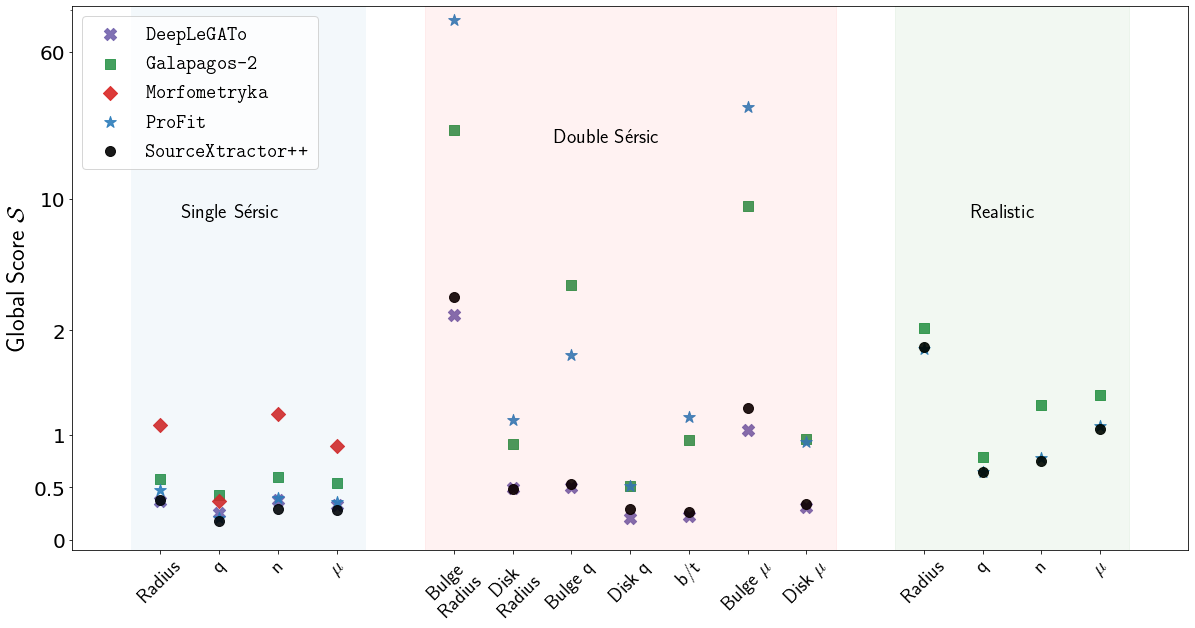}
    \caption{Summary of the global scores $\mathcal{S}_{p}$ for the single-\sersicdot, double-\sersic, and realistic simulations. The $x$-axis shows the different parameters of the different simulations (single-\sersicdot, double-\sersic, and realistic in shaded blue, red, and green, respectively). The global score $\mu_{\mathcal{S}}$ corresponds to the mean of the $\mathcal{S}_{p}$. The $y$-axis represents the corresponding global score $\mathcal{S}$, for each parameter and code, the lower the score the better.}
    \label{fig:global scores}
\end{figure*}

Figure~\ref{fig:global scores} summarises the main trends through the global score $\mathcal{S}$ for every parameter, as well as for each code and type of simulation. We emphasise that while this score is informative, it is not intended as a universal ranking of the different codes. For this reason, we provide an online interactive tool along with this publication, which allows the user to plot different quantities to accommodate for and support choices of each unique science case.

\vspace{0.1cm}
We summarise the main conclusions of our analysis below.

\begin{itemize}
    \item Single-\sersic parameters ($r_{\mathrm{e}}$, $q$, and $n$) of \Euclid VIS galaxies can be estimated with a bias and a dispersion lower than $10\%$ down to a S/N per pixel of approximately $1$. This will include imaging for $400$ million objects by the end of the mission. Despite some differences between the different software codes tools, they all achieve consistent results.

    \item We find that bulge-disc decompositions are often unstable. In particular, the properties of internal structural components (i.e. bulges and discs) can only be recovered in a reliable way ($\mathcal{B}<0.1$ and $\mathcal{D}<0.1$) when they dominate light distributions (\bt $>0.8$ for bulges and \bt $<0.3$ for discs), even for bright objects. Interestingly, we find that the bulge-to-total ratio is well recovered with less than a $10\%$ dispersion, especially by \deepleg and \SEdot. We argue that this is related to the priors assumed by these two codes, which are closer to the true distribution.
    
    \item The different software packages are also tested on galaxies generated with more realistic morphologies. We find the overall performance degrades by a factor between $2$ and $5$ on average, depending on the structural parameter in all the considered metrics, that is to say the bias, dispersion, and number of outliers. We discuss, however, that this factor might be enhanced by some intrinsic biases in the simulations.
    
    \item All codes tend to underestimate the uncertainties in the measurement by a factor of around $2$. This is even true for Bayesian codes such as \profitdot.
    
    \item A simultaneous fit of \Euclid and Rubin-like images results in an improvement of the structural parameters obtained from ground-based data, highlighting the significant synergies between the two projects.
    
    \item Deriving good initial values or priors is important for all fitting parameters, especially during a bulge-disc decomposition (e.g. \citealp{lange2016}) where the constraining power of data is limited.
    
    \item An improved model of the PSF that takes the under-sampling into account can play an important role in improving the accuracy of the shape measurements.
    
    \item While we have not performed an in-depth analysis of the computational time and resources, we can stress that there are non-negligible differences between the various software packages with, for example, \deepleg being very fast in GPUs, while a Bayesian analysis such as \profit requires longer runs. Readers can refer to the appendix of \partone for a more detailed analysis.
 
    \end{itemize}
 
We acknowledge that visual descriptions and classifications of galaxy morphologies are an important ingredient to galaxy evolution studies. While the presented analysis focusses on the parametric description of galaxy morphologies and specifically on comparing codes that measure them, other efforts within the Euclid Collaboration investigate non-parametric and visual descriptions for galaxy classifications with \Euclid. These will be part of forthcoming publications.

\medbreak
 In the following, we review and repeat general conclusions and global results of the EMC, which are also discussed in the companion paper, \partonedot. While this paper focusses on the Euclid Wide Survey, we can expect similar results for the Euclid Deep Survey \citep{laureijs2011} for the same S/N aside from irregularities due to distinct features such as higher levels of contamination or blending. Similarly, we only briefly covered the multi-band abilities in this paper. Both require further analysis. Immediate future work will focus on tests in more realistic environments using Euclid official simulations. One motivation for this challenge was to gauge the pros and cons of available tools to help us identify an appropriate algorithm for the Euclid pipeline. Our study has uncovered that a strategy of combining strengths can lead to improved results, for example using the fast neural network \deepleg to provide priors for \SEdot. We currently investigate this option and work towards the final implementation of the chosen algorithms into the Euclid pipeline. In turn, this challenge has also led some participants to develop and upgrade their software packages. It is important to reiterate that our challenge is based on five independent and valid approaches that differ widely in their techniques and choices (different use of priors, different pre-processing strategies, different approaches for multi-band fitting, etc.) and that we had to make choices that meant not every aspect and every strength or weakness was included in our analysis. For example, we do not include the computation time, required resources, compatibility with the current pipeline, accuracy of uncertainty budget estimates, etc. Naturally, other science cases with other priorities may lead to different results and decisions. We invite interested readers to use the online tool to test and manipulate the dataset according to their science cases. We hope this will be useful for others to make decisions and explore the software packages. 

\begin{acknowledgements} 
\AckEC
\newline
\newline
MHC acknowledges financial support from the State Research Agency (AEI-MCINN) of the Spanish Ministry of Science and Innovation under the grant and “Galaxy Evolution with Artificial Intelligence” with reference PGC2018-100852-A-I00, from the ACIISI, Consejer\'ia de Econom\'{i}a, Conocimiento y Empleo del Gobierno de Canarias and the European Regional Development Fund (ERDF) under grant with reference PROID2020010057, and from IAC project P/301802, financed by the Ministry of Science and Innovation, through the State Budget and by the Canary Islands Department of Economy, Knowledge and Employment, through the Regional Budget of the Autonomous Community.
\end{acknowledgements}


\bibliographystyle{aa}
\bibliography{main.bib}

\begin{thebibliography}{56}
\expandafter\ifx\csname natexlab\endcsname\relax\def\natexlab#1{#1}\fi

\bibitem[{{Bait} {et~al.}(2017){Bait}, {Barway}, \& {Wadadekar}}]{bait2017}
{Bait}, O., {Barway}, S., \& {Wadadekar}, Y. 2017, \mnras, 471, 2687

\bibitem[{{Bertin} {et~al.}(2020){Bertin}, {Schefer}, {Apostolakos},
  {{\'A}lvarez-Ayll{\'o}n}, {Dubath}, \& {K{\"u}mmel}}]{SE1}
{Bertin}, E., {Schefer}, M., {Apostolakos}, N., {et~al.} 2020, in Astronomical
  Society of the Pacific Conference Series, Vol. 527, Astronomical Data
  Analysis Software and Systems XXIX, ed. R.~{Pizzo}, E.~R. {Deul}, J.~D.
  {Mol}, J.~{de Plaa}, \& H.~{Verkouter}, 461

\bibitem[{{Bland-Hawthorn} {et~al.}(2005){Bland-Hawthorn}, {Vlaji{\'c}},
  {Freeman}, \& {Draine}}]{bland2005}
{Bland-Hawthorn}, J., {Vlaji{\'c}}, M., {Freeman}, K.~C., \& {Draine}, B.~T.
  2005, \apj, 629, 239

\bibitem[{{Brennan} {et~al.}(2017){Brennan}, {Pandya}, {Somerville}, {Barro},
  {Bluck}, {Taylor}, {Wuyts}, {Bell}, {Dekel}, {Faber}, {Ferguson},
  {Koekemoer}, {Kurczynski}, {McIntosh}, {Newman}, \& {Primack}}]{brennan2017}
{Brennan}, R., {Pandya}, V., {Somerville}, R.~S., {et~al.} 2017, \mnras, 465,
  619

\bibitem[{{Buitrago} {et~al.}(2008){Buitrago}, {Trujillo}, {Conselice},
  {Bouwens}, {Dickinson}, \& {Yan}}]{buitrago2008}
{Buitrago}, F., {Trujillo}, I., {Conselice}, C.~J., {et~al.} 2008, \apjl, 687,
  L61

\bibitem[{{Buitrago} {et~al.}(2013){Buitrago}, {Trujillo}, {Conselice}, \&
  {H{\"a}u{\ss}ler}}]{buitrago2013}
{Buitrago}, F., {Trujillo}, I., {Conselice}, C.~J., \& {H{\"a}u{\ss}ler}, B.
  2013, \mnras, 428, 1460

\bibitem[{{Ciotti}(1991)}]{ciotti1991}
{Ciotti}, L. 1991, \aap, 249, 99

\bibitem[{{Cole} {et~al.}(2000){Cole}, {Lacey}, {Baugh}, \& {Frenk}}]{cole2000}
{Cole}, S., {Lacey}, C.~G., {Baugh}, C.~M., \& {Frenk}, C.~S. 2000, \mnras,
  319, 168

\bibitem[{{Conselice}(2003)}]{conselice2003_CAS}
{Conselice}, C.~J. 2003, \apjs, 147, 1

\bibitem[{{Conselice} {et~al.}(2003){Conselice}, {Gallagher}, \&
  {Wyse}}]{conselice2003}
{Conselice}, C.~J., {Gallagher}, John~S., I., \& {Wyse}, R. F.~G. 2003, \aj,
  125, 66

\bibitem[{{Cropper} {et~al.}(2010){Cropper}, {Refregier}, {Guttridge},
  {Boulade}, {Amiaux}, {Walton}, {Thomas}, {Rees}, {Pool}, {Endicott},
  {Holland}, {Gow}, {Murray}, {Amara}, {Lumb}, {Duvet}, {Cole}, {Augueres}, \&
  {Hopkinson}}]{cropper2010}
{Cropper}, M., {Refregier}, A., {Guttridge}, P., {et~al.} 2010, in Society of
  Photo-Optical Instrumentation Engineers (SPIE) Conference Series, Vol. 7731,
  Space Telescopes and Instrumentation 2010: Optical, Infrared, and Millimeter
  Wave, ed. J.~{Oschmann}, Jacobus~M., M.~C. {Clampin}, \& H.~A. {MacEwen},
  77311J

\bibitem[{{de Vaucouleurs}(1948)}]{devaucouleurs1948}
{de Vaucouleurs}, G. 1948, Annales d'Astrophysique, 11, 247

\bibitem[{{dos Reis} {et~al.}(2020){dos Reis}, {Buitrago}, {Papaderos},
  {Matute}, {Afonso}, {Amarantidis}, {Breda}, {Gomes}, {Humphrey}, {Lobo},
  {Lorenzoni}, {Pappalardo}, {Paulino-Afonso}, \& {Scott}}]{dosreis2020}
{dos Reis}, S.~N., {Buitrago}, F., {Papaderos}, P., {et~al.} 2020, \aap, 634,
  A11

\bibitem[{Euclid Collaboration:~Bretonnière {et~al.}(2022)Euclid
  Collaboration:~Bretonnière, Huertas-Company, Boucaud, Lanusse, Jullo,
  Merlin, Tuccillo, Castellano, Brinchmann, Conselice, Dole, Cabanac, Courtois,
  Castander, Duc, Fosalba, Guinet, Kruk, Kuchner, Serrano, Soubrie, Tramacere,
  Wang, Amara, Auricchio, Bender, Bodendorf, Bonino, Branchini, Brau-Nogue,
  Brescia, Capobianco, Carbone, Carretero, Cavuoti, Cimatti, Cledassou,
  Congedo, Conversi, Copin, Corcione, Costille, Cropper, Da~Silva, Degaudenzi,
  Douspis, Dubath, Duncan, Dupac, Dusini, Farrens, Ferriol, Frailis,
  Franceschi, Fumana, Garilli, Gillard, Gillis, Giocoli, Grazian, Grupp,
  Haugan, Holmes, Hormuth, Hudelot, Jahnke, Kermiche, Kiessling, Kilbinger,
  Kitching, Kohley, Kümmel, Kunz, Kurki-Suonio, Ligori, Lilje, Lloro,
  Maiorano, Mansutti, Marggraf, Markovic, Marulli, Massey, Maurogordato,
  Melchior, Meneghetti, Meylan, Moresco, Morin, Moscardini, Munari, Nakajima,
  Niemi, Padilla, Paltani, Pasian, Pedersen, Pettorino, Pires, Poncet, Popa,
  Pozzetti, Raison, Rebolo, Rhodes, Roncarelli, Rossetti, Saglia, Schneider,
  Secroun, Seidel, Sirignano, Sirri, Stanco, Starck, Tallada-Crespí, Taylor,
  Tereno, Toledo-Moreo, Torradeflot, Valentijn, Valenziano, Wang, Welikala,
  Weller, Zamorani, Zoubian, Baldi, Bardelli, Camera, Farinelli, Medinaceli,
  Mei, Polenta, Romelli, Tenti, Vassallo, Zacchei, Zucca, Baccigalupi,
  Balaguera-Antolínez, Biviano, Borgani, Bozzo, Burigana, Cappi, Carvalho,
  Casas, Castignani, Colodro-Conde, Coupon, de~la Torre, Fabricius, Farina,
  Ferreira, Flose-Reimberg, Fotopoulou, Galeotta, Ganga, Garcia-Bellido,
  Gaztanaga, Gozaliasl, Hook, Joachimi, Kansal, Kashlinsky, Keihanen,
  Kirkpatrick, Lindholm, Mainetti, Maino, Maoli, Martinelli, Martinet,
  McCracken, Metcalf, Morgante, Morisset, Nightingale, Nucita, Patrizii,
  Potter, Renzi, Riccio, Sánchez, Sapone, Schirmer, Schultheis, Scottez,
  Sefusatti, Teyssier, Tutusaus, Valiviita, Viel, Whittaker, \&
  Knapen}]{bretonniere2022}
Euclid Collaboration:~Bretonnière, H., Huertas-Company, M., Boucaud, A.,
  {et~al.} 2022, \aap, 657, A90

\bibitem[{Euclid Collaboration:~Scaramella {et~al.}(2021)Euclid
  Collaboration:~Scaramella, Amiaux, Mellier, Burigana, Carvalho, Cuillandre,
  Da~Silva, Derosa, Dinis, Maiorano, {et~al.}}]{scaramella2021}
Euclid Collaboration:~Scaramella, R., Amiaux, J., Mellier, Y., {et~al.} 2021,
  arXiv:2108.01201

\bibitem[{{Euclid Collaboration: Schirmer} {et~al.}(2022){Euclid Collaboration:
  Schirmer}, {Jahnke}, {Seidel}, {Aussel}, {Bodendorf}, {Grupp}, {Hormuth},
  {Wachter}, {Appleton}, {Barbier}, {Brinchmann}, {Carrasco}, {Castander},
  {Coupon}, {De Paolis}, {Franco}, {Ganga}, {Hudelot}, {Jullo}, {Lan{\c{c}}on},
  {Nucita}, {Paltani}, {Smadja}, {Strafella}, {Venancio}, {Weiler}, {Amara},
  {Auphan}, {Auricchio}, {Balestra}, {Bender}, {Bonino}, {Branchini},
  {Brescia}, {Capobianco}, {Carbone}, {Carretero}, {Casas}, {Castellano},
  {Cavuoti}, {Cimatti}, {Cledassou}, {Congedo}, {Conselice}, {Conversi},
  {Copin}, {Corcione}, {Costille}, {Courbin}, {Da Silva}, {Degaudenzi},
  {Douspis}, {Dubath}, {Dupac}, {Dusini}, {Ealet}, {Farrens}, {Ferriol},
  {Fosalba}, {Frailis}, {Franceschi}, {Franzetti}, {Fumana}, {Garilli},
  {Gillard}, {Gillis}, {Giocoli}, {Grazian}, {Guzzo}, {Haugan}, {Hoekstra},
  {Holmes}, {Hornstrup}, {K{\"u}mmel}, {Kermiche}, {Kiessling}, {Kilbinger},
  {Kitching}, {Kohley}, {Kunz}, {Kurki-Suonio}, {Laureijs}, {Ligori}, {Lilje},
  {Lloro}, {Maciaszek}, {Maiorano}, {Mansutti}, {Marggraf}, {Markovic},
  {Marulli}, {Massey}, {Maurogordato}, {Mellier}, {Meneghetti}, {Merlin},
  {Meylan}, {Moresco}, {Moscardini}, {Munari}, {Nakajima}, {Nichol}, {Niemi},
  {Padilla}, {Pasian}, {Pedersen}, {Percival}, {Pettorino}, {Pires}, {Poncet},
  {Popa}, {Pozzetti}, {Prieto}, {Raison}, {Rhodes}, {Rix}, {Roncarelli},
  {Rossetti}, {Saglia}, {Sartoris}, {Scaramella}, {Schneider}, {Secroun},
  {Serrano}, {Sirignano}, {Sirri}, {Stanco}, {Tallada-Cresp{\'\i}}, {Taylor},
  {Teplitz}, {Tereno}, {Toledo-Moreo}, {Torradeflot}, {Trifoglio}, {Valentijn},
  {Valenziano}, {Wang}, {Weller}, {Zamorani}, {Zoubian}, {Andreon}, {Bardelli},
  {Boucaud}, {Camera}, {Farinelli}, {Graci{\'a}-Carpio}, {Maino}, {Medinaceli},
  {Mei}, {Morisset}, {Polenta}, {Renzi}, {Romelli}, {Tenti}, {Vassallo},
  {Zacchei}, {Zucca}, {Baccigalupi}, {Balaguera-Antol{\'\i}nez}, {Biviano},
  {Blanchard}, {Borgani}, {Bozzo}, {Burigana}, {Cabanac}, {Cappi}, {Carvalho},
  {Casas}, {Castignani}, {Colodro-Conde}, {Cooray}, {Courtois}, {Crocce},
  {Cuby}, {Davini}, {de la Torre}, {Di Ferdinando}, {Escartin}, {Farina},
  {Ferreira}, {Finelli}, {Fotopoulou}, {Galeotta}, {Garcia-Bellido},
  {Gaztanaga}, {George}, {Gozaliasl}, {Hook}, {Ili{\'c}}, {Kansal},
  {Kashlinsky}, {Keihanen}, {Kirkpatrick}, {Lindholm}, {Mainetti}, {Maoli},
  {Martinelli}, {Martinet}, {Maturi}, {Mauri}, {McCracken}, {Metcalf},
  {Monaco}, {Morgante}, {Nightingale}, {Patrizii}, {Peel}, {Popa}, {Porciani},
  {Potter}, {Reimberg}, {Riccio}, {S{\'a}nchez}, {Sapone}, {Scottez},
  {Sefusatti}, {Teyssier}, {Tutusaus}, {Valieri}, {Valiviita}, {Viel}, \&
  {Hildebrandt}}]{schirmer_2022}
{Euclid Collaboration: Schirmer}, M., {Jahnke}, K., {Seidel}, G., {et~al.}
  2022, \aap, 662, A92

\bibitem[{{Falc{\'o}n-Barroso} {et~al.}(2017){Falc{\'o}n-Barroso}, {Lyubenova},
  {van de Ven}, {Mendez-Abreu}, {Aguerri}, {Garc{\'\i}a-Lorenzo},
  {Bekerait{\'e}}, {S{\'a}nchez}, {Husemann}, {Garc{\'\i}a-Benito}, {Mast},
  {Walcher}, {Zibetti}, {Barrera-Ballesteros}, {Galbany},
  {S{\'a}nchez-Bl{\'a}zquez}, {Singh}, {van den Bosch}, {Wild}, {Zhu},
  {Bland-Hawthorn}, {Cid Fernandes}, {de Lorenzo-C{\'a}ceres}, {Gallazzi},
  {Gonz{\'a}lez Delgado}, {Marino}, {M{\'a}rquez}, {P{\'e}rez}, {P{\'e}rez},
  {Roth}, {Rosales-Ortega}, {Ruiz-Lara}, {Wisotzki}, {Ziegler}, \& {Califa
  Collaboration}}]{falcon2017}
{Falc{\'o}n-Barroso}, J., {Lyubenova}, M., {van de Ven}, G., {et~al.} 2017,
  \aap, 597, A48

\bibitem[{{Ferrari} {et~al.}(2015){Ferrari}, {de Carvalho}, \&
  {Trevisan}}]{ferrari2015}
{Ferrari}, F., {de Carvalho}, R.~R., \& {Trevisan}, M. 2015, \apj, 814, 55

\bibitem[{{F{\"o}rster Schreiber} {et~al.}(2009){F{\"o}rster Schreiber},
  {Genzel}, {Bouch{\'e}}, {Cresci}, {Davies}, {Buschkamp}, {Shapiro},
  {Tacconi}, {Hicks}, {Genel}, {Shapley}, {Erb}, {Steidel}, {Lutz},
  {Eisenhauer}, {Gillessen}, {Sternberg}, {Renzini}, {Cimatti}, {Daddi},
  {Kurk}, {Lilly}, {Kong}, {Lehnert}, {Nesvadba}, {Verma}, {McCracken},
  {Arimoto}, {Mignoli}, \& {Onodera}}]{schreiber2009}
{F{\"o}rster Schreiber}, N.~M., {Genzel}, R., {Bouch{\'e}}, N., {et~al.} 2009,
  \apj, 706, 1364

\bibitem[{{Freeman}(1970)}]{freeman1970}
{Freeman}, K.~C. 1970, \apj, 160, 811

\bibitem[{{Graham} \& {Guzm{\'a}n}(2003)}]{graham2003}
{Graham}, A.~W. \& {Guzm{\'a}n}, R. 2003, \aj, 125, 2936

\bibitem[{{Guy} {et~al.}(2022){Guy}, {Cuillandre}, {Bachelet}, {Banerji},
  {Bauer}, {Collett}, {Conselice}, {Eggl}, {Ferguson}, {Fontana}, {Heymans},
  {Hook}, {Aubourg}, {Aussel}, {Bosch}, {Carry}, {Hoekstra}, {Kuijken},
  {Lanusse}, {Melchior}, {Mohr}, {Moresco}, {Nakajima}, {Paltani}, {Troxel},
  {Allevato}, {Amara}, {Andreon}, {Anguita}, {Bardelli}, {Bechtol}, {Birrer},
  {Bisigello}, {Bolzonella}, {Botticella}, {Bouy}, {Brinchmann}, {Brough},
  {Camera}, {Cantiello}, {Cappellaro}, {Carlin}, {Castander}, {Castellano},
  {Chari}, {Chisari}, {Collins}, {Courbin}, {Cuby}, {Cucciati}, {Daylan},
  {Diego}, {Duc}, {Fotopoulou}, {Fouchez}, {Gavazzi}, {Gruen}, {Hatfield},
  {Hildebrandt}, {Landt}, {Hunt}, {Ibata}, {Ilbert}, {Jasche}, {Joachimi},
  {Joseph}, {Kotak}, {Laigle}, {Lan{\c{c}}on}, {Larsen}, {Lavaux}, {Leclercq},
  {Leonard}, {von der Linden}, {Liu}, {Longo}, {Magliocchetti}, {Maraston},
  {Marshall}, {Mart{\'\i}n}, {Mattila}, {Maturi}, {McCracken}, {Metcalf},
  {Montes}, {Mortlock}, {Moscardini}, {Narayan}, {Paolillo}, {Papaderos},
  {Pello}, {Pozzetti}, {Radovich}, {Rejkuba}, {Rom{\'a}n},
  {S{\'a}nchez-Janssen}, {Sarpa}, {Sartoris}, {Schrabback}, {Sluse}, {Smartt},
  {Smith}, {Snodgrass}, {Talia}, {Tao}, {Toft}, {Tortora}, {Tutusaus}, {Usher},
  {van Velzen}, {Verma}, {Vernardos}, {Voggel}, {Wandelt}, {Watkins}, {Weller},
  {Wright}, {Yoachim}, {Yoon}, \& {Zucca}}]{2022zndo...5836022G}
{Guy}, L.~P., {Cuillandre}, J.-C., {Bachelet}, E., {et~al.} 2022, in Zenodo id.
  5836022, Vol.~58

\bibitem[{{H{\"a}u{\ss}ler} {et~al.}(2013){H{\"a}u{\ss}ler}, {Bamford}, {Vika},
  {Rojas}, {Barden}, {Kelvin}, {Alpaslan}, {Robotham}, {Driver}, {Baldry},
  {Brough}, {Hopkins}, {Liske}, {Nichol}, {Popescu}, \&
  {Tuffs}}]{haeussler2013}
{H{\"a}u{\ss}ler}, B., {Bamford}, S.~P., {Vika}, M., {et~al.} 2013, \mnras,
  430, 330

\bibitem[{{H{\"a}u{\ss}ler} {et~al.}(2022){H{\"a}u{\ss}ler}, {Vika}, {Bamford},
  {Johnston}, {Brough}, {Casura}, {Holwerda}, {Kelvin}, \&
  {Popescu}}]{haeussler2022}
{H{\"a}u{\ss}ler}, B., {Vika}, M., {Bamford}, S.~P., {et~al.} 2022, arXiv,
  arXiv:2204.05907

\bibitem[{{Hubble}(1926)}]{1926ApJ....64..321H}
{Hubble}, E.~P. 1926, \apj, 64, 321

\bibitem[{{Huertas-Company} {et~al.}(2011){Huertas-Company}, {Aguerri},
  {Bernardi}, {Mei}, \& {S{\'a}nchez Almeida}}]{huertas2011}
{Huertas-Company}, M., {Aguerri}, J.~A.~L., {Bernardi}, M., {Mei}, S., \&
  {S{\'a}nchez Almeida}, J. 2011, \aap, 525, A157

\bibitem[{{Huertas-Company} {et~al.}(2008){Huertas-Company}, {Rouan}, {Tasca},
  {Soucail}, \& {Le F{\`e}vre}}]{huertas2008}
{Huertas-Company}, M., {Rouan}, D., {Tasca}, L., {Soucail}, G., \& {Le
  F{\`e}vre}, O. 2008, \aap, 478, 971

\bibitem[{{Kelvin} {et~al.}(2012){Kelvin}, {Driver}, {Robotham}, {Hill},
  {Alpaslan}, {Baldry}, {Bamford}, {Bland-Hawthorn}, {Brough}, {Graham},
  {H{\"a}ussler}, {Hopkins}, {Liske}, {Loveday}, {Norberg}, {Phillipps},
  {Popescu}, {Prescott}, {Taylor}, \& {Tuffs}}]{kelvin2012}
{Kelvin}, L.~S., {Driver}, S.~P., {Robotham}, A. S.~G., {et~al.} 2012, \mnras,
  421, 1007

\bibitem[{{Kennedy} {et~al.}(2015){Kennedy}, {Bamford}, {Baldry},
  {H{\"a}u{\ss}ler}, {Holwerda}, {Hopkins}, {Kelvin}, {Lange}, {Moffett},
  {Popescu}, {Taylor}, {Tuffs}, {Vika}, \& {Vulcani}}]{kennedy2015}
{Kennedy}, R., {Bamford}, S.~P., {Baldry}, I., {et~al.} 2015, \mnras, 454, 806

\bibitem[{Kingma \& Welling(2019)}]{Kingma2019}
Kingma, D.~P. \& Welling, M. 2019, Foundations and Trends® in Machine
  Learning, 12, 307–392

\bibitem[{{Kormendy}(1977)}]{kormendy1977}
{Kormendy}, J. 1977, \apj, 218, 333

\bibitem[{{Kormendy} \& {Kennicutt}(2004)}]{kormendy2004}
{Kormendy}, J. \& {Kennicutt}, Robert~C., J. 2004, \araa, 42, 603

\bibitem[{{K{\"u}mmel} {et~al.}(2020){K{\"u}mmel}, {Bertin}, {Schefer},
  {Apostolakos}, {{\'A}lvarez-Ayll{\'o}n}, \& {Dubath}}]{SE2}
{K{\"u}mmel}, M., {Bertin}, E., {Schefer}, M., {et~al.} 2020, in Astronomical
  Society of the Pacific Conference Series, Vol. 527, Astronomical Data
  Analysis Software and Systems XXIX, ed. R.~{Pizzo}, E.~R. {Deul}, J.~D.
  {Mol}, J.~{de Plaa}, \& H.~{Verkouter}, 29

\bibitem[{{Lang} {et~al.}(2014){Lang}, {Wuyts}, {Somerville}, {F{\"o}rster
  Schreiber}, {Genzel}, {Bell}, {Brammer}, {Dekel}, {Faber}, {Ferguson},
  {Grogin}, {Kocevski}, {Koekemoer}, {Lutz}, {McGrath}, {Momcheva}, {Nelson},
  {Primack}, {Rosario}, {Skelton}, {Tacconi}, {van Dokkum}, \&
  {Whitaker}}]{lang2014}
{Lang}, P., {Wuyts}, S., {Somerville}, R.~S., {et~al.} 2014, \apj, 788, 11

\bibitem[{{Lange} {et~al.}(2016){Lange}, {Moffett}, {Driver}, {Robotham},
  {Lagos}, {Kelvin}, {Conselice}, {Margalef-Bentabol}, {Alpaslan}, {Baldry},
  {Bland-Hawthorn}, {Bremer}, {Brough}, {Cluver}, {Colless}, {Davies},
  {H{\"a}u{\ss}ler}, {Holwerda}, {Hopkins}, {Kafle}, {Kennedy}, {Liske},
  {Phillipps}, {Popescu}, {Taylor}, {Tuffs}, {van Kampen}, \&
  {Wright}}]{lange2016}
{Lange}, R., {Moffett}, A.~J., {Driver}, S.~P., {et~al.} 2016, \mnras, 462,
  1470

\bibitem[{{Laureijs} {et~al.}(2011){Laureijs}, {Amiaux}, {Arduini},
  {Augu{\`e}res}, {Brinchmann}, {Cole}, {Cropper}, {Dabin}, {Duvet}, {Ealet},
  {Garilli}, {Gondoin}, {Guzzo}, {Hoar}, {Hoekstra}, {Holmes}, {Kitching},
  {Maciaszek}, {Mellier}, {Pasian}, {Percival}, {Rhodes}, {Saavedra Criado},
  {Sauvage}, {Scaramella}, {Valenziano}, {Warren}, {Bender}, {Castander},
  {Cimatti}, {Le F{\`e}vre}, {Kurki-Suonio}, {Levi}, {Lilje}, {Meylan},
  {Nichol}, {Pedersen}, {Popa}, {Rebolo Lopez}, {Rix}, {Rottgering},
  {Zeilinger}, {Grupp}, {Hudelot}, {Massey}, {Meneghetti}, {Miller}, {Paltani},
  {Paulin-Henriksson}, {Pires}, {Saxton}, {Schrabback}, {Seidel}, {Walsh},
  {Aghanim}, {Amendola}, {Bartlett}, {Baccigalupi}, {Beaulieu}, {Benabed},
  {Cuby}, {Elbaz}, {Fosalba}, {Gavazzi}, {Helmi}, {Hook}, {Irwin}, {Kneib},
  {Kunz}, {Mannucci}, {Moscardini}, {Tao}, {Teyssier}, {Weller}, {Zamorani},
  {Zapatero Osorio}, {Boulade}, {Foumond}, {Di Giorgio}, {Guttridge}, {James},
  {Kemp}, {Martignac}, {Spencer}, {Walton}, {Bl{\"u}mchen}, {Bonoli},
  {Bortoletto}, {Cerna}, {Corcione}, {Fabron}, {Jahnke}, {Ligori}, {Madrid},
  {Martin}, {Morgante}, {Pamplona}, {Prieto}, {Riva}, {Toledo}, {Trifoglio},
  {Zerbi}, {Abdalla}, {Douspis}, {Grenet}, {Borgani}, {Bouwens}, {Courbin},
  {Delouis}, {Dubath}, {Fontana}, {Frailis}, {Grazian}, {Koppenh{\"o}fer},
  {Mansutti}, {Melchior}, {Mignoli}, {Mohr}, {Neissner}, {Noddle}, {Poncet},
  {Scodeggio}, {Serrano}, {Shane}, {Starck}, {Surace}, {Taylor},
  {Verdoes-Kleijn}, {Vuerli}, {Williams}, {Zacchei}, {Altieri}, {Escudero
  Sanz}, {Kohley}, {Oosterbroek}, {Astier}, {Bacon}, {Bardelli}, {Baugh},
  {Bellagamba}, {Benoist}, {Bianchi}, {Biviano}, {Branchini}, {Carbone},
  {Cardone}, {Clements}, {Colombi}, {Conselice}, {Cresci}, {Deacon}, {Dunlop},
  {Fedeli}, {Fontanot}, {Franzetti}, {Giocoli}, {Garcia-Bellido}, {Gow},
  {Heavens}, {Hewett}, {Heymans}, {Holland}, {Huang}, {Ilbert}, {Joachimi},
  {Jennins}, {Kerins}, {Kiessling}, {Kirk}, {Kotak}, {Krause}, {Lahav}, {van
  Leeuwen}, {Lesgourgues}, {Lombardi}, {Magliocchetti}, {Maguire}, {Majerotto},
  {Maoli}, {Marulli}, {Maurogordato}, {McCracken}, {McLure}, {Melchiorri},
  {Merson}, {Moresco}, {Nonino}, {Norberg}, {Peacock}, {Pello}, {Penny},
  {Pettorino}, {Di Porto}, {Pozzetti}, {Quercellini}, {Radovich}, {Rassat},
  {Roche}, {Ronayette}, {Rossetti}, {Sartoris}, {Schneider}, {Semboloni},
  {Serjeant}, {Simpson}, {Skordis}, {Smadja}, {Smartt}, {Spano}, {Spiro},
  {Sullivan}, {Tilquin}, {Trotta}, {Verde}, {Wang}, {Williger}, {Zhao},
  {Zoubian}, \& {Zucca}}]{laureijs2011}
{Laureijs}, R., {Amiaux}, J., {Arduini}, S., {et~al.} 2011, arXiv, 1110.3193

\bibitem[{{Lintott} {et~al.}(2008){Lintott}, {Schawinski}, {Slosar}, {Land},
  {Bamford}, {Thomas}, {Raddick}, {Nichol}, {Szalay}, {Andreescu}, {Murray}, \&
  {Vandenberg}}]{lintott2008}
{Lintott}, C.~J., {Schawinski}, K., {Slosar}, A., {et~al.} 2008, \mnras, 389,
  1179

\bibitem[{{Lotz} {et~al.}(2004){Lotz}, {Primack}, \& {Madau}}]{lotz2004}
{Lotz}, J.~M., {Primack}, J., \& {Madau}, P. 2004, \aj, 128, 163

\bibitem[{{Mortlock} {et~al.}(2013){Mortlock}, {Conselice}, {Hartley},
  {Ownsworth}, {Lani}, {Bluck}, {Almaini}, {Duncan}, {van der Wel},
  {Koekemoer}, {Dekel}, {Dav{\'e}}, {Ferguson}, {de Mello}, {Newman}, {Faber},
  {Grogin}, {Kocevski}, \& {Lai}}]{mortlock2013}
{Mortlock}, A., {Conselice}, C.~J., {Hartley}, W.~G., {et~al.} 2013, \mnras,
  433, 1185

\bibitem[{Papamakarios {et~al.}(2021)Papamakarios, Nalisnick, Rezende, Mohamed,
  \& Lakshminarayanan}]{papamakarios2021}
Papamakarios, G., Nalisnick, E., Rezende, D.~J., Mohamed, S., \&
  Lakshminarayanan, B. 2021, Journal of Machine Learning Research, 22, 1

\bibitem[{{Pawlik} {et~al.}(2016){Pawlik}, {Wild}, {Walcher}, {Johansson},
  {Villforth}, {Rowlands}, {Mendez-Abreu}, \& {Hewlett}}]{pawlik2016}
{Pawlik}, M.~M., {Wild}, V., {Walcher}, C.~J., {et~al.} 2016, \mnras, 456, 3032

\bibitem[{{Peng} {et~al.}(2002){Peng}, {Ho}, {Impey}, \& {Rix}}]{peng2002}
{Peng}, C.~Y., {Ho}, L.~C., {Impey}, C.~D., \& {Rix}, H.-W. 2002, \aj, 124, 266

\bibitem[{{Peterson} {et~al.}(2015){Peterson}, {Jernigan}, {Kahn}, {Rasmussen},
  {Peng}, {Ahmad}, {Bankert}, {Chang}, {Claver}, {Gilmore}, {Grace}, {Hannel},
  {Hodge}, {Lorenz}, {Lupu}, {Meert}, {Nagarajan}, {Todd}, {Winans}, \&
  {Young}}]{phosim2015}
{Peterson}, J.~R., {Jernigan}, J.~G., {Kahn}, S.~M., {et~al.} 2015, \apjs, 218,
  14

\bibitem[{{Pohlen} \& {Trujillo}(2006)}]{pohlen2006}
{Pohlen}, M. \& {Trujillo}, I. 2006, \aap, 454, 759

\bibitem[{{Robotham} {et~al.}(2018){Robotham}, {Davies}, {Driver}, {Koushan},
  {Taranu}, {Casura}, \& {Liske}}]{robotham2018}
{Robotham}, A.~S.~G., {Davies}, L.~J.~M., {Driver}, S.~P., {et~al.} 2018,
  \mnras, 476, 3137

\bibitem[{{Robotham} {et~al.}(2017){Robotham}, {Taranu}, {Tobar}, {Moffett}, \&
  {Driver}}]{robotham2017}
{Robotham}, A.~S.~G., {Taranu}, D.~S., {Tobar}, R., {Moffett}, A., \& {Driver},
  S.~P. 2017, \mnras, 466, 1513

\bibitem[{{Rowe} {et~al.}(2015){Rowe}, {Jarvis}, {Mandelbaum}, {Bernstein},
  {Bosch}, {Simet}, {Meyers}, {Kacprzak}, {Nakajima}, {Zuntz}, {Miyatake},
  {Dietrich}, {Armstrong}, {Melchior}, \& {Gill}}]{galsim}
{Rowe}, B.~T.~P., {Jarvis}, M., {Mandelbaum}, R., {et~al.} 2015, Astronomy and
  Computing, 10, 121

\bibitem[{{Schreiber} {et~al.}(2017){Schreiber}, {Elbaz}, {Pannella}, {Merlin},
  {Castellano}, {Fontana}, {Bourne}, {Boutsia}, {Cullen}, {Dunlop}, {Ferguson},
  {Micha{\l}owski}, {Okumura}, {Santini}, {Shu}, {Wang}, \& {White}}]{EGG}
{Schreiber}, C., {Elbaz}, D., {Pannella}, M., {et~al.} 2017, \aap, 602, A96

\bibitem[{{S\'ersic}(1968)}]{sersic1968}
{S\'ersic}, J.~L. 1968, {Atlas de galaxias australes} (Cordoba, Argentina:
  Observatorio Astronomico, 1968)

\bibitem[{{Simard} {et~al.}(2011){Simard}, {Mendel}, {Patton}, {Ellison}, \&
  {McConnachie}}]{simard2011}
{Simard}, L., {Mendel}, J.~T., {Patton}, D.~R., {Ellison}, S.~L., \&
  {McConnachie}, A.~W. 2011, \apjs, 196, 11

\bibitem[{{Tal} \& {van Dokkum}(2011)}]{tal2011}
{Tal}, T. \& {van Dokkum}, P.~G. 2011, \apj, 731, 89

\bibitem[{{Trujillo} {et~al.}(2001){Trujillo}, {Graham}, \&
  {Caon}}]{trujillo2001}
{Trujillo}, I., {Graham}, A.~W., \& {Caon}, N. 2001, \mnras, 326, 869

\bibitem[{{Tuccillo} {et~al.}(2018){Tuccillo}, {Huertas-Company},
  {Decenci{\`e}re}, {Velasco-Forero}, {Dom{\'\i}nguez S{\'a}nchez}, \&
  {Dimauro}}]{tuccillo2018}
{Tuccillo}, D., {Huertas-Company}, M., {Decenci{\`e}re}, E., {et~al.} 2018,
  \mnras, 475, 894

\bibitem[{{van de Sande} {et~al.}(2017){van de Sande}, {Bland-Hawthorn},
  {Fogarty}, {Cortese}, {d'Eugenio}, {Croom}, {Scott}, {Allen}, {Brough},
  {Bryant}, {Cecil}, {Colless}, {Couch}, {Davies}, {Elahi}, {Foster},
  {Goldstein}, {Goodwin}, {Groves}, {Ho}, {Jeong}, {Jones}, {Konstantopoulos},
  {Lawrence}, {Leslie}, {L{\'o}pez-S{\'a}nchez}, {McDermid}, {McElroy},
  {Medling}, {Oh}, {Owers}, {Richards}, {Schaefer}, {Sharp}, {Sweet}, {Taranu},
  {Tonini}, {Walcher}, \& {Yi}}]{vandesande2017}
{van de Sande}, J., {Bland-Hawthorn}, J., {Fogarty}, L. M.~R., {et~al.} 2017,
  \apj, 835, 104

\bibitem[{{Vega-Ferrero} {et~al.}(2021){Vega-Ferrero}, {Dom{\'\i}nguez
  S{\'a}nchez}, {Bernardi}, {Huertas-Company}, {Morgan}, {Margalef}, {Aguena},
  {Allam}, {Annis}, {Avila}, {Bacon}, {Bertin}, {Brooks}, {Carnero Rosell},
  {Carrasco Kind}, {Carretero}, {Choi}, {Conselice}, {Costanzi}, {da Costa},
  {Pereira}, {De Vicente}, {Desai}, {Ferrero}, {Fosalba}, {Frieman},
  {Garc{\'\i}a-Bellido}, {Gruen}, {Gruendl}, {Gschwend}, {Gutierrez},
  {Hartley}, {Hinton}, {Hollowood}, {Honscheid}, {Hoyle}, {Jarvis}, {Kim},
  {Kuehn}, {Kuropatkin}, {Lima}, {Maia}, {Menanteau}, {Miquel}, {Ogando},
  {Palmese}, {Paz-Chinch{\'o}n}, {Plazas}, {Romer}, {Sanchez}, {Scarpine},
  {Schubnell}, {Serrano}, {Sevilla-Noarbe}, {Smith}, {Suchyta}, {Swanson},
  {Tarle}, {Tarsitano}, {To}, {Tucker}, {Varga}, \& {Wilkinson}}]{vega2021}
{Vega-Ferrero}, J., {Dom{\'\i}nguez S{\'a}nchez}, H., {Bernardi}, M., {et~al.}
  2021, \mnras, 506, 1927

\bibitem[{{Vulcani} {et~al.}(2014){Vulcani}, {Bamford}, {H{\"a}u{\ss}ler},
  {Vika}, {Rojas}, {Agius}, {Baldry}, {Bauer}, {Brown}, {Driver}, {Graham},
  {Kelvin}, {Liske}, {Loveday}, {Popescu}, {Robotham}, \&
  {Tuffs}}]{vulcani2014}
{Vulcani}, B., {Bamford}, S.~P., {H{\"a}u{\ss}ler}, B., {et~al.} 2014, \mnras,
  441, 1340

\end{thebibliography}


\begin{appendix}

\section{Interactive plot website}\label{app:app1}

We describe here the interactive web page we have created to accompany this paper (link in footnote\footnote{ \url{https://share.streamlit.io/hbretonniere/euclid_morphology_challenge}}). It will allow the reader to create most of the plots we present in this paper but with control over some choices we made for the representation. The tool was constructed using the \texttt{Streamlit} python package. All plots can be computed on the fly, with the common catalogues for each type of simulation (i.e., using the results of the Euclid Morphology Challenge, which were the basis for the analysis presented in the paper). The following is a brief introduction to the tool. We recommend the use of Firefox, Opera, or Chrome browsers, and found that the platform is unstable in Safari. \texttt{JavaScript} has to be authorised for the platform to work. In some cases, the data needs to be download on the server, and it can take several minutes. For any additional information, contact H.Bretonnière.

The main plotting parameters are located on the top of the left panel of the page. First, choose a type of simulation in the dataset option, single-\sersicdot, double-\sersicdot, realistic, or multi-band. 
Depending on the type of simulation, additional options appear, for example the fitting made with a free \sersic index for the bulge component of the double-\sersic simulations. Then, just below the dataset panel, the different types of plots will appear: the summary plots (e.g. Fig.~\ref{fig:summary_ss}), the trumpet plots (e.g. Fig.~\ref{fig:trumpet_ss_re}), the summary score plot (Fig.~\ref{fig:global scores}), or the error prediction (e.g. Fig\ref{fig:ss_uncertainty}). The options vary depending on the selected dataset (e.g., 2D plots are only available for the double-\sersic dataset). 
Next, choose the parameters to analyse and select the different software packages. For each, the user can select or deselect the parameter or software, or click the `all parameters' or `all softwares' button to select them all.
Finally, for the summary and trumpet plots, the user can change the $x$-axis, (i.e. the true parameter) in addition to the number of bins. The available options are the $\VIS$ magnitude, redshift, half-light radius or \sersic index.
According to the options selected in the panel on the left, the figure will change in the panel on the right. Above the plot, there is a check box called \texttt{Demo version}. When selected, only $1\%$ of the catalogue is used to draw the plots. This option should only be used to play with the parameters until the right plot is selected. Then, the box must be un-checked to produce plots including all measurements of the combined catalogue, as was used in the plots of the paper.
The slider button allows easy control over the value of the outlier threshold (see Sect.~\ref{sec:metrics}). Here again, the plot will change accordingly, for example changing the third columns of the summary type plots. 
Another slider control changes the $x$-axis range, and the $y$-axis if we are working on the 2D summary plots. By default, the $y$-axis range is fixed to hard-coded values. The user can undo this limitation by un-checking the \texttt{limit y axis range} button. If only one parameter is selected, three other sliders will appear to control the $y$-axis ranges. 

Finally, below the figure, a checkbox allows one to save the results used to produce the summary plot, as a python dictionary. The dictionary $D$ is structured as follows: 
\begin{equation}
    D = \mathrm{dict}[i][p]\;, 
    \label{eq:dico}
\end{equation}
and contains all the bins of the metric $i$, on a parameter $p$ and software name $s$ where $i=0, 1, \mathrm{or}~ 2$ for the bias, dispersion and outlier fraction, respectively.

The different buttons and options change depending on the dataset or plot type. For example, when plotting the global score, four additional sliders appear to control the weights of the metrics (see Eq.~\ref{eq:score}). The user will also be able to remove the bin weighting, to have an equal weight for all magnitudes. Some of \partonedot's plots can also be plotted by changing the 'select EMC paper' option on the top button of the left panel.

\section{Effect of PSFEx on the \SE fitting}
\label{sec:PSFX}
\begin{figure}[h!]
\centering
    \includegraphics[width=\linewidth]{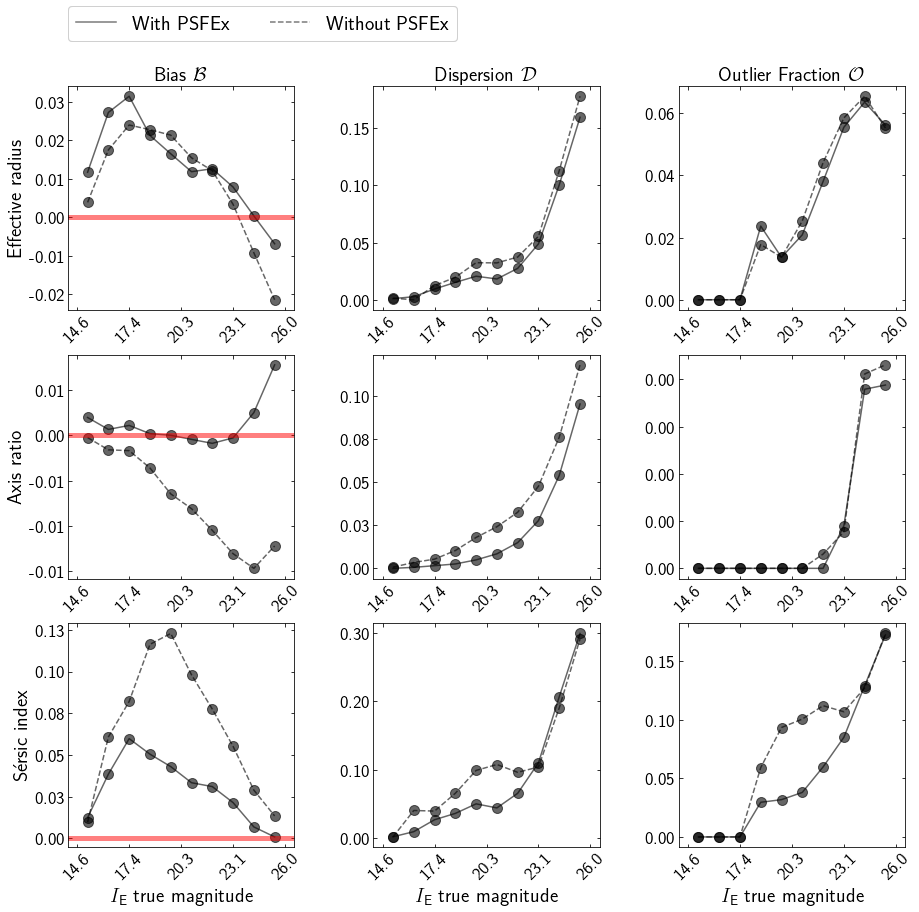}
    \caption{Summary plot with the comparison of the fitting with (solid line) or without (dashed line) the use of PSFEx. With PSFEx (as in the rest of the paper), the results are still slightly different than the one presented earlier because only one field is analysed.}
    \label{fig:psfx}
\end{figure}

The challenge instructions to the participants did not restrict or predict procedure steps, but left maximal freedom to allow for best choices by the experts. This lead to individual decisions and interpretations of the Challenge requirements. One such choice by \SE was to employ the full pipeline, including some important pre-processing steps. In preparation for model fitting with \SEdot, the team made modifications to the data that was provided by the challenge team, in order to treat them in the same way they would treat real data. In particular, the Rubin-like and \Euclid NIR images were re-sampled back to their original pixel scale ($\ang{;;0.2}$ and $\ang{;;0.3}$, respectively). The team also extracted oversampled PSF images using the PSFEx software.
They thus used the provided PSF images in the given sampling and have oversampled PSF images in all bands. This choice was made based on the team's assumption that it was an essential condition to deliver satisfactory results. The outcome of the \SE team's internal checks found that PSF images were sufficiently well oversampled (two times for Rubin-like data, and three times for the \Euclid NIR bands), but not for the $\VIS$ band. They concluded that their own PSF, created by PSFEx, could improve results for morphology and stellar photometry with a $1/0.45$ oversampling factor.
Figure~\ref{fig:psfx} compares outputs that do and do not use this step of the \SE pipeline. Dashed lines (without PSFEx) use the PSF we provided, while the solid lines use the re-sampled PSF after PSFEx was employed. The performance of measuring the axis ratio and \sersic index particularly improve, but also the effective radius.

\section{Performance of metrics at different redshifts}

\begin{figure}
\centering
    \includegraphics[width=\linewidth]{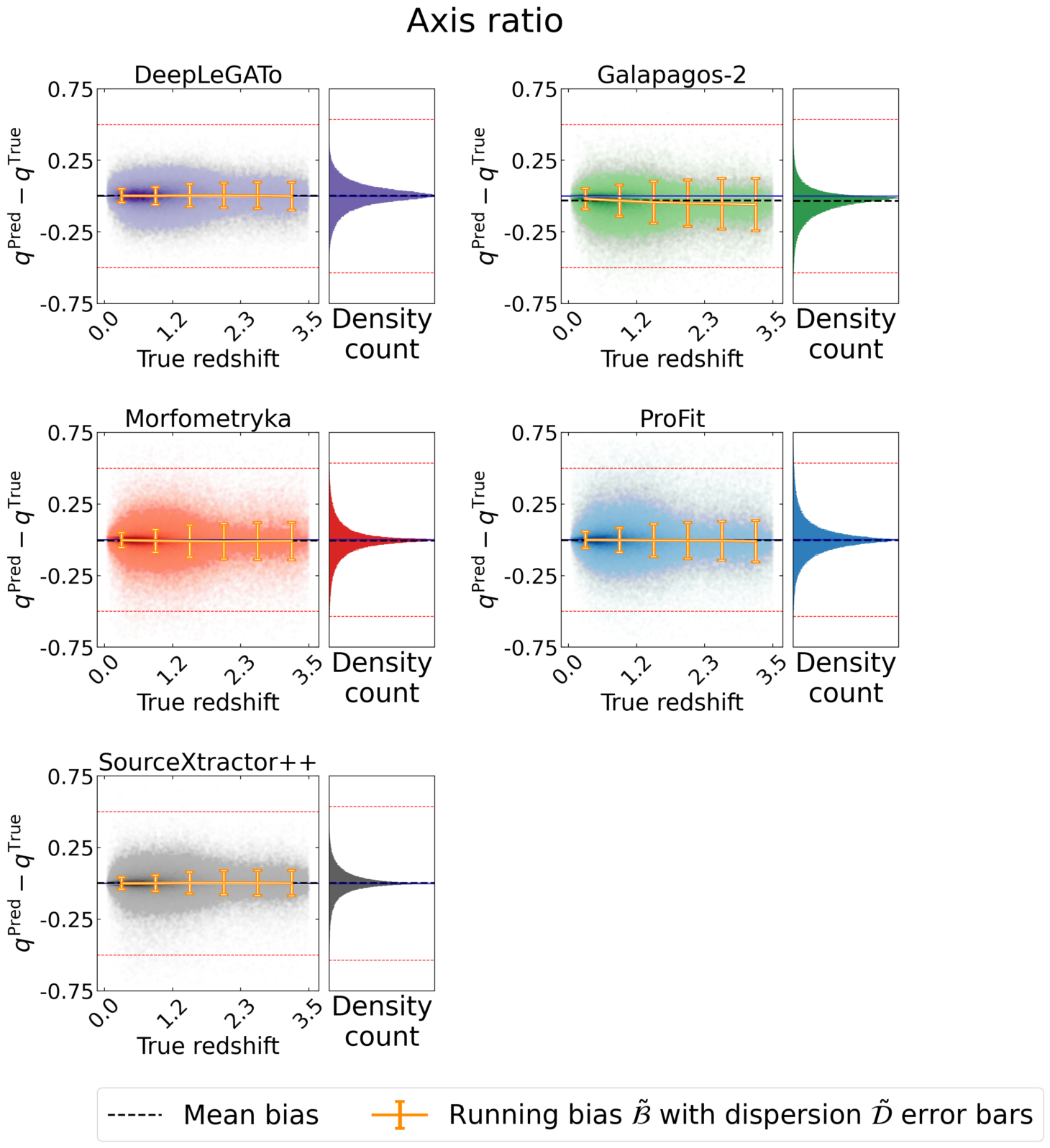}
    \caption{Trumpet plot for the single-\sersic \ba fitting as a function of the true redshift. See Fig.~\ref{fig:trumpet_ss_re} for further information.}
    \label{fig:trumpet_re_z}
\end{figure}

\begin{figure}[h!]
\centering
    \includegraphics[width=\linewidth]{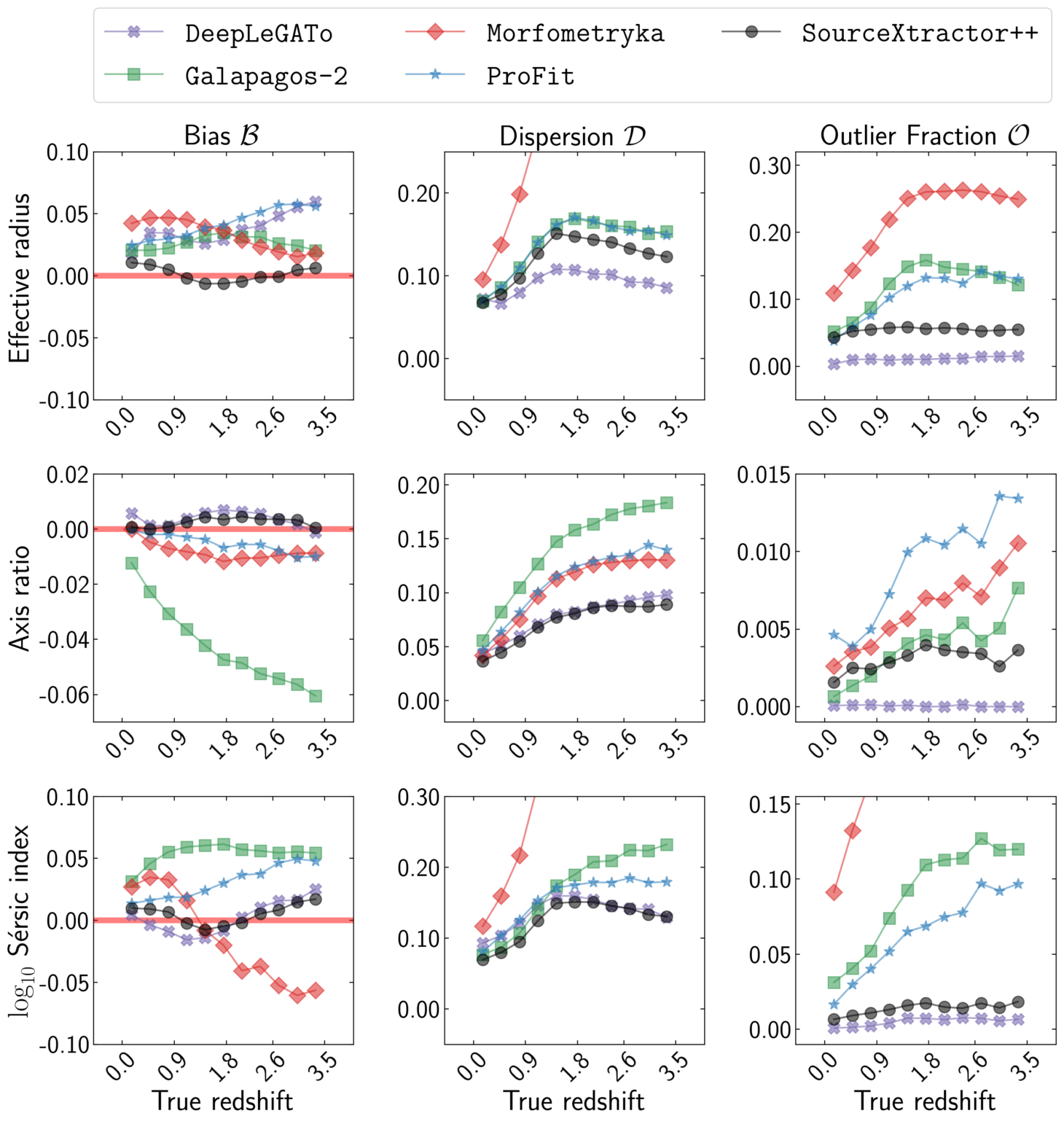}
    \caption{Summary plot for the fitting of the single \sersic simulations regarding the true redshift.}
    \label{fig:summary_redshift}
\end{figure}

The behaviour of all parameters with varying redshifts can be explored in the online tool. Here, we detail the general trend by looking at selected examples for \ba in the single-\sersic simulations. In Figs.~\ref{fig:trumpet_re_z} and \ref{fig:summary_redshift}, we show examples of the scatter- and summary-plots as a function of the true redshift of galaxies (rather than $\VIS$ magnitude, which was shown throughout the paper). We first see that most galaxies are at a redshift smaller than $2$ (see also Fig.~\ref{fig:param_distribution}). In the scatter plot, the dependence on the quality of (the exemplary) \ba prediction and the redshift is less clear than it was with magnitude, without the characteristic trumpet shape. Nevertheless, both the running bias and dispersion demonstrate that the dispersion is higher at higher redshifts. We detect clear trends in all codes in the three metrices, bias, dispersion and outlier fraction (Fig.~\ref{fig:summary_redshift}). We can also see that in general, the metrics get worse from redshift $z\simeq 0$ to $z\simeq2$, and then reach a plateau. 

In \partonedot, we detail in figure ~2 the distribution of redshifts as a function of $\VIS$. This further helps to link the papers presented here with redshift.

\section{Inspection of selected quality flags}\label{sec:flags}
Most participating software packages also include flags with additional information that allow the user to choose to reject objects based on certain criteria. In addition, some codes give more flags than others and therefore lead to more or less restrictive common catalogues. Since we are interested in relative fit qualities, we have decided to use a common catalogue that is as inclusive as possible, that is one that maximises the number of objects to test how the individual codes deal with the same set of difficult objects. This naturally affects the result on a code-by-code basis to some extent. In order to investigate the seriousness of this affect, we here test a number of selected flags.

\begin{figure}
 \includegraphics[width=1\linewidth]{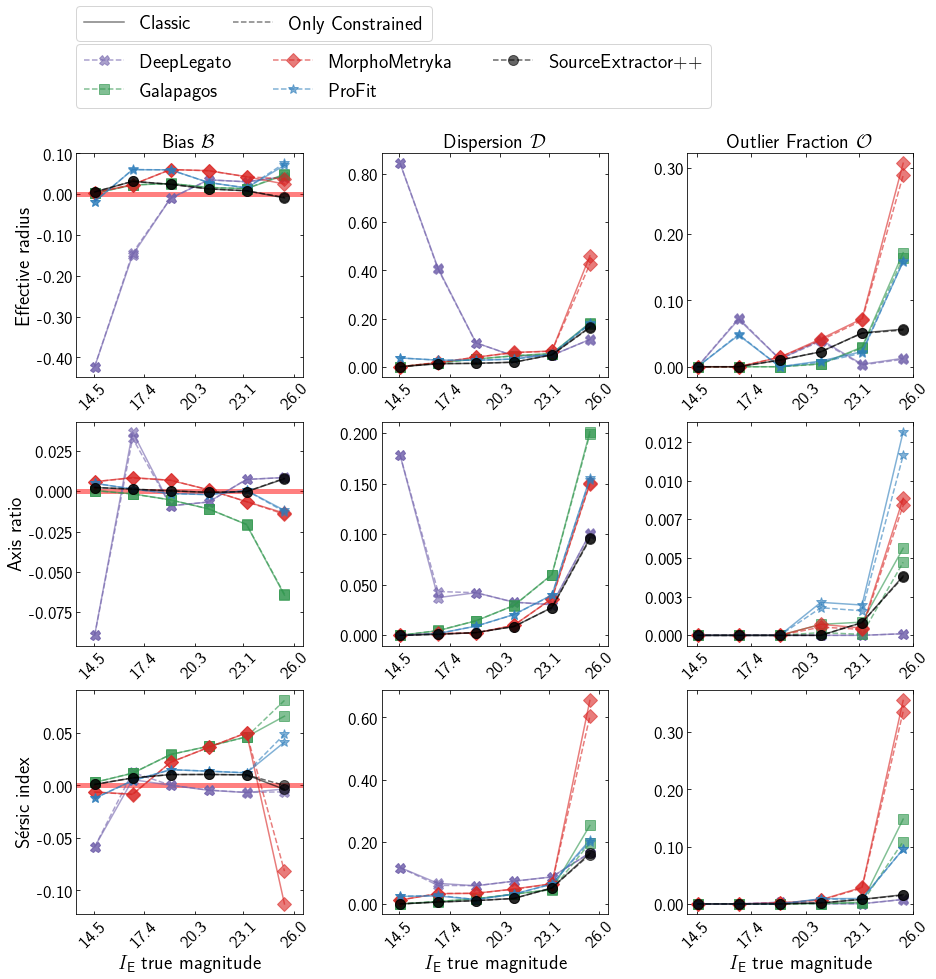}
 \caption{Summary plot comparing the fit on the single-\sersic simulations with or without considering \gala constraint flags.}\label{fig:gala_flag_ss}
\end{figure}

\begin{figure}
 \centering
 \includegraphics[width=1\linewidth]{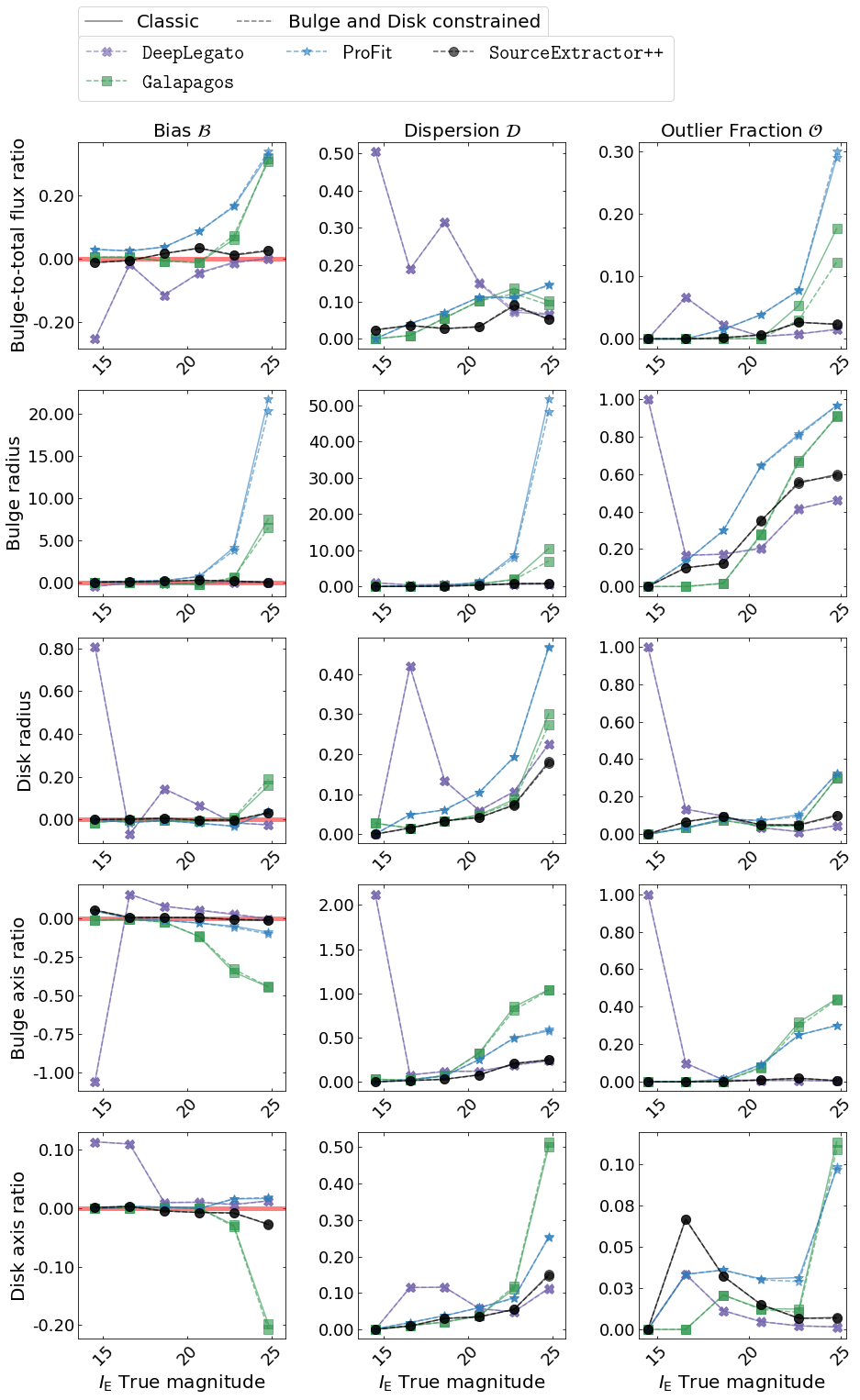}
 \caption{Summary plot comparing the fit on the double-\sersic simulations with or without considering \gala constraint flags.}\label{fig:gala_flag_ds}
\end{figure}

\gala returns five quality flags that link back to whether fits (of one or more components) encountered fitting constraints 
or indicate the relative brightness of the double-\sersic components (see \partone for a summary of all flags). These are thoroughly described and tested in \citet{haeussler2022}. Here, we test the impact of the constraint flags (\texttt{USE\_FLAG\_SS}, \texttt{USE\_FLAG\_BULGE\_CONSTR} and \texttt{USE\_FLAG\_DISK\_CONSTR}) on the summary figures. For the single-\sersic case, we remove all galaxies that are not constrained, which represent $~4\%$ of the common catalogue. These are mainly faint galaxies (Fig.~\ref{fig:gala_flag_ss}), improving the results by few percent in the last magnitude bin. Removing those galaxies also improves the results for \metryka and \profitdot, which indicates that all three codes returned unreliable measurements for these galaxies, but only \gala specified this with a flag. 
For the double-\sersic simulation, presented in Fig.~\ref{fig:gala_flag_ds}, we remove galaxies where the bulge or the disc fits were not constrained. This is the case for $13\%$ of the catalogue, mainly because the bulges were not constrained. This is a non negligible impact for the bulge components, with the impact greater for \gala than for other codes. This indicates that the same galaxies are successfully fit by some codes, and not by others; some of the codes (like \gala and \metrykadot) have additional quality flags in place to inform the user.

\begin{figure}[h!]
\centering
    \includegraphics[width=\linewidth]{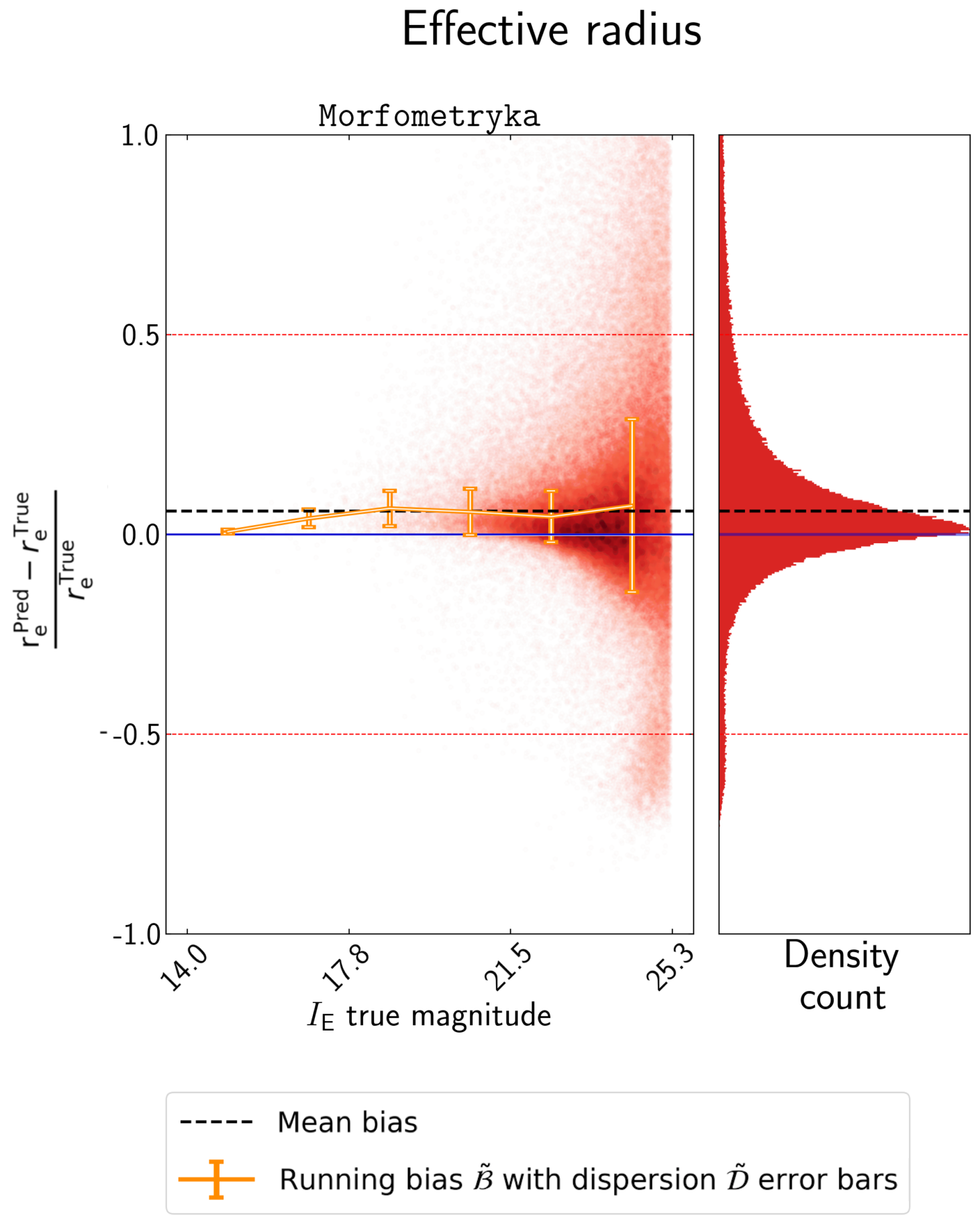}
    \caption{Trumpet plot for the single-\sersic effective radius fitting, removing the objects flagged as ``stars'' by \metrykadot. Wee see that the bi-modality saw in Fig.~\ref{fig:trumpet_ss_re} vanishes here.}
    \label{fig: trumpet_re_no_stars_flag}
\end{figure}

As mentioned in the text, we also studied the impact of the objects flagged as `\texttt{TARGETISSTAR}' by \metrykadot. We show in Fig.~\ref{fig: trumpet_re_no_stars_flag} the trumpet plot of the fitting of the single-\sersic effective radius fitting by \metrykadot, and conclude that taking into account this flag will remove the bi-modality, as explained Sect.~\ref{sec:ss}.

\section{Contact for reproducibility}
We provide information about the precise version of the software packages used for the EMC, along with the configuration files and \texttt{README}s, where available. These can be found in a folder in the same repository that we provide for the interactive plot website\footnote{\url{https://github.com/Hbretonniere/Euclid_Morphology_Challenge}}.

We also detail here the role of the different authors in the production of the paper to give the reader the possibility to contact for further details.
\smallbreak
The simulations were produced in a joint effort by E.~Merlin, M.~Castellano, D.~Tucillo, M.~Huertas-Company and H.~Bretonnière. Contact E.~Merlin for information about the single- and double-\sersic simulations, and H.~Bretonnière for the FVAE simulation.
\smallbreak
Regarding the use of the different codes presented in the challenge:
\begin{enumerate}
    \item \deepleg was run by D.~Tucillo. Contact M.~Huertas-Company for further information.
    \item \gala was run by B.~H{\"a}u{\ss}ler.
    \item \metryka was run by L.~Ferreira, F.Ferrari and F.G.~Lucatelli, with C.~Conselice. Contact L.~Ferreira for further information.
    \item \profit was run by A.S.G.~Robotham.
    \item \SE was run in a joint effort by A.~\'{A}lvarez-Ayll\'{o}n, E.~Bertin, P.~Dubath, R.~Gavazzi, W.~Hartley, D.~Hernandez Lang, M.K{ümmel}, and M.~Schefer. Contact W.~Hartley for further information.
\end{enumerate}

The analysis was conducted and the paper was written by H.~Bretonnière, U.~Kuchner and M. Huertas-Company with input from members of the Euclid morphology working group and challenge participants.

\end{appendix}
\end{document}